\documentclass{aa}              
\usepackage{graphics}
\usepackage{psfig}

\begin{document}

\newcommand{\apss}[2]{Ap\&SS #1, #2}
\newcommand{\aap}[2]{A\&A #1, #2}
\newcommand{\aaps}[2]{A\&A Suppl. #1, #2}
\newcommand{\aj}[2]{AJ #1, #2}
\newcommand{\apj}[2]{ApJ #1, #2}
\newcommand{\apjs}[2]{ApJS #1, #2}
\newcommand{\araa}[2]{A\&AR #1, #2}
\newcommand{\anrev}[2]{ARA\&A #1, #2}
\newcommand{\pasp}[2]{PASP #1, #2}
\newcommand{\mnras}[2]{MNRAS #1, #2}

\newcommand\beq{\begin{equation}}
\newcommand\eeq{\end{equation}}
\newcommand\beqa{\begin{eqnarray}}
\newcommand\eeqa{\end{eqnarray}}
\def\strutup{\rule[+1mm]{0.0cm}{3mm}} 
\def\strutdown{\rule[-2mm]{0.0cm}{3mm}} 
\newcommand{\al}{{et al.}}
\newcommand{\eps}{$\epsilon$}
\newcommand{\teff}{T$_{\mathrm{eff}}$}
\def\kms{km~s$^{-1}$}
\def\vsini{$v\sin i$}
\def\MSun{M$_{\odot}$}

\title{Detached double-lined eclipsing binaries as critical tests of stellar evolution}
\subtitle{Age and metallicity determinations from the HR diagram}

\author{E. Lastennet\inst{1} \and D. Valls--Gabaud\inst{2} \, }

\institute{Observat\'orio Astron\'omico da Universidade de Coimbra, Santa Clara,
           P-3040 Coimbra, Portugal
           \and 
	UMR CNRS 5572, Laboratoire d'Astrophysique, 
           Observatoire Midi-Pyr\'en\'ees, 
           14, Avenue Edouard Belin, 
           31400 Toulouse, France
          }

\offprints{dvg@ast.obs-mip.fr}

\date{Received \rule{2.0cm}{0.01cm} / Accepted \rule{2.0cm}{0.01cm} }

\authorrunning{E. Lastennet \& D. Valls--Gabaud}
\titlerunning{Detached binaries in the HR diagram}

\abstract{ 
Detached, double-lined spectroscopic binaries which are also eclipsing 
provide the most accurate determinations of stellar mass, radius, 
temperature and  distance-independent luminosity for each of their  
individual components, and hence constitute a stringent test of single-star 
stellar evolution theory. 
We compile a large sample of 60 non interacting, well-detached systems mostly 
with typical errors 
smaller than 2\% for mass and radius and smaller than 5\% for effective temperature, 
and compare them with the properties predicted by stellar evolutionary
tracks from a minimization method.  
To assess the systematic errors introduced  by a given set of tracks, we 
compare the results obtained using three widely-used independent sets of
tracks, computed with different physical ingredients (the Geneva, Padova and
Granada models).  
We also test the hypothesis that the components of these systems are coeval and  
have the same metallicity, and compare the derived ages and metallicities with 
the ones obtained by fitting a single isochrone to the system. Overall, there
is a good agreement among the different determinations, and we provide a  
comprehensive discussion on the sub-sample of systems which either
present problems or have estimated metallicities.   
Although within the errors the published tracks can fit most of the systems, 
a large degeneracy between age and metallicity remains. 
The power of the test is thus limited because the metallicities of most of the 
systems are unknown. 
\keywords{stars:  evolution 
       -- stars:  fundamental parameters 
       -- stars:  binaries  
       -- binaries: eclipsing
       -- stars:  HR diagram  
       -- Galaxy: evolution 
       -- stars:  statistics 
}
}

\maketitle

\section{Introduction}

The theory of stellar evolution is one of the most successful in astrophysics, 
and has been widely applied to phenomena ranging from stellar oscillations 
to the evolution of galaxies. 
While open and globular clusters have been the classical tests of the 
theory because their colour-magnitude diagrams (CMD) constrain the shape 
of an isochrone in detail, they are not ideal probes, since, among other 
things, they are contaminated by field stars, have unresolved
binaries, 
suffer dynamical evolution, 
may present differential extinction, and may be distant enough to prevent 
very accurate determinations of their distances (see Andersen \&  Nordstr\"{o}m 1999 
for a recent review on the observational requirements for 
 open clusters). 
Furthermore, Hipparcos measures show that there is a very fuzzy  
correlation, if any, between the main sequence position and metallicity 
(van Leeuwen 1999, Lebreton et al. 1999, and references therein), casting some doubts --at
least as far as the metallicity scaling is concerned-- on current stellar evolution models.

So far relatively less attention has been paid, in comparison, to binary 
systems. Yet these systems provide the only way to measure directly 
stellar masses, and, if the components do not interact strongly and suffer 
mass transfer, they should have independent evolution and be representative 
of single stars. 
Visual systems are hampered by the lack of radial velocity measures 
(but see Pourbaix 2000), and
hence mass ratios are measured indirectly (e.g. Fernandes {\al} 1998 for
a few examples). Interferometric systems, although promising, are too few
to provide a good sample (e.g. Hummel {\al} 1995, 2001, Torres {\al} 2002).
In this context, double-lined eclipsing binaries provide the most accurate  
determinations of stellar radii, masses and temperatures, and hence are ideal
stringent tests of the theory, as first pointed out by Str\"omgren (1967) (see
also Andersen 1997). 
Testing evolutionary tracks is important not only for the understanding of 
the properties of the stars, but also to assess the uncertainties in, for 
example, the interpretation of the spectro-photometric properties of galaxies, 
from low to high redshifts, or the chemical evolution of our Galaxy. \\
\indent Although only a handful of systems are extremely well-measured (see 
below), with typical errors in masses and radii of less than 2 per cent (Andersen 
1991), the sample of systems is bound to increase in the near future, 
with dedicated observing programs (at the Danish 50cm SAT, see Clausen, 
Helt \& Olsen 2001; Elodie at OHP, e.g. Kurpinska-Winiarska \& Oblak 2000), 
the analysis of Hipparcos data (e.g., Fabricius et al. 2002), 
the systematic detections of thousands of systems 
as a by-product of microlensing surveys (see, e.g., Alcock {\al} 1997, 
and Ferlet, Maillard \& Raban 1997 for an overview), and 
with future space missions such as DIVA, SIM and GAIA. 
In addition, these systems have accurate absolute dimensions and provide another 
distance indicator, independent of any distance calibration (e.g. Paczy\'nski 1997, 
Fitzpatrick {\al} 2002).  
However, in practice, the power of the test using these systems is limited by the 
small volume in parameter space sampled by most of the published theoretical tracks. 
For instance, many tracks adopt a single helium abundance $Y$ for a given metallicity 
$Z$, and do not consider any possible dispersion, while others fix the overshooting 
parameter or the mixing length parameter to prescribed values. Even though 
these values provide good fits to the colour magnitude diagrams of clusters,
it is likely that these parameters change with  stellar mass or 
metal abundance. The extant (and growing) sample
of binaries will yield extremely important constraints on these parameters
(and their possible correlations) when tracks with a wider range of properties
will be available.  \\
\indent The aims of this paper are three-fold. First, we want to test
whether current evolutionary tracks can account for the properties of
these extremely well measured systems. This is a requisite before
using them for further inferences.
Even though a given set of tracks may seemingly give good results, and provide 
good fits, it will not produce an absolute determination because of
the underlying hypotheses used in the computation of the tracks.
For instance the determination
of the Helium abundance relies critically on the effective temperature
scale, and so different tracks, using different scales,
 may yield different abundances or ages. 
To assess the possible systematic effects
introduced by a given set, we
take three widely-used sets of tracks which have been computed independently
and with different physical ingredients. For the same data set, which
includes only the best measured stars currently available, are these sets
consistent with each other? 

Secondly, we want to test to which extent stellar evolution is a predictive
theory, in the sense that given some observables it can predict the remaining
ones, if the predictions are unique. This relies on the 
so-called Vogt-Russell theorem\footnote{We did not 
find any convincing reason to add Russell's name to
Vogt's theorem, given that  even though Russell may had derived it independently
of Vogt, he only mentioned it, briefly, in his textbook published in 1927 
(Russell, Dugan \& Stewart 1927) where  full credit is given to Vogt.} 
(Vogt 1926) which states
that stars of the same chemical composition (and hence same opacity and energy
generation rate) have radii, effective temperatures and luminosities solely
determined by their masses. Although this theorem has been widely criticized,
mainly because counterexamples were found (albeit in rather contrived situations, 
see e.g. Lauterborn 1972, 1973; K\"ahler 1978), 
it remains the basis of stellar evolution theory  
and is {\sl implicitely} assumed in  all analyses of stellar evolution data. 
As far as we know, no tests have been carried out to assess {\sl empirically} 
the validity of this theorem. So, given, say, a luminosity and effective 
temperature, can the tracks provide a unique value for the mass and the radius?
Are these values consistent with the measured ones? Allende Prieto \& Lambert (1999)
found that one set of tracks was able to reproduce, assuming solar metallicity,
radii and masses with errors smaller than 8\%, and effective temperatures
within 2\% for mainly main sequence stars. In a different study, Young et al. (2001)
found agreement with the measured radii within 3\%, and temperatures
within 4\% but luminosities could only be reproduced to 11\% with their
set of tracks. Are these figures the same for
other published tracks or for stars with lower gravities?

Furthermore, and this is the third and main goal of the paper, can these sets of tracks
predict the ages and metallicities of these systems? 
Proper testing of the accuracy of stellar dating has not yet been done before 
in detail (see e.g. von Hippel {\al}, 2001 for a summary). 
The components of these detached 
systems provide an ideal sample not only to properly calibrate the stellar
tracks, but also to assess the uncertainties in the stellar ages used for
the study of the evolution of our Galaxy. For instance  Edvardsson {\al} (1993)
used a set of tracks from VandenBerg (1985) with a fixed set of physical
parameters, and succeeded in reaching an accuracy of about 0.1 dex in the
relative dating of F dwarfs. We want to assess whether this accuracy is robust 
to changes in the evolutionary tracks. In this context it is curious to observe
that there has been no  systematic test to see whether the components of these systems
are indeed coeval. Popper (1997) observed that some secondary stars between 0.7 and 1.1
$M_\odot$ appeared systematically older (by at least a factor of two)
than their primaries. Are the ages derived for each component consistent with
them being on a single isochrone? Are their metallicities the same? 
What are the uncertainties (including systematics) in the absolute ages? \\

\indent This paper presents the sample, the method and the detailed fits, 
discusses the implications of the results on the coevality 
of the components, the ages and metallicities inferred, and their relation to 
the chemical evolution of the Galaxy. 
When relevant we also present comparisons with the results obtained by Pols 
{\al} (1997), Ribas {\al} (2000), and Young {\al} (2001), hereafter referred 
to as P97, R00 and Y01 respectively. 
 Section \S2 briefly describes the physical
properties of the tracks  used in the analysis of the data, which are
presented in Section \S3. Section \S4 deals with specific issues related to
some interesting binary systems, and Sect \S5 discusses the general results 
and provide new photometric [Fe/H] constraints. 
The conclusions are  presented in Section \S6.

\section{Evolutionary tracks}
\label{section:tracks}
Ideally tracks with a wide range of physical parameters (not only $Y$ and $Z$ but 
also overshooting and convection parameters) should be used to properly
test the different correlations between the parameters and the inferred ages and
metallicities of the systems, so that proper statistical uncertainties can be assessed. 
Previous analyses (e.g. P97, Allende Prieto \& Lambert 1999, R00, Y01) have
relied on a given set, so cross-comparisons and systematics are difficult to assess. 
It is unfortunate that there is presently no published set of tracks that fully 
fulfill this condition\footnote{ 
Grids of stellar evolutionary models for various values of initial helium 
abundances are available: Claret 1995, Claret \& Gim\'enez 1995, Claret 1997
and  Claret \& Gim\'enez 1998 (referred to as CG models) 
for metallicities $Z$$=$0.004, 0.01, 0.02 and 0.03. However, overshooting and convection 
parameters are fixed in these models.} 
and we are limited to use different sets. 
In this study, we  adopted three widely-used tracks computed by (1) the Geneva group
(Schaller {\al} 1992,  Schaerer {\al} 1993ab, Charbonnel {\al} 1993,  Mowlavi {\al} 1998) 
which will be referred to collectively as the  Geneva tracks, (2) 
 the Padova group (Bressan {\al} 1993, Fagotto {\al} 1994abc)\footnote{
We do not use the more recent Padova tracks (Girardi {\al}, 2000) because they 
do not include models for stars more massive  than 7 {\MSun} and $Z$ larger 
than 0.03, both relevant here.}, and (3) the Granada group (Claret \& Gim\'enez 1992, 
hereafter CG92) evolutionary stellar models for a smaller range in  metallicity 
than the Geneva and Padova ones. 
Table \ref{tab:tracksummary} summarizes the main physical ingredients that 
characterize these different tracks. 
We wish to emphasize that our main concern here is to present and validate a 
general method whichever the models used, but not to test all the available  
(or the most recent) tracks: we only choose three different sets in order 
to understand and quantify possible systematic effects introduced by a
given set. Kovaleva (2001) did a similar analysis using a subset of 43
systems
and the Geneva and Padova tracks.
We also describe in Table \ref{tab:tracksummary} the older set from Hejlesen 
(1980ab, H80) because these tracks were often used previously for some of the 
systems studied here. They have a large mixing length theory (MLT) parameter ($\alpha$= 2.0), 
do not include overshooting or mass loss, and use the outdated Cox \& Stewart (1970ab) opacities. 
For all these reasons we will only mention results derived from the H80
models as an historical guideline, because there is no reason to probe their
validity (see comments on H80 related problems in e.g. Andersen {\al} 1984).  
Some important points to note are the following :

\begin{table*}[ht]
\caption[]{Basic properties of the Geneva, Padova, Granada and H80 tracks 
\label{tab:tracksummary}}
\begin{flushleft}
\begin{center}
\begin{tabular}{lcclcclcclcc}
\hline\noalign{\smallskip}
  & \multicolumn{3}{c}{Geneva} &  \multicolumn{3}{c}{Padova} &  
\multicolumn{3}{c}{Granada} & \multicolumn{2}{c}{H80 $^a$}\\ 
\noalign{\smallskip}
\hline\noalign{\smallskip}
  &    $Y$      &     $Z$     &   \multicolumn{1}{c}{ Ref. } 
  &    $Y$      &     $Z$     &   \multicolumn{1}{c}{ Ref. } 
  &    $Y$      &     $Z$     &   \multicolumn{1}{c}{ Ref. } 
  &    $Y$      &     $Z$      \\ 
\noalign{\smallskip}
\noalign{\smallskip}
\hline\noalign{\smallskip}
 &         &           &           &   0.230   &   0.0004   &      P94a  &           &          &         &  0.2996   &  0.0004   \\ 
 & 0.243   &   0.0010  &  G92      &           &            &            &           &          &         &           &           \\ 
 & 0.252   &   0.0040  &  G93b     &   0.240   &   0.0040   &      P94b  &           &          &         &  0.296    &  0.0040   \\
 &         &           &           &           &            &            &           &          &         &  0.196    &  0.0040   \\
 & 0.264   &   0.0080  &  G93a     &   0.250   &   0.0080   &      P94b  &           &          &         &           &           \\
 &         &           &           &           &            &            &   0.267   &  0.0100  &  CG92   &  0.290    &  0.0100   \\
Chemical    & 0.300    &   0.0200  &  G92      &   0.280    &   0.0200   &   P93  &  0.280   &  0.0200 &  CG92     &  0.380    &  0.0200  \\ 
composition &          &           &           &            &            &           &          &         &           &  0.280    &  0.0200  \\
 &         &           &           &           &            &            &           &          &         &  0.180    &  0.0200   \\
 &         &           &           &           &            &            &   0.321   &  0.0300  &  CG92 &  0.270    &  0.0300   \\
 & 0.340   &   0.0400  &  G93c     &           &            &            &           &          &         &  0.360    &  0.0400   \\ 
 &         &           &           &           &            &            &           &          &         &  0.260    &  0.0400   \\
 &         &           &           &   0.352   &   0.0500   &   P94a  &           &          &         &           &           \\ 
 & 0.480   &   0.1000  &  G98      &   0.475   &   0.1000   &   P94c  &           &          &         &           &           \\ 
\noalign{\smallskip}
\hline
\noalign{\smallskip}
 Opacities & \multicolumn{3}{c}{ RI92, K91, IRW92, } &
\multicolumn{3}{c}{ IRW92, H77, CS70 $^c$} &
\multicolumn{3}{c}{ RI92, A92 ($=$AF94), } &
\multicolumn{2}{c}{ CS70}  \\
            & \multicolumn{3}{c}{ IR96, AF94 $^b$} &
\multicolumn{3}{c}{} &
\multicolumn{3}{c}{H77, W90 $^d$} &
\multicolumn{2}{c}{ }  \\
\noalign{\smallskip}
\hline
\noalign{\smallskip}
MLT $\alpha$ & 
\multicolumn{3}{c}{1.6} &
\multicolumn{3}{c}{1.63 [1.5 for $Z$$=$0.1]} &
\multicolumn{3}{c}{1.5} &
\multicolumn{2}{c}{2.0}  \\
\noalign{\smallskip}
\hline
\noalign{\smallskip}
$\Lambda_c$ & 
\multicolumn{3}{c}{0.2} &
\multicolumn{3}{c}{0.2} &
\multicolumn{3}{c}{0.2} &
\multicolumn{2}{c}{none}  \\
\noalign{\smallskip}
\hline
\noalign{\smallskip}
Mass loss & 
\multicolumn{3}{c}{yes} &
\multicolumn{3}{c}{yes} &
\multicolumn{3}{c}{yes} &
\multicolumn{2}{c}{none}  \\
\noalign{\smallskip}
\hline
\noalign{\smallskip}
\noalign{\smallskip}
\end{tabular}
\end{center}
$^a$
Chemical composition ($Y$, $Z$) from Hejlesen (1980b). \\
$^b$
For $Z$ = 0.001, 0.008 and 0.020, opacities are from Rogers \& Iglesias 1992 (RI92). 
At low temperatures, from 6000 K to 2100 K, they have completed the tables of RI92 with the 
atomic and molecular opacities of Kurucz 1991 (K91). For $Z$ = 0.004 and 0.040, the OPAL 
radiative opacities (Iglesias {\al} 1992, IRW92) are used. These tables are also completed 
at low temperatures with K91. For $Z$ = 0.1, opacities are from Iglesias \& Rogers 1996 (IR96), 
along with the opacities from Alexander \& Ferguson 1994 (AF94) at low temperatures.\\
$^c$
Because the OPAL tables of Iglesias {\al} 1992 (IRW92) do not extend below 6000 K and 
above $10^8$ K, the opacities at lower and higher temperature, respectively, are taken from the 
LAOL tabulations by Huebner {\al} 1977 (H77) and Cox \& Stewart 1970a,b (CS70). In general, 
the OPAL are significantly higher than the LAOL in two temperature ranges (a few $10^5$ K and 
$\sim$ $10^6$ K).\\
$^d$
Opacities are taken from RI92, but for {\teff} $<$ 6000 K or {\teff} $> 10^8$ K they have 
used Alexander 1992 (A92, published later as Alexander \& Ferguson 1994,  
AF94) opacities and {\it Los Alamos Opacity Library}  -- LAOL -- (Huebner {\al} 1977 [H77], 
Weiss {\al} 1990 [W90]) respectively. \\
\end{flushleft}
\end{table*}

\begin{itemize}  
\item Masses. For the 6 metallicities available in the Geneva tracks, the mass range is from 0.8 
to 120 \MSun\ (0.8 to 60 \MSun\ for $Z =$ 0.1), while for the 6 metallicities in 
the Padova tracks the mass range is 0.6 to 120 $M_{\odot}$ 
(restricted to the range 0.6 to 9 \MSun\ for $Z =$ 0.1). 
The CG92 grids cover  masses from 1.0 to 40 \MSun\ with 3 different metallicities. The CG92 upper 
mass limit is large enough for the purpose of this work because the most massive component is a 
32 \MSun\ star (namely \object{DH Cep}).  

\item Mass loss. The Geneva tracks use the mass loss rates given by De Jager {\al} (1988)  and
Nieuwenhuijzen \& De Jager (1990) for Pop. I stars. The Padova tracks set a mass loss rate only 
for stars more massive than 12 \MSun\  (from De Jager {\al} 1988) with a metallicity 
dependence given by the prescription of Kudritzki {\al} (1989). 
No mass loss by stellar winds is taken into account for the $Z=$ 0.1 Padova tracks. 
Mass loss is also included in the CG92 tracks following Nieuwenhuijzen \& De Jager (1990) 
and Reimers (1977) for cool giant stars.
 
\item Nuclear rates. The main difference between the Geneva and Padova tracks is the  
$^{12}C(\alpha,\gamma)^{16}O$ reaction which occurs during the helium burning phase, 
causing differences in the blue loop extension in the HR diagram. 
The Padova tracks include the nuclear reaction rates of Caughlan \& Fowler (1988)
which are somewhat smaller than the rates used by the Geneva group (Caughlan {\al} 1985).

\item Overshooting. Although the nominal overshooting  parameter $\Lambda_{c}$ of the Padova 
tracks is 0.5,  this is just due to a different definition, and, as Fagotto {\al} (1994a) 
explain, this value is equivalent to  an overshooting distance of $\Lambda_{c}$(Geneva) 
$=$ 0.2 pressure scale heights.  
Claret \& Gim\'enez (1992) have also adopted $\Lambda_{c}$ $=$ 0.2.  
To test the influence of overshooting on the results we also use Geneva tracks without 
overshooting.   
However these models 
cover only the mass range [0.8--1.25] \MSun\ and hence limit the application
of the test to the few systems in this mass range. 


\end{itemize}

\subsection{Ages}
One important definition which is not only semantic is the age of a star. Although some
prescriptions use the start of the collapse of the parent cloud, we adopt here the convention 
of assigning a `zero age' to the main sequence. The reason for this is twofold. First, the 
collapse phase is still poorly understood, and accretion processes before or during the 
Hayashi phase could well modify the time of arrival on the main sequence. Second, it
seems simpler to follow the traditional definition of a ZAMS, a 'Zero-Age'
Main Sequence precisely when the radius of the star reaches a minimum
value\footnote{The usual prescription based on the H burning does not
apply for instance to massive stars, see e.g. Bernasconi \& Maeder (1996).}. 
To avoid any possible confusion, we hence define an `age' $\tau$ as the time elapsed 
{\it after} reaching the ZAMS, in logarithmic units :
\beq
\tau = \log (t - t_{\rm ZAMS})
\label{eq:age}
\eeq
where $t$ is the actual (unknown) age and $t_{\rm ZAMS}$ the age when the star reaches 
the ZAMS (also unknown in absolute terms), both measured in years.
 We stress again that the actual age is 
meaningless unless a well defined  reference epoch is agreed on, and all the 
physical processes along the pre-MS evolution are fully understood and quantified.


 The Padova and CG92 tracks do follow the convention of adopting  $t_{\rm ZAMS} \equiv $ 0, 
while the Geneva tracks do not. 
The dependence with mass and metallicity of the ratio $t_{\rm ZAMS}/t_{\rm TAMS}$ 
(where TAMS is the Terminal Age Main Sequence i.e. central hydrogen exhaustion) for 
these tracks shows that for all practical purposes the non zero $t_{\rm ZAMS}$ constitutes a small 
correction of about 2\% for evolved systems, but that is important for the youngest systems.
All the ages derived 
in the present work are post-ZAMS, i.e. we explicitely substract $t_{\rm ZAMS}$ while dealing 
with the Geneva tracks. The H80 tracks seem to set  $t_{\rm ZAMS}=0$ for all masses.

\subsection{Metallicities}
\label{section:Z} 
The set of tracks used in this paper do not have helium abundances $Y$ independent 
of the metallicity $Z$ (except those from H80), but rather follow the usual 
prescription of increasing $Y$ for larger values of $Z$, roughly like
$Y = 0.23 + (3.0 \pm 1.0) Z$ (see e.g., Timmes {\al} 1995). As shown in 
Table~\ref{tab:tracksummary},  
the Geneva, Padova and CG92 tracks adopt
different enrichment prescriptions. This means that the metallicity $Z$ 
derived from a given set cannot strictly be compared with the $Z$ derived
from another set, since the different helium abundance will effectively 
mimic a slightly different $Z$. This effect has to be kept in mind when
comparing the derived metallicities from the fits.


Usually, a simple relation is adopted between $Z$ and [Fe/H]:  
[Fe/H]$=$ log($Z$/$X$)$-$log($Z_{\odot}$/$X_{\odot}$)
(where we adopt solar abundance from Grevesse, 1997, priv. comm.: 
$Z_{\odot}$$=$0.017 and $X_{\odot}$$=$0.713). 
However for low metallicity stars, the abundance of heavy elements cannot be 
simply derived from spectroscopic measurements of [Fe/H] because Oxygen (O)
and $\alpha$ elements (Ca, Ne, Mg, S, Si) are overabundant in comparison 
to Iron (Fe). Therefore, because we may have to consider low metallicities  
for some stars in our sample, throughout this paper we consider the 
correction proposed by Pagel (1996),  
\beq 
Z_{\rm corrected} = Z \, \times \left[ 0.638 f_{\alpha} + 0.362 \right]
\label{eqn:fit2pagel}
\eeq
where f$_{\alpha}$ is a correction factor for O and the $\alpha$ elements:  
\beq
f_{\alpha} = \left\{ \begin{array}{lcrrl}
10^{0.4}                    & {\rm for} &        & {\rm [Fe/H]} & \leq -1  \\
10^{0.4 \times {\rm [Fe/H]}^{2} } & {\rm for} & 
-1 < & {\rm [Fe/H]} & < + 0  \\
1                         & {\rm for} &        & {\rm [Fe/H]} & \geq + 0  \\
\end{array} \right.
\eeq                                                               

\section{Methodology}

\subsection{The sample}
\label{section:sample} 
We have compiled a comprehensive list of 60 double-lined eclipsing binaries (EBs) 
which have the most accurate masses, radii and temperatures found in the literature. 
The core of the sample is the large catalogue from Andersen (1991, hereafter A91), 
which is supplemented with 11 systems\footnote{
 \object{RT And} (Popper 1994), 
 \object{AD Boo} (Lacy 1997b), 
 \object{SW CMa} (Lacy 1997c), 
 \object{AH Cep} (Holmgren {\al} 1990),
 \object{CG Cyg} (Popper 1994),
 \object{Y Cyg} (Simon, Sturm \& Fielder 1994), 
 \object{CM Dra} (Viti {\al} 1997),  
 \object{AG Per} (Gim\'enez \& Clausen 1994),
 \object{V505 Per} (Marschall {\al} 1997), 
 \object{V3903 Sgr} (Vaz {\al} 1993, Vaz {\al} 1997), 
 \object{V526 Sgr} and \object{YY Sgr} (Lacy 1997a).}. 
 We also add three other systems with slightly lower  accuracy 
criteria\footnote{  
 \object{AR Aur} (Nordstr\"om \& Johansen 1994a, mass accuracy about 4\%), 
 \object{V477 Cyg} (Gim\'enez \& Quintana 1992, mass accuracy: M$_A$$\simeq$ 7\% 
and M$_B$$\simeq$ 4\%), and 
 \object{DH Cep} (Hilditch {\al} 1996, mass accuracy about 5\%).}  
which can be used for {\it a posteriori} testing. 
A few systems have more recent and accurate measures than those
listed by A91.  We have used the updated parameters for 
5 systems and for the EW Ori secondary component\footnote{
 \object{V539 Ara}      (Clausen 1996), 
 \object{$\beta$ Aur}   (Nordstr\"om \& Johansen 1994b), 
 \object{GG Lup}    (Andersen {\al} 1993),
 \object{EW Ori} B (Popper 1997), 
 \object{AI Phe}   (Milone {\al} 1992), 
 and 
 \object{DM Vir}   (Latham {\al} 1996).}.  
It is worth noting that we kept all the systems from the Andersen list 
even EK Cep 
whose secondary component is in a pre-MS phase (which is not included 
in the stellar tracks that we used) because our method presented in 
\S\ref{section:method} can provide interesting results from the primary alone.   

\begin{figure}[htb]
\psfig{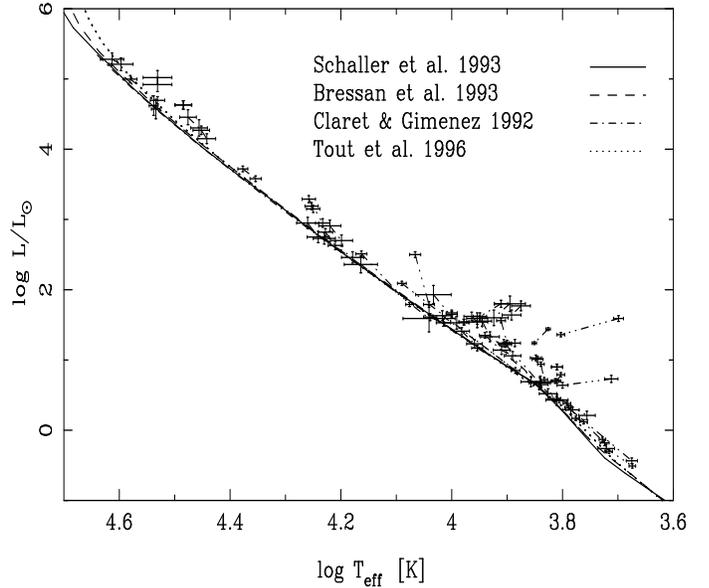}
\caption{HR diagram for the sample of 58 binary systems with
components more massive than 0.6 \MSun.  ZAMS sequences computed by 
different authors for $Z=$0.02 are also indicated for comparison. }
\label{fig:hrdtotal}
\end{figure}

One possible concern is that these systems may have undergone mass transfer
episodes in the past. Although in some special cases strong constraints
can be placed on the evolution of the system (e.g. Daems, Waelkens \& Mayor 1997),
in general this confuses the issue. However in our sample this is unlikely to be 
the case :
(1) We have computed the Roche radii of the components and checked that they
were much larger than the stellar radii (Lastennet 1998, from the formula of 
Eggleton 1983), and 
(2) most of the systems are  still between the ZAMS and the TAMS,  
and their radii should not have evolved significantly.
 The hypothesis of independent evolution without phases of mass transfer 
seems acceptable for every system in the sample, except 
perhaps for \object{DH Cep}.  
This system is an ellipsoidal variable, and the lack of eclipses makes any 
determination of the inclination very uncertain. 
Although Hilditch {\al} (1996) and Penny {\al} (1997) find that the system is 
detached, Burkholder {\al} (1997) argue that the radii of the components are, 
according to their solution, very close to the Roche values.
The uncertainties in the inclination of the orbit makes this system
much less constraining.

Figure \ref{fig:hrdtotal} shows the distribution of the sample on the
HR diagram, where it is readily apparent that the systems are mostly
sampling the evolutionary phases close to the main sequence. Also indicated
are the ZAMS computed by different authors at $Z=0.02$, and the different prescriptions 
for overshooting and mass loss result in slightly different sequences for
massive stars. Note that these sequences are very different (by about 0.05
dex in $T_{\rm eff}$ and 0.5 dex in $L/L_\odot$ in the range of 
$\log T_{\rm eff}$ from 4.0 to about 4.6) from the widely used sequence 
tabulated by Schmidt-Kaler (1982), which is no longer appropriate (Lastennet 1998).

Hereafter, each binary is associated with a number between 
square brackets (e.g. $\beta$ Aur [27]) corresponding to its entry 
in all the tables and figures throughout this paper. 

\subsection{Effective temperatures and luminosities} 
\label{section:temperature}
The individual stars in the sample have typical relative errors smaller 
than 5\% for the effective temperature\footnote{Except in 6 less constraining 
systems where $\sigma$({\teff)}/{\teff} is between 5 and 9.6\%: V1031 Ori [28], 
V451 Oph [31], GG Lup B [36], EM Car [45], DH Cep [53] and SW CMa [58].} 
and relative errors smaller than 10\% for the luminosity. 
The interest of the luminosity is that it follows directly from the 
temperature and radius (Stefan's law), and hence is independent of distance. 
Since analyses of these systems combine spectroscopy and 
photometry, the {\teff}s are likely to be more reliable than those 
in most of studies of single field stars. 

For these reasons, these 2 parameters were used directly from the compilation 
of Andersen (1991) by different authors (e.g. Lastennet, Lejeune  
\& Valls-Gabaud 1996, or Pols {\al} 1997). 
However, because {\teff}s (and therefore luminosities) are far more 
indirect and inhomogeneous quantities among the fundamental parameters, 
some efforts have been done to revise their determinations from 
photometric methods (e.g. Jordi {\al} 1997, Lastennet {\al} 1999a,  
Ribas {\al} 2000, and Lastennet, Cuisinier \& Lejeune 2002).  

We decided however to keep the original {\teff} values from the references 
given in \S\ref{section:sample}  for simplicity because a comparison of the {\teff}s 
used by Ribas {\al} (2000) and the ones used in this paper show a very 
close agreement for the 41 systems in common ($\sim$66\% of our sample), 
the great majority (60 out of 82 stars) showing a difference below 2\%. 
The {\teff} difference is less than 4\% for all the components, 
except for 11 stars: both components of PV Cas [32],  
RS Cha [14], AI Hya [20], QX Car [42] and IQ Per [33], and GZ CMa A [23].  
Since our results are not reliable for the  first two (bad fit for [32]
and  pre-MS stars for [14]) the {\teff} disagreement has no influence on 
our conclusions, however the results derived for the latter systems ([20], [23], 
[33] and [42]) have to be taken with caution because the discrepancy in 
{\teff} would affect the luminosity as well, giving slightly different solutions. 
As suggested by Lastennet {\al} (1999a, hereafter L99a), homogeneous determinations 
are highly preferred and to pick up some {\teff}s in R00  
and others in A91 and L99a  
would probably confuse even more the issue. Therefore for the purpose of this
work and before a  reliable homogeneous {\teff} determination for {\it all the
systems of  the sample}, we decide to keep the original {\teff}s  (see
\S\ref{section:sample}). 

\subsection{Fitting isochrones to binary systems}
\label{section:method}
There has been a variety of methods dealing with the problem of
fitting isochrones to a given system. 
Besides the early attempts by Lastennet, Lejeune \& Valls-Gabaud (1996), 
Pols {\al} (1997) assumed that both components
are coeval and of the same metallicity and proceeded to minimize a
$\chi^2$ functional which takes the observed masses, radii and temperatures 
and infers the best fit masses, common age and metallicity. 
Ribas {\al} (2000) compute isochrones for each component, fixing each mass 
and gravity to the observed values, and proceed to compute a 
$\chi^2$ which minimizes the weighted difference
between observed and predicted temperatures, as well as the difference
between the predicted ages of each component. Because the masses are
fixed
to the observed values, the $\chi^2$ depends only on two remaining
parameters,
metallicity and helium content. In a recent paper, Young {\al} (2001)
use
yet another formulation. First, they fix the metallicity to the solar
value, and run specific evolutionary tracks of stars whose masses are 
given by the observations. Their $\chi^2$ functional minimizes the
(sum of the squares of) residuals of luminosities and radii. Coevality
is assumed, so that the  $\chi^2$ depends only on age. 

A common property of these formulations is that it is not clear 
whether the observed quantities are used to their best discriminant
power. Also the intrinsic correlations between these quantities
are not taken into account, and yet the very method of deriving these
values, by relying on both photometric light curves and spectroscopic
analyses, produces correlations among them. Another problem is that,
for instance, masses are fixed (except in P97) to the mean
observed values, so that even the observed dispersions in mass are not taken 
into account. Obviously
the dispersion in mass should have some effect on the final fits, but
this is not considered (except in the P97 formulation).
Although one could argue that using $M$, $R$ and $T_{\rm eff}$ (as P97 do)
is the best way to proceed (no derived --and hence less precise--
quantities, such as luminosity or gravity, are used), P97 must assume
coevality and same chemical composition. Stellar evolution theory is
not therefore used to predict {\it a priori} values for each component. 
There cannot be possible discrepancies (which may well be the case, but 
it cannot be tested by this method). Likewise, 
R00 minimize the age difference between components, and derive a
global metallicity and helium abundance for the systems. 

In a first formulation,  we want then to test whether ages and
metallicities
can de derived {\it for each component independently}. As a first
step, we shall use here only
 the temperatures and
luminosities of each of the components of the system. As it will be
shown later
on (see section \S\ref{section:comparison}) this does not make
a difference in the final results. 
Note that because luminosity and temperature are not derived
independently, the error bars in the HR diagram are highly correlated
and strictly the error ellipse is always rotated (see e.g. 
Casey {\al} 1998, for a brief discussion).  
To test the hypothesis that each component $i$ ($i=A,B$) of a given
system
 has the same age and metallicity, we derive them by minimizing a
 functional defined by 
\beqa
\chi^2 (M_i, t_i, Z_i)  & = &  \left(\frac{\log L(track) - \log
L_i}{\sigma(\log L_i)}\right)^2   \nonumber \\
 & & +  \left(\frac{\log T_{\rm eff}(track) - \log T_{{\rm eff } i}}{\sigma(\log T_{{\rm eff }  i})}\right)^2  
\eeqa
where the $\sigma$ are the 68\% confidence level dispersions.
The 3 sets of evolutionary tracks described in \S2 are used to compute the
isochrones by interpolating bilinearly at equal evolutionary phases 
in metallicity and age, and hence derive the values of $L(track)$ and $T_{\rm eff}(track)$.

Because there are 3 parameters and only 2 data values, the problem is
degenerate and one could find an infinity of solutions that fit the
data. Among them we seek the minimal values using the {\tt MINUIT} package from the
CERNLIB library (James 1994) and then explore the $\chi^2$ contours
on a grid centred on the minimum minimorum.  

\begin{figure*}[htb]
\psfig{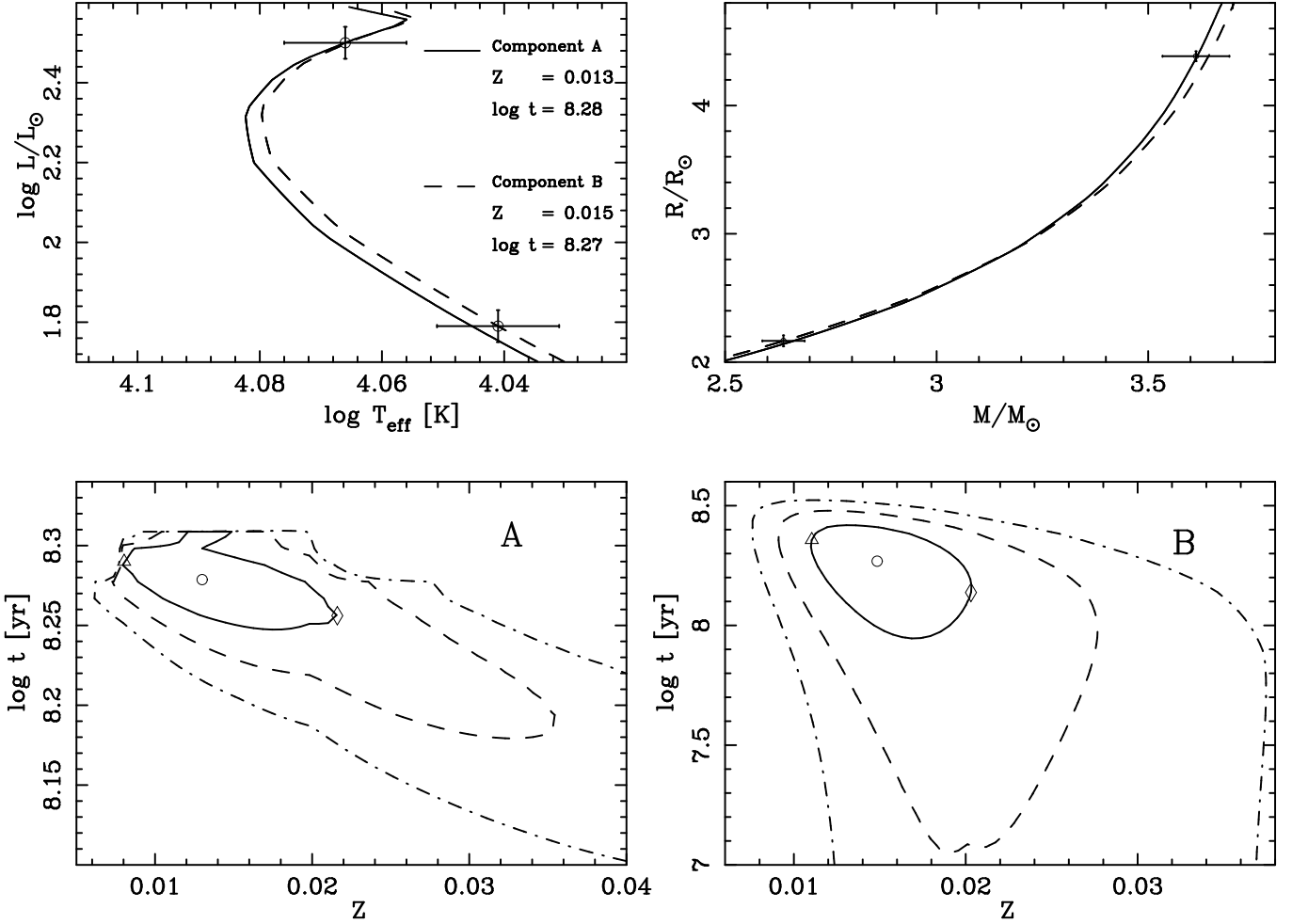}
\caption{{\it Upper left panel: } Zoom on the classical 
log {\teff}-log(L/L$_{\odot}$) HR diagram around the 
components of $\chi^{2}$ Hya [34] (Andersen 1991). 
In this example, an isochrone from the Geneva tracks is fitted to each
component independently, without requiring equal ages or metallicities. The 
best fit isochrones are shown for each component (solid line : component A,
dashed : component B). 
The isochrones derived from the HR diagram are also reported on the mass-radius 
diagram ({\it upper right panel}) and they show a very good fit to the very 
accurate masses and radii, even though these quantities were {\it not}
used for the fit.   
{\it Lower panels: } Confidence regions in the (metallicity, age) plane at 1$\sigma$ 
(solid line), 2$\sigma$ (dashed line) and 3$\sigma$ (dot-dashed line) confidence 
levels for the primary (left panel) and the secondary (right panel) components.
The triangles and diamonds indicate the ($Z, \log t$) values used in Fig.~\ref{fig:chi2hyaab2}. 
}
\label{fig:chi2hyaab1}
\end{figure*}

It turns out that, in practice, the mass is extremely well
 constrained and agrees with the measured 
value (see \S\ref{section:mass}). 
This implies that a nearly unique solution $(t_i, Z_i)$ ought to be found. In 
practice  a whole set of solutions is found, which gives the {\sl empirical} 
limitation to Vogt's theorem. 
Let us illustrate the procedure taking as an example the $\chi^2$ Hya [34] 
system. 
Figure \ref{fig:chi2hyaab1} shows the best fitting isochrones --using the
Geneva tracks-- for each component : they are not identical. The extent to which 
they are statistically compatible depends on the confidence intervals for
the metallicity and age of each of the components. The lower panels present the
$\chi^2$ contours corresponding to the 1, 2 and 3$\sigma$ bounds in this
reduced distance, that is, 
the $\chi_{min}^2+$ 1, 4, and 9 contours respectively. The degeneracy of the
solutions is well illustrated here, as lower metallicity tracks fit a given
component just as well at a slightly larger age, or a higher metallicity
at a younger age. Note that at 1$\sigma$, a factor of 2 in metallicity
is allowed, whereas the uncertainty in the age is about 0.05 dex for
component A and 0.3 for component B. The difference stems of course
from the fact that A is more evolved, and hence better constrained
than B.

\begin{figure}[h]
\psfig{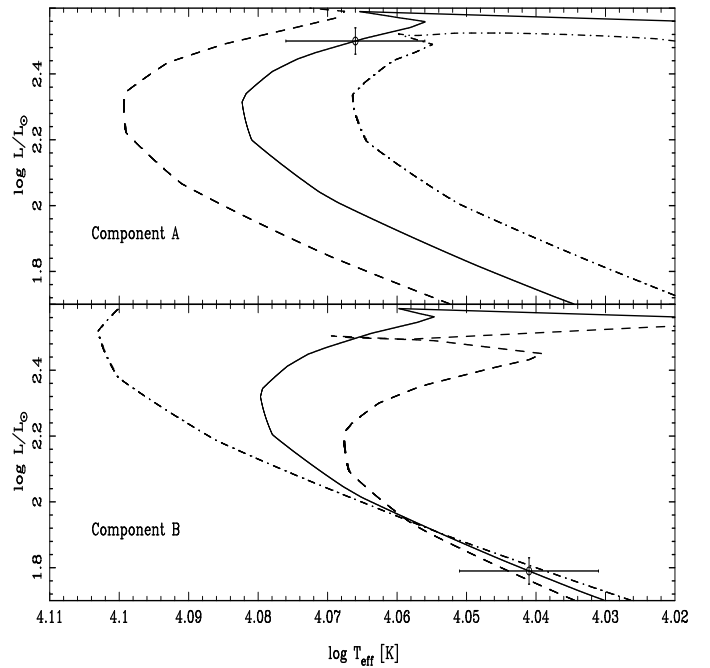}
\caption{$\chi^2$ Hya [34], individual components, Geneva tracks.  For each component, 
the best fitting isochrones (from Fig.~\ref{fig:chi2hyaab1}) along with the isochrones
corresponding to the 1$\sigma$ level are indicated (dashed line :
triangle; dot-dashed line : diamond, see Fig.~\ref{fig:chi2hyaab1}).}
\label{fig:chi2hyaab2}
\end{figure}

Note also that the 1 and 2$\sigma$ contours on Fig.~\ref{fig:chi2hyaab1} 
are closed : within them the Vogt theorem is clearly violated, but since their 
areas are relatively small, one could argue that, {\em in practice}, the 
theorem is valid since a measure of the metallicity of one of the components 
would be enough to constrain very accurately its age.
Nevertheless, how different are isochrones distant 1$\sigma$ from the best solution ?
Two examples are  shown in Figure~\ref{fig:chi2hyaab2}, where it is clear that 
even though the best fitting values for A and B are not the same 
(Fig.~\ref{fig:chi2hyaab1}), at the 1$\sigma$ level they are statistically
compatible. This is also apparent from the fact that the 1$\sigma$ contours 
of both components overlap : the ages and the metallicities of each component 
appear to be statistically compatible with each other.

\begin{figure}[h]
\psfig{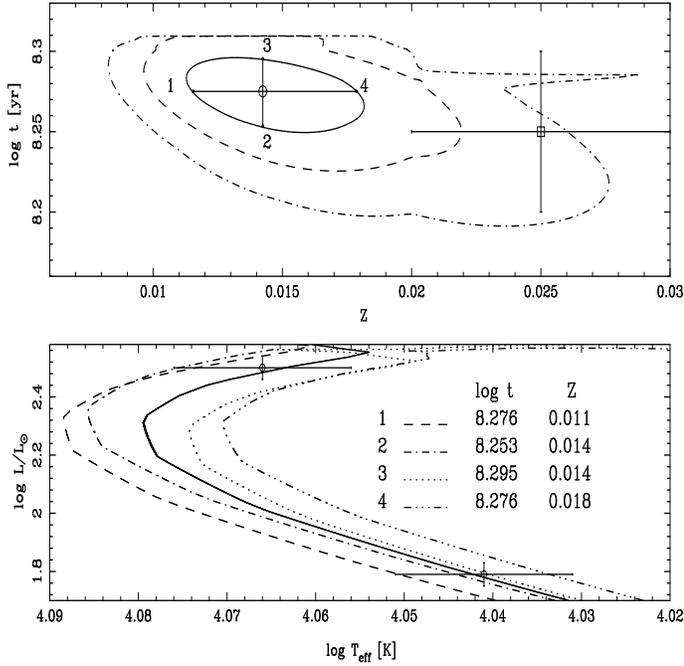}
\caption{$\chi^2$ Hya  system [34], Geneva tracks. 
{\it Upper panel: }  Confidence regions in the (metallicity, age) plane at 1$\sigma$ 
(solid line), 2$\sigma$ (dashed line) and 3$\sigma$ (dot-dashed line) confidence levels for 
both components. For the sake of completeness, 
the best fit obtained  by Clausen \& Nordstr\"{o}m (1978) 
using the H80 models is also shown (square with error bars). 
{\it Lower panel: } Corresponding isochrones on the  HR diagram (error bars 
from Andersen 1991) as in Fig.\ref{fig:chi2hyaab1}. The best fit isochrone obtained 
with the Geneva tracks and defined by ($\log t, Z$) $=$ (8.275, 0.014) is indicated by 
the solid line. The 4 points on the 1$\sigma$ contour above correspond to the 4 
isochrones showed in the lower panel. }
\label{fig:chi2hyasys}
\end{figure}

It remains to be seen whether the ages and metallicities of each
component are the same as the ones inferred from the fitting of a 
{\it single} isochrone to the {\sl combined} A+B system. 
Fitting the best isochrone to the  combined system is done in a
similar way, but in this case the $\chi^2$ functional is obviously 
\beqa
\chi^2 (t_{\rm sys}, Z_{\rm sys})  & = &  
\sum_{i=A}^{B} \left[ \left(\frac{\log L(track) - \log
L_i}{\sigma(\log L_i)}\right)^2 \right.  \nonumber \\
 &  &+ \left.  \left(\frac{\log T_{\rm eff}(track) - \log T_{{\rm eff } i}}{\sigma(\log T_{{\rm eff } i})}\right)^2 \right] 
\eeqa
In this case the masses are no longer parameters (they are left free to
take
any value, just like the radii) and with 4 data values and 2 
free parameters, we expect to find a $\chi^2$ distribution with 2 degrees of 
freedom
Using again {\tt MINUIT} to find the true minimum value, we form the $\chi^2$ 
grid in the metallicity--age plane and compute the 1, 2 and 3$\sigma$ bounds.
Figure \ref{fig:chi2hyasys} shows a few isochrones within the 1$\sigma$ 
contour, and they are all compatible, at the 1$\sigma$ level, with the solutions 
found for each component (Fig.~\ref{fig:chi2hyaab1}). Also illustrated are some 
isochrones chosen to show the effect of varying age, for a fixed $Z$, and changing 
metallicity, for a given age. 
In the case of $\chi^2$ Hya, metallicity effects are clearly more important,
and therefore a measure of their metallicities would constrain extremely
well their common age within 0.05 dex (that is, about 11\% at 1$\sigma$). Note
also that the solution found by Clausen and Nordstr\"om (1978) using the
Hejlesen (1980) tracks is  close to the 3$\sigma$ contour (Fig.~\ref{fig:chi2hyasys}) 
and is therefore highly unlikely, as expected with models using the outdated
Cox \& Stewart (1970) opacities (cf. \S\ref{section:tracks}).


The irregularly-shaped contours are due to the different evolutionary
speeds and short lived phases, and may complicate the interpretation in
some cases.
It is important to note, also, that since the contours are somewhat diagonal, 
the degeneracy between age and metallicity is lifted as soon as a measure of 
the metallicity is made. The contours for the system are much smaller than
the contours for the individual components, as could be expected from the
addition of information.

The results of the fitting of, on one side, each component of the
system, and, on the other hand, of the system on a single isochrone,
are given in Table~\ref{tab:results}, for all the sets of
tracks which we have tested here. 

\subsection{Masses and radii: the most stringent tests of the method} 
\label{section:mass}
To what extent these fitted isochrones are able to {\it predict} the
observed values of the masses and radii? 
These are truly predictions from the theory, since the fit involved 
{\it only} luminosities and temperatures.  
Are the derived metallicities consistent with the observations?
Metallicities will be discussed for the few cases where constraints
exist in section \S\ref{section:feh}, and here we analyse the
predictions made by the fits for the entire sample to the 
very accurate masses and radii available  
(typical errors are smaller than 2\% for mass and radius, except for 3 
systems where typical mass errors are between 4 and 7\%).\\

\begin{figure}[h]
\psfig{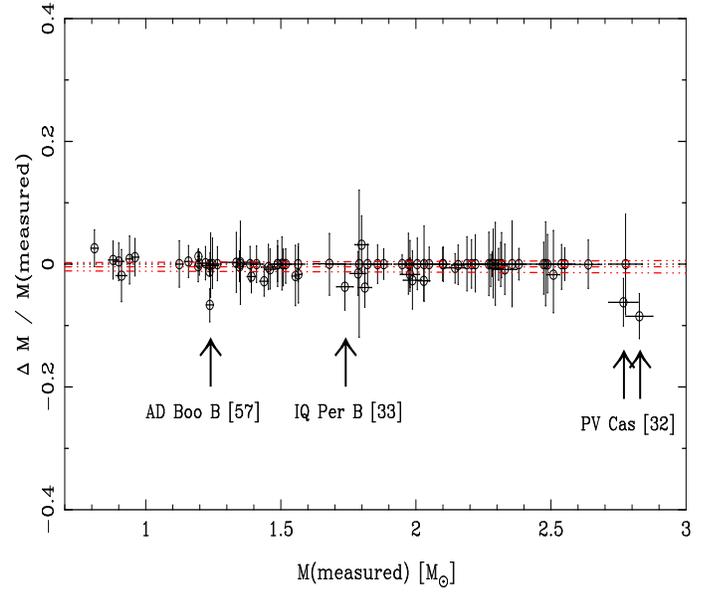}
\caption{Relative error in mass,
$\Delta M / M_{\rm measured} = (M_{\rm derived} -  M_{\rm measured}) 
/  M_{\rm measured}$, inferred from the fitting of Geneva
isochrones to the position of each component on the HR diagram. 
The dotted line is  $\Delta$$M$ $=$ 0 and the dot-dashed lines are 
linear regressions to the data points.}
\label{fig:mass}
\end{figure}

\begin{figure}[h]
\psfig{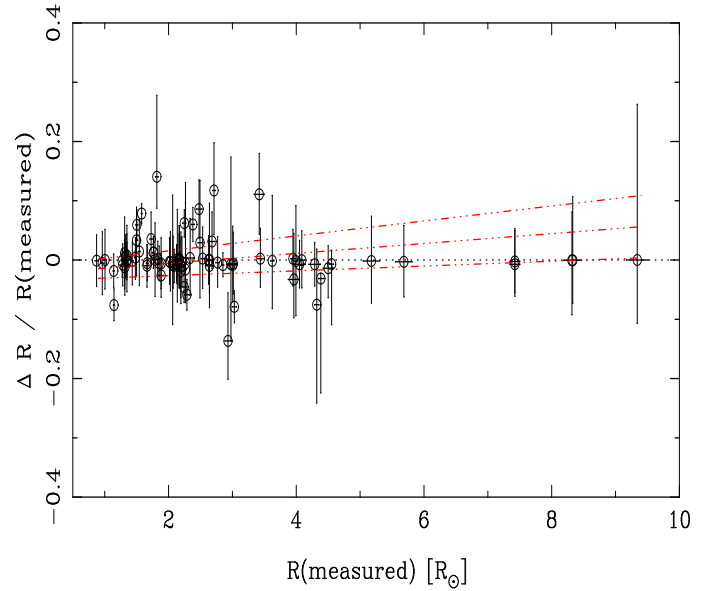}
\caption{   
Relative error in radius,
$\Delta R / R_{\rm measured} = (R_{\rm derived} -  R_{\rm measured}) 
/  R_{\rm measured}$, inferred from the fitting of Padova 
isochrones to the position of each component on the HR diagram. 
The dotted line is  $\Delta$$R$ $=$ 0 and the dot-dashed lines are 
linear regressions to the data points.} 
\label{fig:radius}
\end{figure}

As clearly shown on Fig.~\ref{fig:mass} and \ref{fig:radius}, 
the correct mass an radius are predicted for each individual star 
from isochrone fitting in the (log {\teff}, log L/L$_{\odot}$) 
HR diagram.  
The relative error for the derived masses of each component is given
in Fig.~\ref{fig:mass}, for the Geneva tracks. 
The other tracks yield virtually the same results. 
Note again that the inferred mass comes from
the fitting of an isochrone to the position of that component in the 
(log {\teff}, log L/L$_{\odot}$) diagram, without considering any other
constraint. A linear regression yields $\Delta M / M_{\rm measured} =
(M_{\rm derived} -  M_{\rm measured}) /  M_{\rm measured} = 
(-4.2 \pm 5.7) \, 10^{-3} - (6.35 \pm 1478.3) \, 10^{-6}$ $M_{\rm measured}$, 
that is, both slope and systematic trend consistent with zero. The average
relative error in mass is $-$0.54\%, with a dispersion of 1.7\%. This
is a remarkable achievement for the theoretical tracks, since 
over this mass range we are probing
stars which are almost fully convective to fully radiative stars. 

Only a few components (4 out of 116) present a significant 
discrepancy between the predicted mass and the observed one, given
the error in the observed value, in  Fig.~\ref{fig:mass}. These are  
component B of AD Boo [57], both components of PV Cas [32] and IQ Per [33]. 
AD Boo B yields in fact a bad fit, and  
one cannot expect to infer a correct mass or radius from a bad fit. 
However, a more recent determination of the secondary component mass 
(M$_B$$=$1.20$\pm$0.03 {\MSun}, Popper 1998) 
than the one we used (1.237$\pm$0.013 {\MSun}, Lacy 1997b)
gives a better agreement with our theoretical predictions from the Padova 
(1.188$\pm$0.024 {\MSun}) or Geneva models (1.155$^{+0.020}_{-0.022}$ {\MSun}),  
and even an excellent agreement with the Granada tracks: 1.205$\pm$0.025 {\MSun}. 

The problem is the same for PV Cas [32], which is badly fitted by the
tracks we have used. Pols {\al} (1997) also obtained bad fits, even at the
upper limit in $Z$ they explored ($Z=$0.03), and despite the fact that
we explore a  larger range (up to $Z$$\geq$0.04) we still do not
manage to get a correct fit.  Popper (1987) failed also to fit it 
with 3 other sets of models: the predicted {\teff} is too hot by $\sim$1700 K. 
The revised R00 {\teff}s, lower by about 700 K, would give a better
agreement with solar metallicity models, but as suggested by Young {\al}
(2001) this system may be still in a pre-MS phase, which is not 
taken into account in the tracks studied here. 

The third, IQ Per [33] doesn't produce a too bad fit, and, as we 
discussed in \S\ref{section:temperature}, adopting the R00 {\teff}s 
would give a better agreement.  

As far as the radii are concerned, Fig.~\ref{fig:radius} shows the
same trend as for the masses. A linear regression yields 
$\Delta R / R_{\rm measured} = (-2.2 \pm 1.2) \, 10^{-2} + (0.8 
\pm 0.4) \, 10^{-2} R_{\rm measured}$. 
The slightly positive trend is produced by a few outliers which 
however are not separated by the measured values by more than 1$\sigma$. 
Overall the mean relative error is 0.03\% with a dispersion of 
3.8\%, again a remarkable achievement 
 for the theoretical tracks, since we are 
probing stars which are almost fully convective to fully radiative stars.  

Hence, the positions of these components in the HR diagram  
yield extremely precise and accurate predictions on their 
radii and masses, with dispersions of below 4\% around a mean error 
below 0.5\% both in radius and mass.

\begin{figure*}[htb]
\psfig{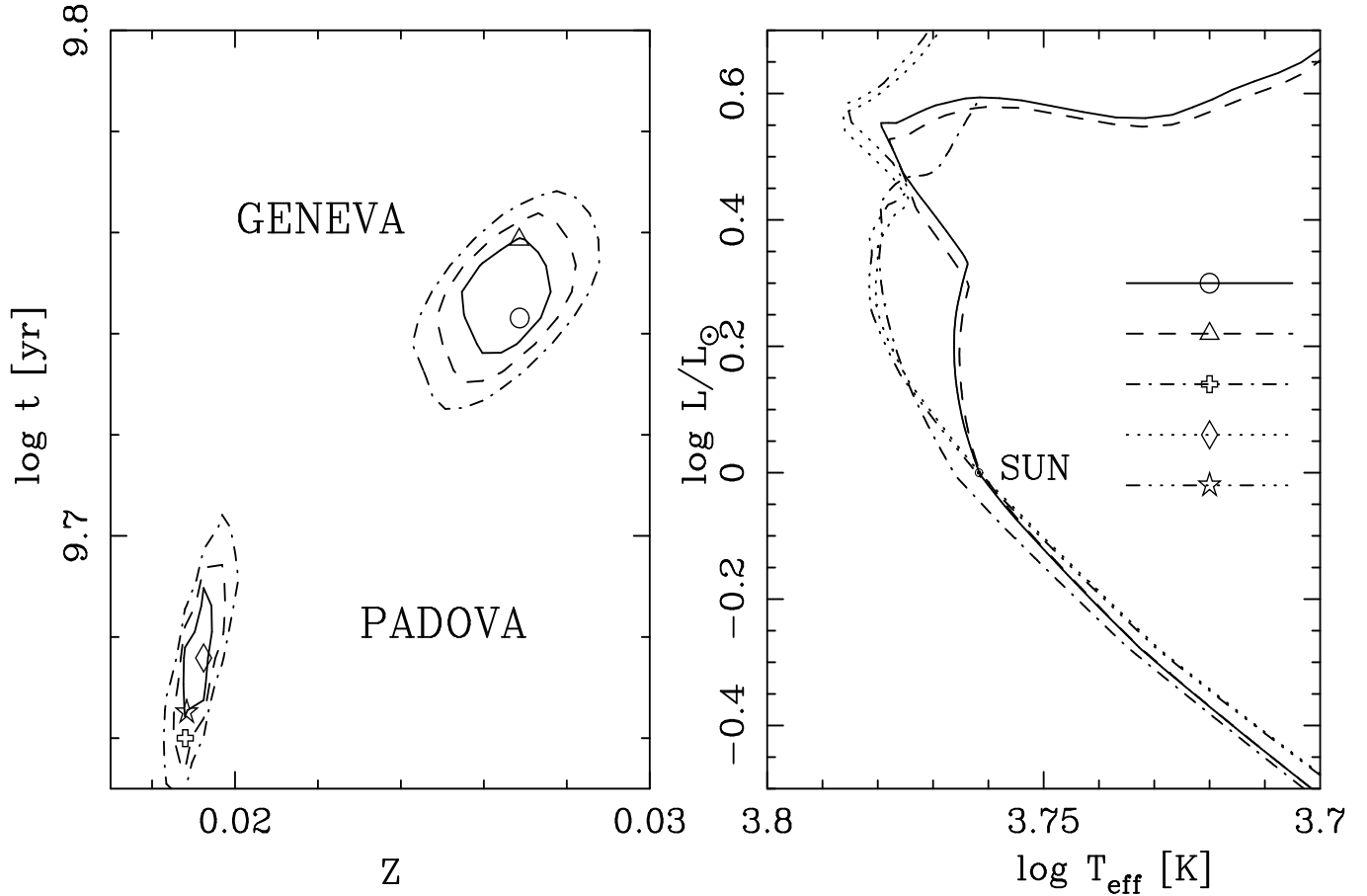}
\caption{Using the \object{Sun} as a test case for the Padova (diamond and open star) 
and Geneva (circle and triangle) tracks. 
The adopted temperature and luminosity are 5777 $\pm$ 2.5 K and
(3.844 $\pm$ 0.01) $\times$ 10$^{33}$ erg s$^{-1}$ (Weiss \& Schlattl 1998). 
A bad fit (open cross) obtained with the Geneva models but with parameters close   
to the $Z$-$\log t$ Padova solutions is also shown for comparison.}
\label{fig:sun}
\end{figure*}

\subsection{Tests and limitations}
We should however be cautious before taking these results at face, 
 because one has to realize that these errors are only a
combination of {\it internal} errors plus a fraction of {\it external}
ones.
Not all the systematic errors have been taken into account, nor
the covariance terms in the internal, measurement errors.  
 We will briefly
mention here four important effects that one has to keep in mind
when using these results:  
the Helium abundance, the availability of tracks 
with a wider range of $Z$, overshooting, and the improved accuracy of the measures.

(1) One obvious test case of a single component is the nearest 
star, the \object{Sun}, for which the
effective temperature is known to 0.04\% and the luminosity to 0.26\%.
Figure~\ref{fig:sun} shows the results of fitting the Sun as a (single)
component using the Padova and the Geneva tracks. 
The very significant difference, at first sight very surprising, 
can mainly be ascribed to the larger He abundance used by the Geneva group 
in comparison to the Padova tracks as shown in Tab.~\ref{tab:tracksummary} 
(Charbonnel \& Lebreton 1993). The effects of
the equation of state and opacities (Yildiz \& Kiziloglu 1997)
or diffusion (Morel {\al} 1997, Weiss \& Schlattl 1998) are
comparatively  much smaller. 
In the context of our work, this
difference is a simple caveat not to use the published tracks blindly, and
also calls for the calculation of a large grid of models with uncorrelated
variations in the main physical parameters. The derived ages and
metallicities are therefore track-dependent, and the values
for $\sigma($Z$), \sigma(\tau)$ that we quote are necessarily {\sl lower}
limits because the systematic effects introduced by the tracks cannot
be taken into account. The best we can do is to compare the results obtained
with different tracks to assess the statistical robustness of the 
values derived with this procedure. 

(2) For systems where the components are quite close to each other in the
HR diagram, the constraints coming from the fitting of the system 
are weaker since there is effectively only one star as a constraint, 
hence increasing the  degeneracy. Note that  
even if the contours produced by each component 
are similar, the best fitting values can be very different.
Sometimes this difference arises from the fact that the best value
is close to the boundary in metallicity allowed by the tracks. 
In several cases (see Table~\ref{tab:results}) the fitting indicates large
metallicities that can only be reached with the Padova and Geneva
tracks.
The best fits obtained are not necessarily good fits. 

(3) The 3 sets of tracks used here have the same prescription for
core overshooting (see Tab.~\ref{tab:tracksummary}). 
Would tracks without overshooting produce worse or
better fits? For instance, P97 argue that a few systems can only be
fitted with models which include overshooting. We have used Geneva tracks without
overshooting for this test, and 
as shown in Tab.~\ref{tab:results}, we find no significant
differences between the Geneva models with and without overshooting. 
However, the power of the test is limited by the small mass range 
(and hence the number of
systems) in common for these two sets of models, and by the fact that most of 
these systems yield bad fits (see \S\ref{section:badfits}). We cannot
conclude,
on the basis of these tracks and this small sample, whether
overshooting is preferred or not. This is yet another source of
systematic uncertainty that is difficult to quantify. 
 
(4) To what extent the age--metallicity degeneracy can be partially lifted by increasing
the accuracy of the measures? A good example is provided by 
 \object{V539 Ara} [41] (\S\ref{section:V539Ara} and Fig.~\ref{fig:v539ara}), where the measures 
from Clausen (1996) are a factor of about two better than the previous
ones from Andersen (1983). The degenerate areas in the ($t,Z$) plane 
are substantially reduced,
although large metallicity solutions are allowed. See also the
case of  \object{$\beta$ Aur} [27] (\S\ref{section:betaAur} and
Fig.~\ref{fig:figbetaaur}) for a similar example. Once again, the most
contraining systems are those with evolved components, even though the
precision in the parameters could be worse.

\section{Analysis of individual systems}

We have applied the techniques developed in \S3 to the entire sample of 60 systems. 
The results are summarized in 
Table \ref{tab:results}
for 58 systems (from [2] to [59]) 
using the 3 different sets of tracks described in \S2. 
The two remaining systems ([1a] and [1b]) are discussed in \S\ref{section:yygem}.  
The column marked 'T' indicates the track used for the fit. Note also
that all ages were corrected for the finite age of the ZAMS in the
case of the Geneva tracks, following Eq.~(1). Although the best
fit values may appear discrepant, it is essential to use the minimum
and maximum values tabulated (at the $n$-th confidence level, as indicated) 
to test for the statistical significance
of the differences. 

It is beyond the scope of this paper to analyse each binary system in
detail, hence in the following we only discuss a few systems which either 
can be compared with previous studies or present problems. Except pre-MS stars
(\S\ref{section:pms})  were good fits cannot be expected with post-MS models,
bad fits were obtained in the HR diagram for stars less massive than 1.2
{\MSun}, EW Ori [3] being the  most massive one. They are mainly discussed in
\S\ref{section:badfits}. 

\subsection{Binaries with (at least) a pre-MS component: EK Cep [17] and RS Cha [14]}
\label{section:pms}

As expected, none of the three sets of tracks are able to fit the system 
EK Cep [17], because the secondary is a pre--main--sequence 
star\footnote{As suggested
by  Popper (1987), and confirmed by Mart\'{\i}n \& Rebolo (1993), and 
Claret, Gim\'enez \& Mart\'{\i}n (1995).} 
and the tracks used do not include these phases. 
Nevertheless, since our method can use information for individual stars, 
we can predict from the primary alone a metallicity, which turns out
to be around solar. \\
RS Cha [14] is also identified as a pre-MS (Mamajek {\al} 2000): 
pre-MS tracks achieve a good fit (Palla \& Stahler 2001). 
RS Cha will not be discussed any further due to 
the uncertainty on its {\teff} (see \S\ref{section:temperature}). \\
Finally, as suggested by Nordstr\"om \& Johansen (1994a, hereafter NJ94a), 
AR Aur B [48] may also be a pre-MS as well. Tab~\ref{tab:results} predicts a 
metallicity slightly sub-solar ($Z$$\sim$0.014$\pm$0.003) for the system, 
in agreement with the NJ94a study. 
We derived from synthetic photometry (L99a) a lower limit on
the metallicity of the primary ($Z$$>$0.017), only marginally consistent 
with the theoretical results. This discrepancy may come from the chemical
peculiarities of both components (Zverko {\al} 1997, and references
therein).    

\begin{figure}[htb]
\psfig{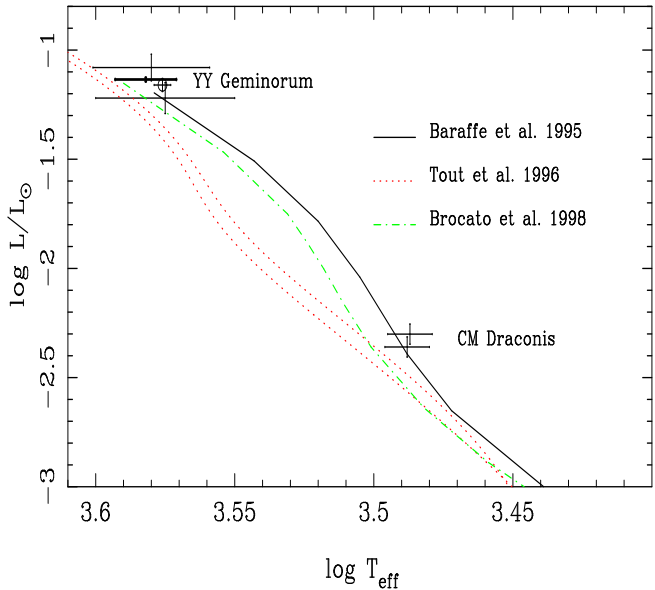}
\caption{YY Gem [1a] and CM Dra [1b]. Data points for CM Dra come from
Viti {\al} (1997). The circle gives the position of YY Gem from Andersen (1991,
both components at the same position), to be compared with Leung 
\& Schneider (1978, error bars for each component) and Torres \& Ribas 
(2002, bold error bars, both components at the same position). 
The isochrones of Baraffe {\al} (1995)  and Brocato {\al} (1998) 
are for a solar metallicity and an age of 10 Gyr. 
Two analytical ZAMS ($Z$$=$0.02 and 0.03) from Tout {\al} (1996) are also shown for 
comparison. }
\label{fig:figYYGemCMDra}
\end{figure}

\subsection{Binaries with VLM star components (M$<$0.7 {\MSun}): 
\object{YY Gem} [1a] and \object{CM Dra} [1b]}
\label{section:yygem}

Neither YY Gem [1a] nor CM Dra [1b] provide a test for 
Geneva, Padova and Granada tracks, since 
their components are very low mass (VLM) stars, with masses below  0.6 {\MSun}. 
For these two systems only, we have used instead 
the evolutionary tracks from  Baraffe {\al} (1995) which are computed  at
$\log t =$ 10 yrs. 
Assuming the position of YY Gem from 
A91\footnote{The comprehensive study 
of YY Gem by Torres \& Ribas (2002) provides a new position in the HR
diagram which however leads to no significant differences.}
(see Fig.~\ref{fig:figYYGemCMDra}), 
these tracks yield a best fit for YY Gem at $Z=$ 0.019. 
This is consistent with the solution found by
Chabrier \& Baraffe (1995), indicating also that YY Gem [1a] could be anywhere
between the upper limit of 10 Gyr and the end of the pre-MS contraction
phase. Since this upper limit may be too old (Torres \& Ribas,  
2002 estimated a younger age of $\sim$ 370 Myr), we refer the interested 
reader to the detailed analysis of YY Gem by Torres \& Ribas (2002). 
We also note that using the older {\teff}-L parameters from Leung and Schneider (1978)
(also shown on Fig.~\ref{fig:figYYGemCMDra}) leads to virtually no constraint 
due to the large error 
bars\footnote{ As shown by Torres \& Ribas (2002), the apparently 
good fit in the HRD for YY Gem using the Baraffe models is deceiving 
because they fail to predict the correct radius of low mass stars for the measured mass, 
by 10\% or more. At least in this particular case, the use of L and {\teff} gives the 
wrong impression. This illustrates the importance of the prescription proposed in this paper: 
1) to fit the classical HRD  {\it and} then 2) to check the predicted masses and radii.}.\\
Torres \& Ribas (2002) adopted {\teff}$=$3820$\pm$100 K 
(see Fig.~\ref{fig:figYYGemCMDra}) from several empirical calibrations. 
However, adopting the colour index values listed in their Tab. 6 for V$-$I$_C$, V$-$K and 
(R$-$I)$_C$, assuming solar abundances and the very accurate surface gravity of YY Gem 
from Torres \& Ribas (2002), and neglecting the reddening\footnote{YY Gem 
is a nearby system ($\sim$16 pc) according to its Hipparcos parallaxe, so we assume a 
colour excess E(B$-$V)$=$0.}, the BaSeL models (Lejeune {\al} 1997, 1998) 
indicate cooler {\teff}s (between 3605 and 3642 K to be compared with the coolest 
temperature that  Torres \& Ribas (2002) obtain: 3719 K from V$-$I$_C$).  
Since the BaSeL models provide very good agreement with empirical calibrations in these 
colours (cf. Fig. 2 of Lejeune {\al}, 1998), we suspect that the {\teff}s of YY Gem 
still need further revisions,  but the temperature 
issue for cool stars is still controversial to give a definitive {\teff} . \\ 
In the case of CM Dra [1b], whose metallicity is compatible with the solar value (Gizis 1997) 
or lower (Viti {\al} 1997, see also Metcalfe {\al} 1996), its position in the HR diagram 
is very well constrained (Viti {\al} 1997) for such a VLM binary, and indicates a 
metallicity close to solar with the models of Baraffe {\al} 1995 (see Fig.~\ref{fig:figYYGemCMDra}).
When using other sets of tracks (from Brocato {\al} 1998, or the analytical ZAMS of Tout {\al}  
1996), the results are substantially different and a $Z=$ 0.02
 composition is
clearly inconsistent 
with the data (see Fig.~\ref{fig:figYYGemCMDra}). 
A larger sample is needed in these range of masses  to confirm these tests, 
and the discovery of an M-dwarf eclipsing binary, GJ 2069A
(Delfosse {\al} 1999),  with components of about 0.4 \MSun\ (which places GJ
2069A between CM Dra and YY Gem), is a first step towards better
observational constraints in the lower mass end of the main sequence 
(see S\'egransan
{\al} 2000 for a recent list of accurate masses of VLM stars).  
  
\subsection{Systems with at least one component in the 0.7-1.1 {\MSun} mass
range:  
 \object{RT And} [47], 
 \object{HS Aur} [2], 
 \object{CG Cyg} [46], and 
 \object{FL Lyr} [4]
} 
\label{section:badfits}

We obtain systematically worse fits for all these systems\footnote{Some of their 
components are not massive enough to be fitted with the Granada tracks, so only 
the Geneva and Padova tracks are used here.}. 
\begin{figure*}[htb]
\psfig{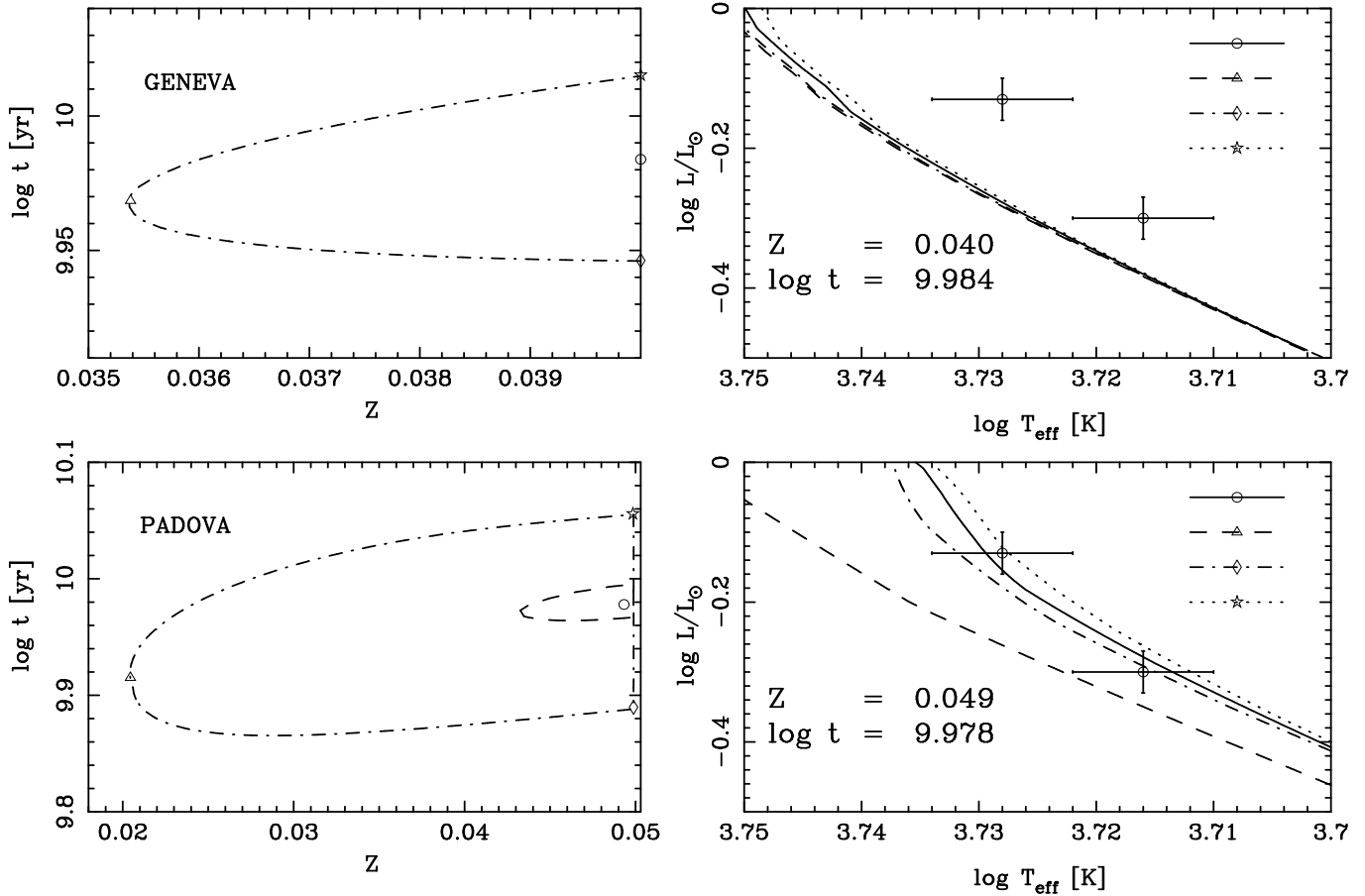}
\caption{HS Aur system [2]. {\it Left  panels : } Confidence regions in the (metallicity, 
age) plane. {\it Right panels :} HR diagrams. The top panels use
the Geneva tracks, while the lower ones use the Padova ones. The best solutions obtained 
are indicated by circles. Keeping the same metallicity as the best solution, we also show 
two other points at 3$\sigma$ from the best solution, with respectively a smaller age (diamond) 
and a larger age (open star). Another point (triangle) is the solution with the lowest metallicity 
allowed at a 3$\sigma$ confidence level. The Geneva isochrones clearly do not fit the HS Aur system, 
no matter what age or metallicity are used. In contrast, the Padova tracks allow a successful 
fit at about 2.5 times the solar metallicity.  }
\label{fig:fighsaur}
\end{figure*}
The bad fits however present two different aspects. While both
components of HS Aur [2] 
can not be fitted (or at least only by very enriched models), 
the systems RT And [47], CG Cyg [46]  
and FL Lyr [4] have a secondary component far too cool to be matched by the 
same isochrone as the primary\footnote{
The same difficulty appears with the Cambridge group models by fitting 
simultaneously their effective temperatures, masses and radii  
(see Pols {\al}, 1997).}. 
This may reveal a different behaviour. 
Under the optimistic hypothesis that the location of both components of HS Aur in 
the HR diagram are perfectly reliable, Fig.~\ref{fig:fighsaur} suggests that the models are 
unable to correctly predict its properties: 
either we have strong difficulties to fit both components of HS Aur (Geneva models) 
or we obtain very questionable metal-rich solutions (Padova models). 
For illustrative purposes, Fig.~\ref{fig:fighsaur} shows that the Geneva tracks 
do not produce confidence areas at the 1$\sigma$ level, and most solutions are
extremely far away (3$\sigma$) from the position of the system. In contrast,
the Padova tracks succeed in fitting the system, but at a metallicity
which is more than 2.5 
times solar ($Z$$\simeq$ 0.04-0.05). This anomaly was also found by Pols 
{\al} (1997): they obtained high $Z$ values for all systems with stars below 1 
{\MSun}.   
The Padova tracks indicate that HS Aur is at the same time 
quite old and metal rich 
(Fig.~\ref{fig:fighsaur}, lower panels): $\sim$[Fe/H]$=$0.4-0.5. 
It remains to be seen whether a spectroscopic analysis confirms this 
predicted large metallicity. \\
Since the results of Pols {\al} (1997) and the present study rely on
the same  
{\teff} determination, a revision may well be needed (in particular the {\teff}s of 
the secondary components) before definitive conclusions.  
Unfortunately, while L99a, R00, and Lastennet, Cuisinier \& Lejeune (2002) carefully 
re-derive {\teff}s for many EBs from photometric calibrations, none of these 
works study the stars in question because individual $uvby$ photometry would be 
necessary. \\
We now have to examine \object{FL Lyr} [4], \object{CG Cyg} [46] and
 \object{RT And} [47]. 
Besides  a {\teff} revision, possible explanations of the disagreement may 
come from mass transfer and starspot activity because these 3 systems are 
known to be active stars.   
As already discussed in \S\ref{section:sample} none of these stars overflows its 
Roche lobe (see also Lastennet {\al}, 2002 for more details). 
Yet in  RT And, the face-to-face position of the spots on the 
surface of both components
may indicate the possibility of a mass transfer from the primary to the secondary 
component through a magnetic bridge connecting both active regions (Pribulla {\al} 
2000). If this is confirmed, the RT And components could not be assumed anymore to evolve 
like single stars, and we could not expect to reproduce their evolution with the models 
considered in this paper. Enhanced surface activity is also present in FL Lyr and CG Cyg. 
These spots lead to quite distorted light curves and consequently less well-determined 
photometric elements, and hence may provide 
 an explanation of the disagreement we have found. \\
If the disagreements appear to remain even after further analyses, they will give 
additional examples to the dilemma pointed out by Popper (1997) who
remarked that 
for systems with masses between 0.7 and 1.1 $M_\odot$ it was  
difficult to place both two components on the same isochrone. The same
happens  in a Mass-Radius 
diagram as shown by Clausen {\al} (1999). \\
Other binaries show similar problems in this mass 
range\footnote{The Padova isochrones do not match the Hyades eclipsing binary V818 Tauri 
in the mass-radius diagram (thus, without using any information on {\teff}), and it 
appears difficult to match the mass of the secondary of the nearby visual binary 85 Peg.},  
as briefly reviewed by Lastennet {\al} (2002).   
However, the systematic differences between models and binaries of this mass range 
are not observed in CD Tau C, a solar mass companion of the triple system CD Tau. 
Ribas {\al} (1999) obtained a perfect fit of the three components with a 
single isochrone, and the issue is still open to debate. 
New accurate data for low mass 
eclipsing binaries should help to work out the discrepancies (Clausen {\al}, 2001; 
see also Lastennet {\al}, 2002 and references therein). 

\subsection{Systems with both components in the 1.1-3. {\MSun} mass range}
\subsubsection{\object{AI Phe} [5]}  
  
\begin{figure}[htb]
\psfig{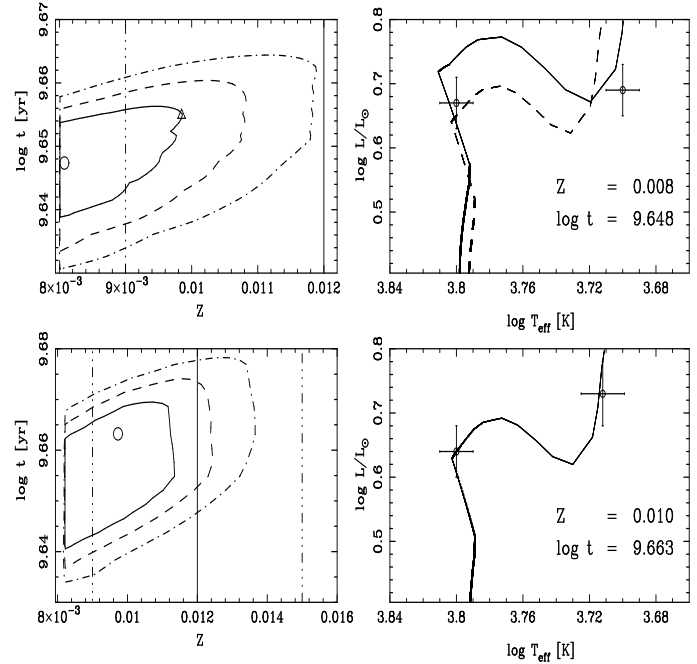}
\caption{\object{AI Phe} [5], Geneva tracks. The upper panels
use the parameters from Andersen {\al} (1988), while the lower ones
take the more recent  data by Milone {\al} (1992).  
Circles (solid line isochrones) are the best solutions. In the upper
panel the triangle (dashed line isochrone) is  at 1$\sigma$  from the
best solution. The spectroscopic metallicity ($Z=$0.012$\pm$0.003) 
from Andersen {\al} (1988) is  indicated with vertical lines.  }
\label{fig:AIPhe}
\end{figure}

AI Phe is one of the most interesting systems in the sample since the coolest component 
is also away from the main--sequence, and observational constraints on $Z$ are available. 
A spectroscopic determination of metallicity from high resolution CCD spectra (Andersen {\al} 1988) 
gives [Fe/H] $= -0.14\pm0.1$, i.e. $Z = 0.012\pm0.003$. Using tracks from VandenBerg (1983) they 
derived an age of $4.1\pm0.5 \; \times \; 10^{9}$ yr from  isochrones  with an helium abundance 
$Y = 0.27\pm0.02$\footnote{This abundance is much smaller than the one found by VandenBerg 
\& Hrivnak (1985), $Y =$ 0.33--0.44, although for a similar age ($t =$ 3--4 $\times$ $10^9$ yr). 
VandenBerg \& Hrivnak (1985) adopted 
[Fe/H] $= 0.17\pm0.20$ (from the observed $m_1$ index, 
the Crawford (1975) calibration for F$-$type stars and the ($\delta$, [Fe/H]) relationship of 
Crawford \& Perry (1976)), 
which is clearly far too large in comparison to the
determination  of Andersen {\al} (1988).}.   
Our 3 sets of tracks can only fit the system at slightly lower metallicities, as shown in 
Fig.~\ref{fig:AIPhe} for the case of
 the Geneva tracks. These results are consistent with P97 and R00. 
The combination of the updated parameters for the system from Milone {\al} (1992) with
the heavy element abundance from Andersen {\al} (1988) constrain extremely
well the system, with uncertainties in age smaller that 0.010 dex at
1$\sigma$ and 0.014 at 3$\sigma$. The best fit solutions are entirely compatible with both
data sets. 
The results obtained with the Granada tracks ($Z =$ 0.011$\pm$0.001, 9.64 $< \log t < $ 9.67 yr, all 
1$\sigma$) and with the Padova ones ($Z=$ $0.008^{+0.003}_{-0.000}$ and 9.59 $< \log t <$ 9.63, 
1$\sigma$) are both in good agreement with the ones deduced from Geneva and the 
metallicity range from Andersen {\al} (1988), although the range allowed in age is a bit larger. 

\subsubsection{\object{UX Men} [6] }

\begin{figure}[htb]
\psfig{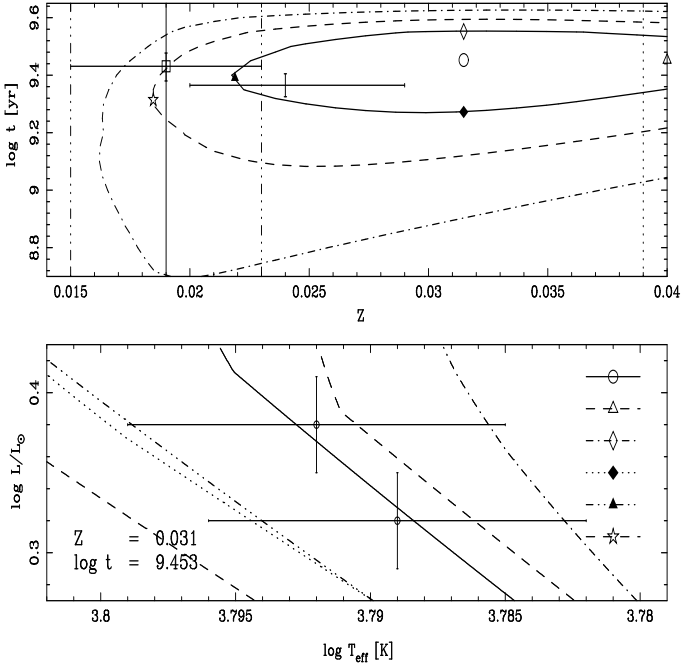}
\caption{\object{UX Men}  system [6], Geneva tracks. Best fit (open circle), 1$\sigma$ points 
(triangles and diamonds), and a 2$\sigma$ point (open star) situated within the observed metallicity range. 
The vertical lines mark the observed value of $Z$ at 0.019$\pm$0.004 (Andersen {\al} 
1989), while the best age (open square) by Andersen {\al} (1989) was derived using the 
VandenBerg (1985) models computed for $Z$ = 0.019. Results from Pols {\al}
(1997) (error bars at $Z=$0.024) and the extrapolated solution of Ribas {\al} (2000) 
(vertical dotted line at $Z=$0.039) are also shown for comparison. The last
one is unreliable because well beyond the upper $Z$-limit covered by their 
models ($Z$$=$0.03).} 
\label{fig:ux.men}
\end{figure}

This system was studied by Andersen {\al} (1989) who derived a best fit isochrone of 
age (2.7$\pm$0.3)$\times$$10^9$ yr and $Y$ = 0.27$\pm$0.01 using the VandenBerg (1985) 
tracks fixed at $Z$ = 0.019, the observed value.
There are several measures of the metallicity\footnote{  
(1) Clausen \& Gr{\o}nbech (1976) find a small value, [Fe/H] $= -0.15$ 
from uvby$\beta$ photometry; 
(2) Kobi \& North (1990) obtain [Fe/H] $= -0.04$;  
(3) From their own Str\"{o}mgren uvby$\beta$ observations and
correcting for reddening, Andersen {\al} (1989) deduce from  Nissen's (1981) calibrations a 
metallicity of [Fe/H] $= -0.05\pm$0.15; 
(4) Andersen {\al} (1989) also obtained from high resolution CCD 
spectra  [Fe/H] = 0.04$\pm$0.10 (i.e. $Z$ = 0.019$\pm$0.004 for 
an assumed $Z_{\odot} =$ 0.0169, VandenBerg 1985).}.
Since they are consistent with each other (except that the Clausen \& 
Gr{\o}nbech value was not corrected for reddening), we adopt
in Fig.~\ref{fig:ux.men} the more precise spectroscopic determination 
(Andersen {\al} 1989). 

\begin{figure*}[htb]
\psfig{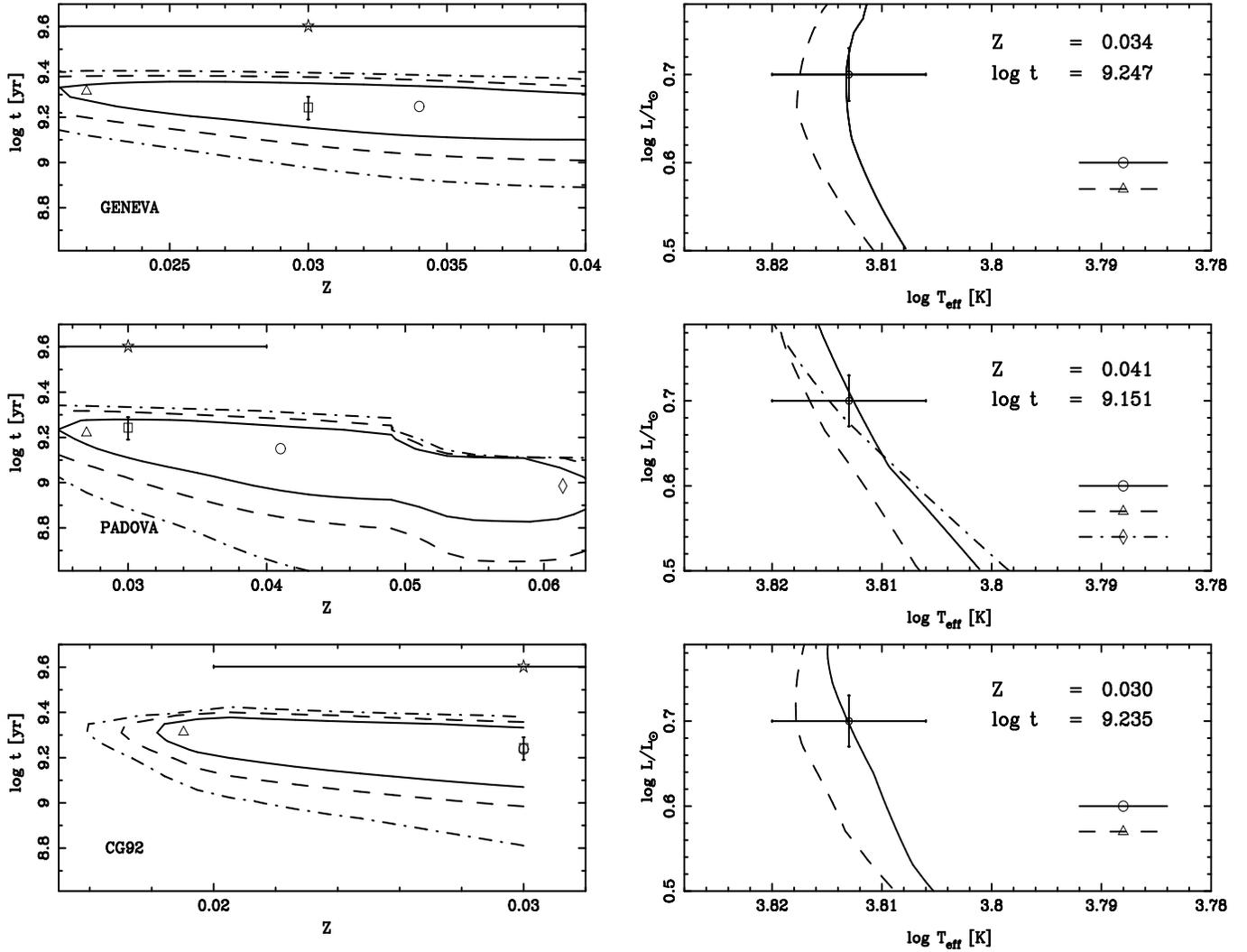}
\caption{\object{DM Vir}   system [8]. Confidence regions using the
Latham {\al} (1996) data and the Geneva ({\it upper panels}),  Padova 
({\it middle panels}) and Granada ({\it lower panels}) tracks. 
Best fit values (circles) are indicated in each panel, along with the 
corresponding isochrone (solid lines) in the HRD. 
Other isochrones, associated with the positions marked by triangles or 
diamonds on the left panels, are also shown in the HRDs. 
The fit obtained by Andersen, Clausen \& Nordstr\"{o}m (1984b) 
with the H80 tracks (open star) and by Latham {\al} (1996) 
with the Granada tracks (square) are also indicated.}
\label{fig:DMVir}
\end{figure*}
The Geneva tracks (with overshooting) give two coeval components with the same 
metallicity\footnote{The primary has ($\log t, Z)_{A} =$ (9.42, 0.033) whereas 
the secondary has ($\log t, Z)_{B} =$ (9.44, 0.030) (see Table~\ref{tab:results} 
for the uncertainties).}. 
Even though the ages are similar to the one derived by Andersen {\al} (1989), the metallicity 
derived from the best fit is 50\% larger than the spectroscopic/photometric value\footnote{An 
even more metallic abundance is suggested by the R00 extrapolated solution: $Z$$=$0.039.}. 
This is also supported by the fitting of the combined system (Fig.~\ref{fig:ux.men}). 
The Padova tracks give 
a very similar result, ($\log t, Z$) = (9.257, 0.034), but the Granada models and P97 yield a 
slightly smaller metallicity (($\log t, Z$) = (9.405, 0.027) and (9.365, 0.024) respectively). 
However, at the 1$\sigma$ confidence level a value as small as $Z=$0.022 (0.025) is 
allowed by the Geneva (Padova) models. We suspect that a slight increase in the Helium abundance 
may reconcile the range inferred from its position on the HR diagram
with the spectroscopically determined metallicity. Alternatively, a revision of the {\teff} 
may be another explanation. This will be discussed further in \S\ref{section:feh}. 

\subsubsection{\object{DM Vir} [8]}

Accurate data from Latham {\al} (1996) have superseded the values
listed by A91 for this system. Figure~\ref{fig:DMVir} shows the best fits 
obtained with the Geneva (upper panels), Padova (middle panels) and 
Granada tracks (lower panels), and illustrates the danger of using a single set of 
tracks to get all the possible solutions. 
The limited $Z$ range in the Granada tracks lead Latham {\al} (1996) to a best fit 
at the $Z$ limit of the Granada models, but which is by no means
unique, 
as shown in Fig.~\ref{fig:DMVir}. A similar good fit was also found by Pols {\al} 
(1997) at $Z$$\sim$0.03, which is the upper limit of the models they used. This system was 
not selected by R00. \\ 
Although the age is very well constrained by all sets 
of tracks (about 0.1dex at 1$\sigma$, see Fig.~\ref{fig:DMVir}, left panels), 
solutions with metallicities ranging from 0.025 to 0.06 are not 
statistically rejected at the 1$\sigma$ level\footnote{
This was also found by Andersen, Clausen \& Nordstr\"{o}m (1984b)
using the H80 models : solutions as separated as  
($\log t =$ 9.60, $Z =$ 0.02, $Y =$ 0.18) and 
($\log t =$ 9.60, $Z =$ 0.04, $Y =$ 0.26) being good fits, the solar 
metallicity solution implying however a rather unlikely low helium 
abundance.}.
Since $m_1$(DM Vir) $-$ $m_1$(Hyades) $=$ 0.001 (Andersen, Clausen \& 
Nordstr\"{o}m 1984b), 
this suggests that DM Vir has a metal content similar to the one in the
Hyades. 
Latham {\al} (1996) also take [Fe/H] $=$ 0.12$\pm$0.12, although a detailed 
spectroscopic analysis is needed. This would imply (with $Y_{\odot}=$ 0.27) 
$Z =$ 0.023$\pm$0.006, yet the best fits obtained here with the 3 sets of tracks
favour much larger  metallicities, especially the Padova tracks. 
Hence, if this metallicity is confirmed, 
the contours of Fig.~\ref{fig:DMVir} show that the Padova tracks would fail to fit
the system in the HR diagram, while the Geneva and Granada ones would be consistent, provided 
the He abundance is as high as the one used by these models (cf.
Tab.~\ref{tab:tracksummary}). 

\subsubsection{\object{RZ Cha} [10]}

\begin{figure}[htb]
\psfig{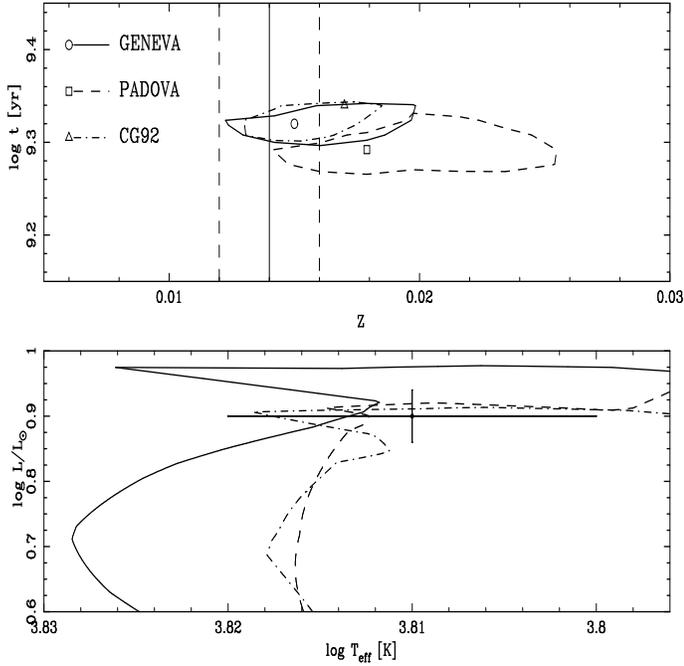}
\caption{RZ Cha system [10]. The upper panel superposes the 1$\sigma$ contours from the 3 
sets of tracks, along with the best fits from Geneva (open circle), Padova (square) 
and Granada (triangle) tracks. The metallicity range found by J{\o}rgensen \& Gyldenkerne (1975), 
corrected for the updated Hyades metallicity, is indicated by the vertical lines. 
The best fit isochrones from the three sets of tracks are shown in the lower panel. }
\label{fig:rz_cha}
\end{figure}

The components of RZ Cha are known to be both evolved and older than 2 Gyr 
(J{\o}rgensen \& Gyldenkerne 1975).  
Andersen {\al} (1975) 
found $\log t $ = 9.301 with a preliminary version of H80 tracks, while we obtain 
($\log t, Z) = (9.32^{+0.02}_{-0.03}$, $0.015^{+0.004}_{-0.003}$) with
the Geneva tracks. 
As illustrated in Fig.~\ref{fig:rz_cha}, the contours resulting from the fit with the 
Padova and Granada models are also very 
similar\footnote{$(\log t, Z)=$ ($9.30^{+0.04}_{-0.03}$, $0.018^{+0.008}_{-0.003}$) and 
$(\log t, Z)=$ ($9.34^{+0.03}_{-0.07}$, $0.017^{+0.002}_{-0.004}$) for Padova and Granada 
models respectively. These results are in a very good agreement with
P97 (their
 overshooting models). 
For the sake of completeness, we note that this system was not studied by R00.}. 
Yet these solutions for $Z$ are systematically smaller than the value of [Fe/H] = $-$0.02 
$\pm$ 0.15\footnote{Derived from $\Delta m_1$ = $m_1$(RZ Cha) $-$
$m_1$(Hyades) = 0.017 and choosing  [Fe/H]$_{\rm Hyades} = $ 0.20 $\pm$ 0.15,
J{\o}rgensen \& Gyldenkerne (1975).}, 
which is almost solar metallicity, $Z=$ 
0.020$^{+0.008}_{-0.006}$.  In fact, the Hyades metallicity adopted by
J{\o}rgensen \& Gyldenkerne was  an overestimation. 
More recent determinations (e.g. Cayrel de Strobel {\al} 1997, Perryman {\al} 1998) 
cluster around [Fe/H]$_{\rm Hyades} =$ 0.14 $\pm$ 0.05, so that  we derive for RZ Cha : 
[Fe/H] $=$ $-$0.08 $\pm$ 0.05 or $Z=$ 0.014 $\pm$ 0.002.
This metallicity is in very good agreement with the best fits obtained by the 3 sets of tracks, 
the Geneva tracks being slightly better (see Fig.~\ref{fig:rz_cha}). The extremely good constraint 
on the age (less than 0.04 dex at 1$\sigma$) arises from its evolved position on the HR diagram.   
Despite the fact that both components of RZ Cha are on the same point in the HR diagram , 
hence a less than optimal case (Fig.~\ref{fig:rz_cha}), 
it is precisely because its position lies by the TAMS, that one can derive good
constraints on both metallicity and  age.  

\subsubsection{\object{PV Pup} [12]}

\begin{figure}[htb]
\psfig{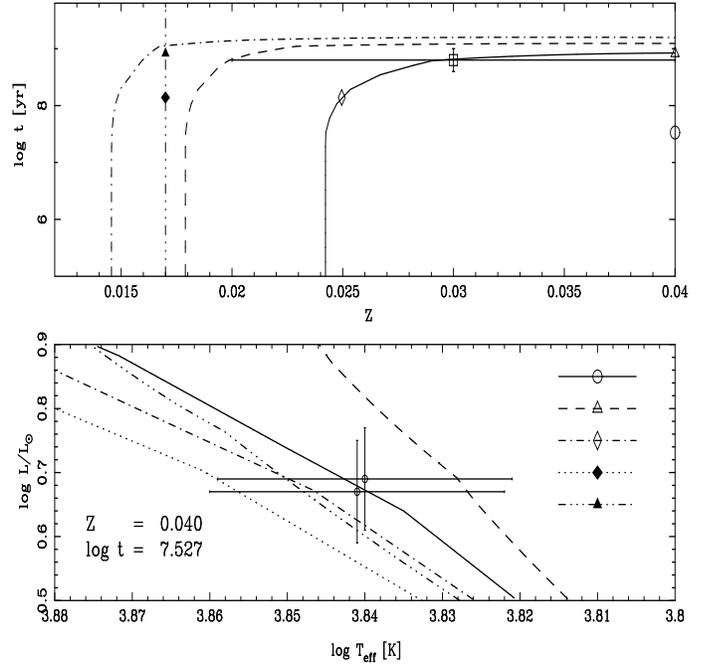}
\caption{PV Pup [12], Geneva tracks. Best fit (open circle at ($\log t, Z) =$ (7.53, 0.04)), 
1$\sigma$ points (triangle at $Z =$ 0.04, and diamond at $Z$ around 0.025) and 
the VA84 solution (square with error bars) with the H80 models are shown. 
The $Z$ value of VA84 (derived from the Nissen (1981) calibrations) 
is $Z =$ 0.017 (vertical line). 
Two other points (filled triangle and filled diamond) are also shown 
at $Z =$ 0.017, with the same ages as the  (open triangle) and (open diamond)
isochrones respectively. }
\label{fig:PVPupG}
\end{figure}

\begin{figure}[htb]
\psfig{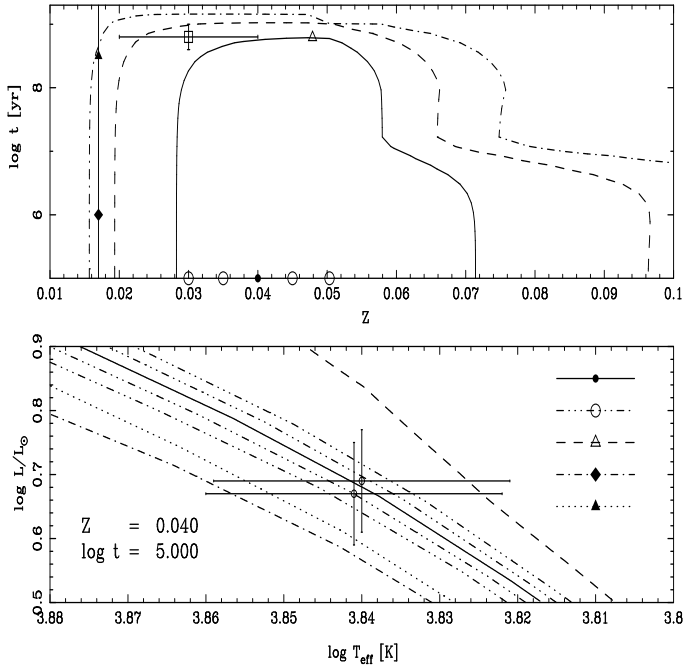}
\caption{PV Pup [12], Padova tracks. Isochrones for the best fit (filled circle at 
($\log t, Z) =$ (5., 0.04)) and for points at 1$\sigma$ away (triangle), and for 
four coeval points (open circles) with the best age solution but with different 
metallicities ($Z =$ 0.030, 0.035, 0.045 and 0.050). 
The VA84 solution (square with error bars) with the H80 models is also shown, 
along with the $Z=0.017$ (vertical line) derived from Str\"{o}mgren indexes (VA84 with 
Nissen (1981) calibrations). Two other points (filled triangle and filled diamond) are 
also shown on the $Z =$ 0.017 line.}
\label{fig:PVPupP}
\end{figure}

The proximity of the two non-evolved components of this system in the HR
diagram does not allow  to define  a very accurate solution: the closer 
the two points on the HR diagram, the larger the contours.  
The Geneva tracks give a best fit at ($\log t, Z) =$ (7.53, 0.040), but metallicities 
as low as $Z = 0.015$ and  ages as large as $\log t =$ 9.1  are not excluded at the 
3$\sigma$ confidence level (Fig.~\ref{fig:PVPupG}). As indicated by Vaz \& Andersen (1984),
the H80 models allow  good fits at ($\log t, Z, Y) =$ (9., 0.02, 0.18) or at (8.60, 0.04, 0.26), 
yet another example of the age$-$metallicity degeneracy. 
Nissen's (1981) calibration of the Str\"{o}mgren $\beta$ and $\delta$m$_1$ indexes 
gives $Z =$ 0.017 (this value is a vertical 
line in the iso--$\chi^2$ diagrams in Figs.~\ref{fig:PVPupG} and \ref{fig:PVPupP}). 
For clarity, Fig.~\ref{fig:PVPupG} shows the results with the Geneva models and 
Fig.~\ref{fig:PVPupP} with Padova models. They both indicate that better fits
are obtained with larger metallicities (in agreement with Pols {\al}
1997: $Z>$0.028), yet isochrones at $Z =$ 0.017 are only  2$\sigma$ away from
the best fits. The revised R00 {\teff}s are the same ones that 
we  use, so our high-metallicity results are unchanged. 
Alternatively, the simultaneous ({\teff}, [Fe/H]) solutions derived from  
Str\"omgren photometry for both components of PV Pup (Lastennet {\al}
1999a) give weak constraints on the metallicity (but an
upper limit of [Fe/H]$<$0.24, i.e. $Z<$0.029, at 2-$\sigma$) and a
rather 
 well-defined range of {\teff} (log {\teff} between 3.86 and 3.88). This would shift
the position of PV Pup in the HR diagram towards larger values of {\teff} (and 
hence L) and  consequently would exclude all the high-$Z$ solutions. 

\subsubsection{\object{TZ For} [18]}

\begin{figure*}[htp]
\psfig{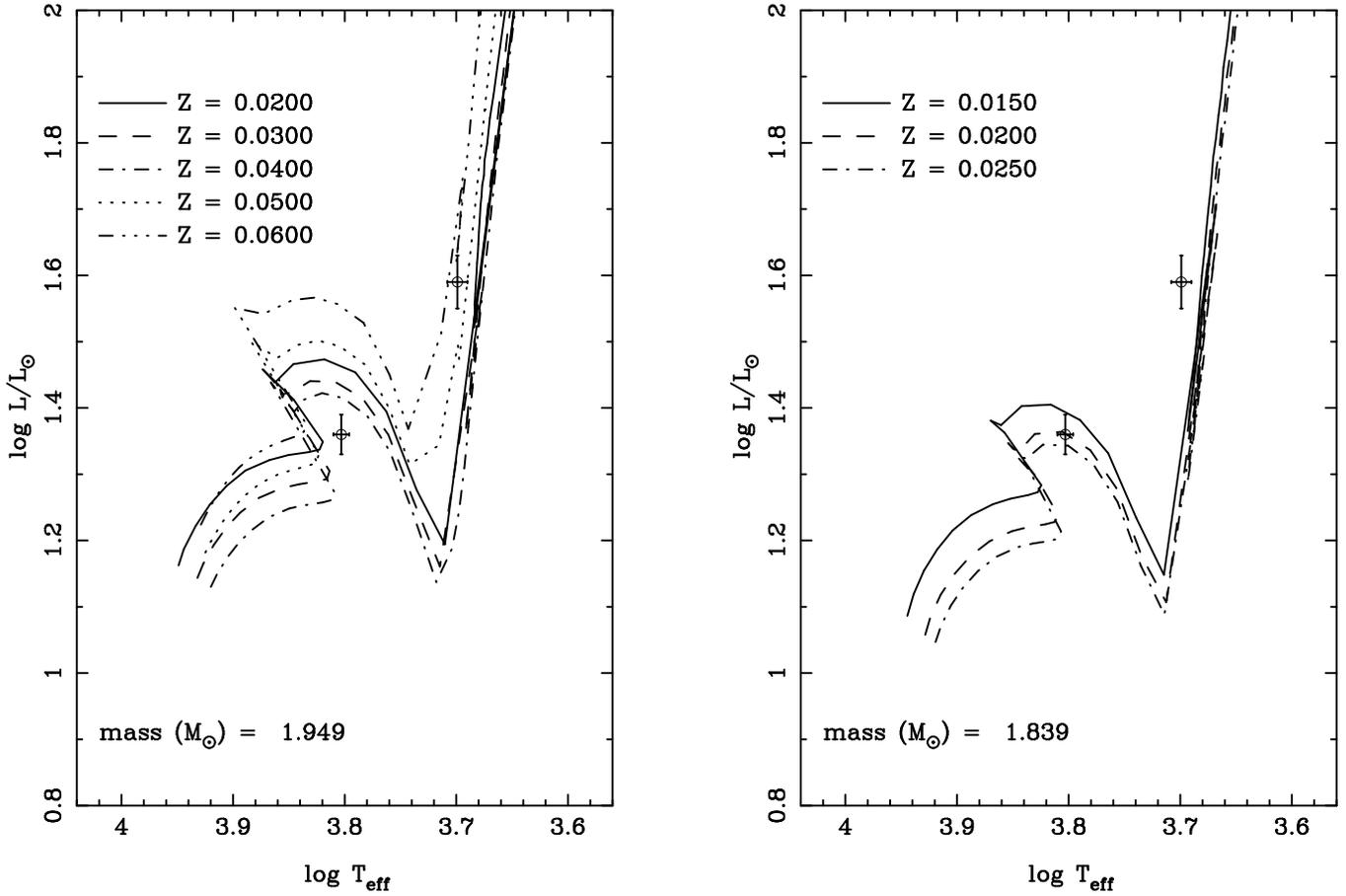}
\caption{TZ For system [18].
The hottest component of TZ For, TZ For B, is a star whose mass is known
to 1\% accuracy (1.949$\pm$0.022 $M_{\odot}$).
Geneva tracks of several metallicities (with a mass equal to the TZ For B mass) are shown,
but none of them fits this component ({\it left panel}).
In order to fit TZ For B with the observed metallicity, 
the mass has to be reduced by 5$\sigma$ ({\it right panel}).
 }
\label{f:tzfor}
\end{figure*}

TZ For is a rare occurrence of a system with a sub$-$gi\-ant and a giant, 
and hence potentially one of the most constraining systems. 
Moreover, there are spectroscopic data for this binary:  
Andersen {\al} (1991) give [Fe/H] $= +0.1\pm$0.15\footnote{  
i.e. $Z = 0.021^{+0.009}_{-0.006}$ (with $Z_{\odot} =$ 0.0169, VandenBerg 1985) 
or $Z = 0.023^{+0.010}_{-0.007}$ (for $Z_{\odot} =$ 0.0189, Maeder \& Meynet 
1988).}. 
The primary of TZ For is too evolved (core helium burning phase, as suggested by 
Claret \& Gim\'enez, 1995a and Pols {\al}, 1997) to be matched by
Granada 
models, hence 
only results derived for the B component are given in Tab.~\ref{tab:results}. 
The agreement around solar metallicity is quite 
satisfactory for the Geneva models, but seems difficult to match with the
Padova tracks: the lower limit of the Padova  models at the 1$\sigma$ level is
only marginally  consistent with a solar metallicity. 
In spite of the better agreement obtained with the Geneva models, a remark 
has to be made. 
The masses of the TZ For system are known with  great accuracy (better than
3\% for the primary and almost 1\% for the secondary) and  provide stringent 
tests: we should derive the same masses with the theoretical tracks. Thus,
if we consider TZ For B which is the strongest test and fix the mass of the 
Geneva tracks to its measured value, Fig. \ref{f:tzfor} clearly shows that none of
the  tracks succeeds in fitting this star whatever the metallicity
used (cf. {\it left panel}).
Alternatively, if we fix the metallicity to its observed value (Andersen
{\al}  1991), the mass of the tracks have to be reduced by 5$\sigma$ to
reproduce the mass of TZ For B (see {\it right panel}). 
This shows that the Geneva models do predict a metallicity consistent 
with the spectroscopic value, but if the metallicity is fixed they do not
predict the correct mass any more with the required precision.   

\subsubsection{\object{VV Pyx} [19]} 
\label{section:vv.pyx}

\begin{figure}[htb]
\psfig{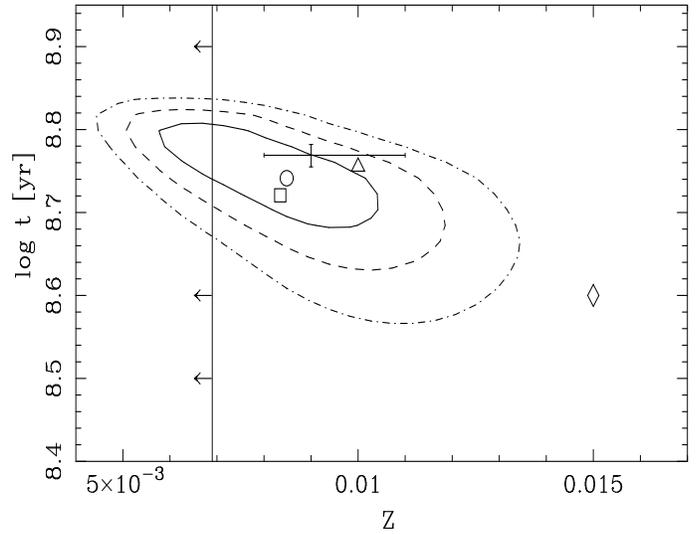} 
\caption{VV Pyx system [19]. Comparison between the best fits obtained
using the Geneva (circle), Padova (square)
and Granada (triangle) tracks, superimposed on the confidence regions 
of the Geneva tracks. The best solutions of Andersen, Clausen \& Nordstr\"{o}m (1984a) 
using the H80 models (diamond) and Pols {\al}, 1997 (error 
bars) are also shown. The photometric constraint on [Fe/H] (Lastennet {\al}  1999a)
is shown as a solid line with arrows. 
} \label{fig:VVPyx}
\end{figure}

The best isochrone solution found with the Padova tracks gives 
$(\log t, Z) =$ (8.721, 0.008) while the Geneva tracks 
predict $\log t =$ 8.753$\pm$0.04 also at the same metallicity. The Granada tracks give 
solutions in good agreement with Padova and Geneva: $(\log t, Z) =$ (8.76$\pm$0.04, 
0.010$\pm$0.001), see Fig.~\ref{fig:VVPyx}. It is unfortunate that there is no metallicity 
indications, because Andersen, Clausen \& Nordstr\"{o}m (1984a) derived a 
solution with a metallicity twice as large ($\log t, Z) =$ (8.6, 0.015) using
the H80 models. The difference in the solutions  certainly arises from the
different physical ingredients of these tracks with respect to the sets we
have used, as explained by Andersen {\al} (1984). 
A measure of [Fe/H] would therefore be an ideal test here, even though spectra
may be contaminated by the visual companion, a main$-$sequence A5$-$A7 star,
itself perhaps a spectroscopic binary (Andersen, Clausen \& Nordstr\"{o}m 1984a). 
The results we derived from the HR diagram would be unchanged with the  
{\teff}s of Ribas {\al} (2000), because the revision is tiny  
($\Delta${\teff}$=$22 K). An attempt to 
derive [Fe/H] from the synthetic BaSeL photometry (Lastennet {\al} 1999a)
suggests a metallicity markedly smaller than solar ([Fe/H]$<-$0.45, i.e.
Z$<$0.007), but the fit of Str\"omgren colours and $\log g$ is bad. While this
bad fit may reveal a problem of the BaSeL models, incorrect colours may be an
alternative explanation, supported by the possible contamination of
the visual companion.   

\subsubsection{\object{V1647 Sgr} [22] }

\begin{figure}[htb]
\psfig{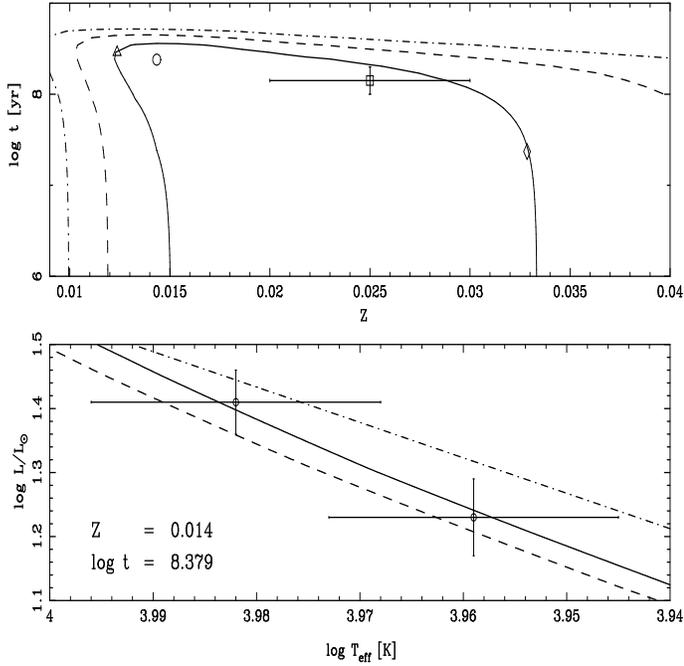}
\caption{V1647 Sgr system [22], Padova tracks. 
{ \it Upper panel: } $\chi^2$ contours in the (metallicity, age) plane.
Note the shift in metallicity between the best solution from Padova tracks (open circle 
at ($\log t, Z) =$ (8.38, 0.014)) and the best fit by Andersen \& Gim\'enez (1985) 
(square with error bars) using the H80 models. 
The two other points (triangle and diamond) define the two corresponding 
isochrones (dashed line and dot--dashed line respectively) in the lower panel. 
{\it Lower panel: } Best fitting Padova isochrone  (solid line), along with
two isochrones at 1$\sigma$. }
\label{fig:figv1647}
\end{figure}

Andersen \& Gim\'enez (1985) used the H80 models to derive best fits  
between ($\log t, Z, Y) =$ (8., 0.03, 0.27) and (8.30, 0.02, 0.23).
With the Geneva  tracks, we get an age consistent with these results ($\log t \sim$ 8.24), 
but a solar metallicity ($Z=0.017$), even though a larger metallicity 
(say $Z=0.023$) at a slightly smaller age is not excluded by the $\chi^2$-contours. 
The Padova models agree with these values (see Fig.~\ref{fig:figv1647}): 
$\log t \leq$ 8.40 and $Z =0.014^{+0.019}_{-0.002}$ as well as the 
Granada tracks : $\log t \in$ 
[7.50, 8.62]  and $Z = 0.016^{+0.004}_{-0.002}$. 
Andersen \& Gim\'enez (1985) quote a possible observed range for $Z$ between 0.02 and
 0.04, which is consistent with the 3 sets of tracks used here, although
smaller values are permitted. If the lower limit of  $Z \geq$ 0.02 is confirmed,
the Granada tracks may not be able to fit the properties of
this system. 

Note however that  the rotation velocities of the components are
not negligible  ($v \sin i)_{A} = 80\pm5$ km s$^{-1}$ and $(v \sin i)_{B} = 
70\pm5$ km s$^{-1}$ (Andersen \& Gim\'enez 1985) and that isochrones
incorporating rotational effects would give slightly younger ages 
and larger metallicities.  It has also been pointed out that the
parameters of this system may have been perturbed by the presence
of its bright visual companion, and this has not been taken into
account in the error budget (Ribas {\al} 1998).

\begin{table*}[hbt]
\caption[]{Comparison of age ($\log t$, in yrs.), metallicity ($Z$), 
and helium abundance ($Y$) determinations obtained for YZ Cas [25].} 
\label{tab:YZCas}
\begin{flushleft}
\begin{center}
\small
\begin{tabular}{lr@{$\pm$}lr@{$\pm$}lr@{$\pm$}lc}
\hline\noalign{\smallskip}
\multicolumn{1}{c}{  Tracks       }      & 
\multicolumn{2}{c}{  $\log t$      }      &
\multicolumn{2}{c}{  $Z$            }      &
\multicolumn{2}{c}{  $Y$$^{(\dag)}$ }      &  
\multicolumn{1}{c}{ Reference     }        \\ 
\noalign{\smallskip}
\hline\noalign{\smallskip}
Mengel {\al} (1979)  &  8.544&0.004          &  0.027&0.003        &  0.30&0.02         &  (1)  \\
Hejlesen (1980a,b)     & \multicolumn{2}{c}{8.613$^{+0.040}_{-0.044}$ } &  0.03&0.01   &  0.30&0.04  & (1) \\
Mengel {\al} (1979)    &  8.8&0.1           &  0.010&0.005        &  0.30&0.05          & (2) \\
Hejlesen (1980a,b)       &  8.5&0.1           &  0.015&0.005        &  0.335&0.050        & (2)  \\
de Loore {\al} (1977)  &  8.65&0.10         &  0.015&0.005        &  0.335&0.050        & (2) \\
Celikel \& Eryurt-Ezer (1989) & \multicolumn{2}{c}{8.77$\pm$0.02} & \multicolumn{2}{c}{0.021$^{(\star)}$} 
& \multicolumn{2}{c}{0.240$^{(\star)}$}& (3) \\
Pols {\al} (1998) STD  &  \multicolumn{2}{c}{8.66$\pm$0.02} &  
\multicolumn{2}{c}{0.014$^{+0.004}_{-0.003}$}  &  \multicolumn{2}{c}{0.268$^{+0.008}_{-0.006}$} & (4) \\
Pols {\al} (1998) OVS  &  \multicolumn{2}{c}{8.70$\pm$0.02} &  
\multicolumn{2}{c}{0.016$^{+0.006}_{-0.004}$}  &  \multicolumn{2}{c}{0.272$^{+0.012}_{-0.008}$} & (4) \\
CG$^{(a)}$  &  \multicolumn{2}{c}{8.654$\pm$0.014} &  
\multicolumn{2}{c}{0.016$\pm$0.004}  & \multicolumn{2}{c}{0.270$\pm$0.026}  & (5) \\
CG92$^{(b)}$                 &  \multicolumn{2}{c}{8.73$^{+0.07}_{-0.19}$} &  
\multicolumn{2}{c}{0.017$^{+0.013}_{-0.004}$}  &  \multicolumn{2}{c}{0.276$^{+0.045}_{-0.006}$}  & (6) \\
Geneva                       &  \multicolumn{2}{c}{8.65$^{+0.10}_{-0.17}$} &  
\multicolumn{2}{c}{0.018$^{+0.009}_{-0.004}$}  &  \multicolumn{2}{c}{0.293$^{+0.020}_{-0.011}$}  & (6)  \\
Padova                       &  \multicolumn{2}{c}{8.62$^{+0.09}_{-0.19}$} &  
\multicolumn{2}{c}{0.020$^{+0.016}_{-0.007}$}  &  \multicolumn{2}{c}{0.280$^{+0.035}_{-0.017}$}  & (6) \\
\noalign{\smallskip} \hline \noalign{\smallskip}
Str\"omgren photometry (B comp.) & \multicolumn{2}{c}{ } & \multicolumn{2}{c}{0.018-0.020}  
&  \multicolumn{2}{c}{ } & (5) \\
\noalign{\smallskip}  
\hline
\hline
\noalign{\smallskip}
\noalign{\smallskip}
\end{tabular}
\end{center}
\scriptsize
$^{(a)}$ Claret (1995), Claret \& Gim\'enez (1995), Claret (1997), Claret \& Gim\'enez (1998); 
$^{(b)}$ Claret \& Gim\'enez (1992); $^{(\star)}$ Assumed composition. \\
(1) Lacy (1981); (2) De Landtsheer \& De Gr\`eve (1984); (3) Celikel \& Eryurt-Ezer (1989); 
(4) Pols {\al} (1997); (5) Ribas {\al} (2000): [Fe/H]$=$0.03, i.e. $Z$$=$0.018
(assuming $Z_{\odot}$$=$0.017, Grevesse 1997, priv. comm.) 
or $Z$$=$0.020 (assuming  $Z_{\odot}$$=$0.0188, Schaller {\al} 1992); (6) This
work. $^{(\dag)}$ Only the CG models allow an independent determination of $Y$
and $Z$. Otherwise, $Y$ is derived from a fixed law once $Z$ is determined. 
\end{flushleft}
\end{table*}

\subsubsection{\object{YZ Cas} [25] }
 
\begin{figure}[htb]
\psfig{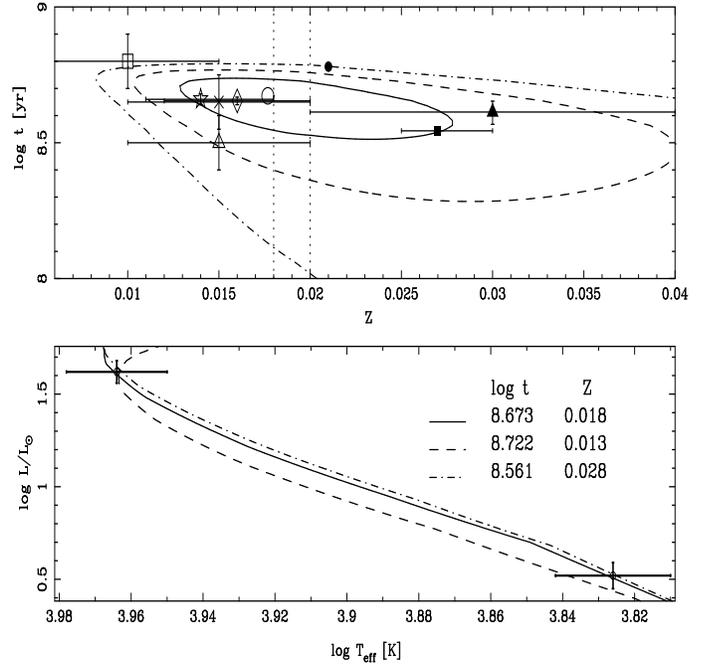}
\caption{YZ Cas [25], Geneva tracks 
(open circle symbol for the best fit, and 1-, 2- 3-$\sigma$ contours). 
For comparison, the results of previous studies are indicated: 
Lacy (1981) using the MDSG79 tracks (filled square) and the H80 models
(filled triangle);
De Landtsheer \& De Gr\`eve (1984) with the MDSG79 (square),  
the H80 (triangle), and the De Loore {\al} (1977) models (cross);
Celikel \& Eryurt-Ezer (1989) (filled circle).  
More recent studies from Pols {\al} (1997) (open star), and Ribas {\al}
(2000) (diamond) are also shown. 
A metallicity estimate from photometry is indicated with vertical dotted lines.
See Table~\ref{tab:YZCas} for further details.  
}
\label{fig:YZCas}
\end{figure}

This system has been studied a lot previously, and Fig.~\ref{fig:YZCas} 
and Table~\ref{tab:YZCas} summarize the results obtained so far.
Except for the measures of Lacy (1981),  
all sets of models (whatever the stellar parameters used) agree in
supporting a 
solar metallicity ($Z$ range between 0.015 and 0.020). As quoted in Tab.~\ref{tab:YZCas} 
a solar composition is supported by photometric determination on the B 
component, a F2V star (Ribas {\al}, 2000). 
Even though its components are not very evolved, YZ Cas is a stringent test of 
the models because of the relatively large difference in its masses. 
The metal abundance determination of R00 is therefore a strong support to the 
quality of the tracks in the $\sim$ 1.3-2.3 {\MSun} range.\\

Apart from this very good agreement between models and data, we have to 
mention that there is still an open problem concerning the primary component, 
YZ Cas A, a metallic Am star. 
While a disagreement may be expected between the atmospheric metallicity 
($Z$$_{\rm atmos.}$) and its intrinsic composition ($Z$, as derived from the
models) due to an enhancement of the metal abundance in its external layers, 
Lastennet, Valls-Gabaud \& Jordi (2001) found some discrepancy between the 
$Z$$_{\rm atmos.}$ determinations. 
While photometric methods suggest that $Z$$_{\rm atmos}$ is 
solar\footnote{
$Z$$_{\rm atmos}$$=$0.017$^{+0.012}_{-0.007}$ (Lastennet, Valls-Gabaud \& Jordi, 2001) 
and $Z$$_{\rm atmos}$$\leq$0.022 (L99a). 
}, 
$Z$$_{\rm atmos}$ derived from IUE observations by De Landtsheer \& Mulder
(1983) is still high ($Z$$_{\rm atmos.}$$=$0.036$\pm$0.005) despite the attempt
of  Lastennet, Valls-Gabaud \& Jordi (2001) to revise some of the assumptions
made by the IUE-based spectral  study. 
More detailed analyses are needed to understand this disagreement. 

%

\subsubsection{\object{$\beta$ Aur} [27]}  
\label{section:betaAur}  

\begin{figure}[h]
\psfig{figure=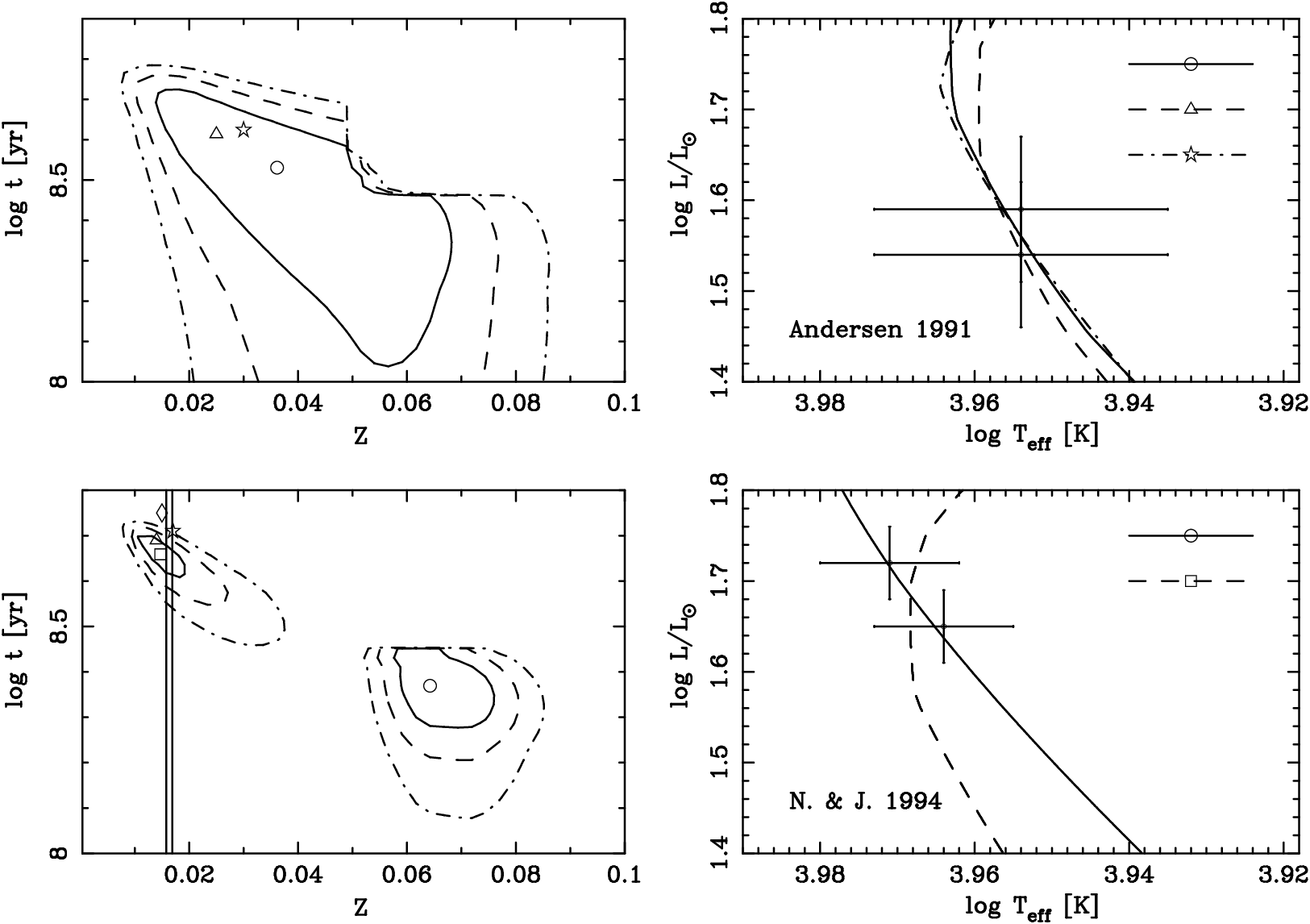,width=9.cm,height=8.8cm,rheight=9.truecm,angle=-90.}
\caption{$\beta$ Aur [27], contours with Padova models.  
{\it Upper panels : } assuming the A91 values in the HR diagram, best fits 
with Padova (circle), Geneva (triangle) and Granada (open star) tracks. Note that the Granada value 
is at the upper limit available in this set ($Z=$ 0.03).
{\it Lower panels : } Using the updated parameters from NJ94b, the 
best fit, high metallicity, Padova isochrone (open circle) is consistent with the best fit 
isochrones from Geneva (triangle) and Granada models (open star), even though their metallicities 
differ by a factor of 3. Also shown is a Padova isochrone (dashed line) of solar metallicity 
(open square). The diamond  represents the NJ94b solution (using the Granada tracks). 
The observed constraints on metallicity (Toy 1969) between [Fe/H] =$-$0.03 
and [Fe/H] $=$ 0  are indicated with vertical lines. 
}
\label{fig:figbetaaur}
\end{figure}

In this system we can compare the effects of increased precision in the measures of its 
parameters. Fig.~\ref{fig:figbetaaur} uses the $\log T_{\rm eff}$ and $\log L$ parameters 
listed by Andersen (1991) (upper panels) and the more recent ones\footnote{Not modified by 
the {\teff} revision of R00.} from Nordstr\"{o}m \& Johansen (1994b,  hereafter NJ94b) in the lower 
panels. The former values give  $Z =$ 0.025 and $\log t $= 8.614 using the
Geneva tracks, similar to the best fit values found for each component 
separately\footnote{
($\log t, Z$)$_{A}$ $=$ 8.587, 0.029 and ($\log t, Z$)$_{B}$ $=$ 8.640, 0.023.}. 
These results are inconsistent with the  constraints in
metallicity given by Toy (1969) : $-$0.03 $\leq$ [Fe/H] $\leq$ 0, and
the disagreement is even worse with the  Padova tracks, since they
give best fit models at 
$Z =$ 0.035 and $\log t =$ 8.537, although at the 1$\sigma$ level they
may be compatible. 
 
The NJ94b analysis yields  
 two components which are well separated in the HR diagram and the constraints 
in the ($Z$, $\log t$) diagram are much
better, as the lower panels of  Fig.~\ref{fig:figbetaaur} show.
The uncertainty regions are greatly reduced and the solutions are shifted towards smaller 
metallicities. 
Nordstr\"{o}m \& Johansen (1994b) obtained $Z =$ 0.015 and an associated age of 
$\log t =$ 8.75\footnote{As also noted by Nordstr\"{o}m \& Johansen (1994b),
this system is about 8 times younger than the Sun, and yet has a similar metallicity.}.
The solutions obtained here with the Geneva ($Z =$ $0.014^{+0.006}_{-0.004}$), Padova 
($Z =$ $0.066^{+0.010}_{-0.059}$, see Fig.~\ref{fig:figbetaaur}, lower panels) and Granada 
($Z =$ $0.017^{+0.002}_{-0.005}$) tracks are all
in good agreement and consistent with the metallicity constraint of
 Toy (1969). Note however 
that there is an {\it island} of solutions at high metallicity and smaller age that remains with the 
Padova tracks, giving rise to isochrones that are statistically indistinguishable from the lower 
$Z$ ones (see the left bottom panel in Fig.~\ref{fig:figbetaaur}). This {\it island} is in principle 
excluded because the implied metallicity is much larger than the observed one,
although a modern spectroscopic determination would be useful to
confirm Toy's (1969) values.  

\subsubsection{\object{V1031 Ori} [28] }
 \label{section:V1031Ori}

 All the models converge towards the same solution for this system. 
 Andersen, Clausen \& Nordstr\"{o}m (1990) used two sets of tracks and derived 
 ($\log t, Z, Y$) = (8.74, 0.024, 0.27)  with VandenBerg's (1985) models and 
 ($\log t, Z, Y$) = (8.85, 0.02, 0.28) with the older generation of Geneva models 
 (Maeder \& Meynet 1988, 1989). The updated Geneva tracks give 
 a best fit at ($\log t, Z$) = (8.771, 0.021), while the Padova tracks indicate  
 ($\log t, Z$) = (8.747, 0.020), in agreement with Pols {\al} (1997), 
 ($\log t, Z$) = (8.79, 0.023), and Ribas {\al} (2000), 
 ($\log t, Z$) = (8.846, 0.016). 
 Granada models favour coeval solutions with a larger metallicity ($Z \approx$ 
 0.027), although at the 1$\sigma$ confidence level there is no disagreement
 with the other sets of tracks, and with the photometric constraint 
 derived from the BaSeL models on V1031 Ori B ($Z$$<$0.029, Lastennet {\al}
 1999a). Nevertheless, another photometric determination (Ribas {\al}, 
 2000) suggests a sub-solar  metallicity ($Z$$\sim$0.010) which would 
 imply a systematic problem in all the above-mentioned theoretical models 
 for main-sequence stars around 2.3--2.5 {\MSun}.  
 The CG models used by Ribas {\al} (2000) would be marginally consistent 
 while the others are clearly ruled out. 
 Once again, a detailed spectroscopic analysis would be needed before 
 further inferences.  

\begin{figure}[htb]
\psfig{figure=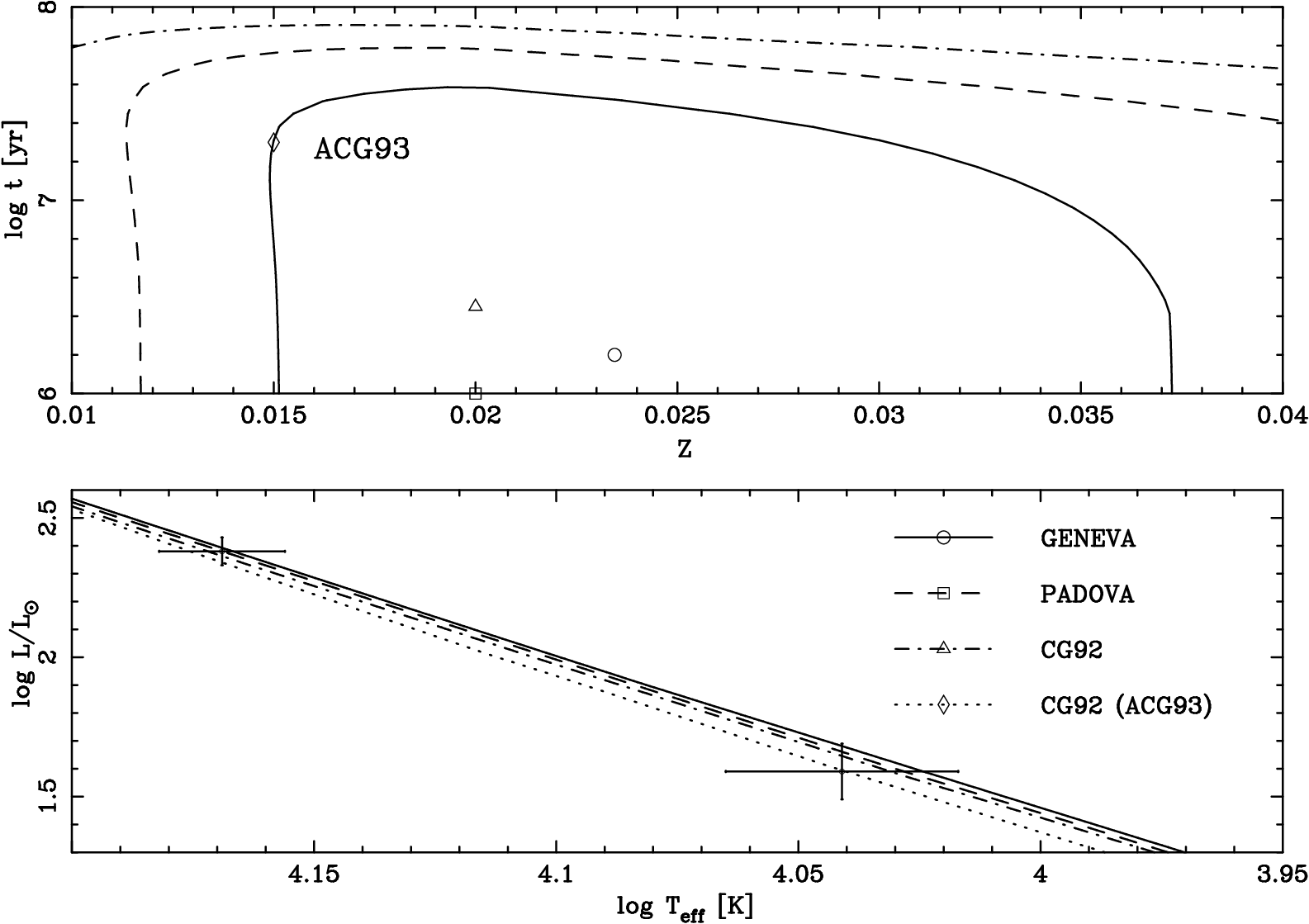,width=9.cm,height=8.8cm,rheight=9.truecm,angle=-90.}
\caption{GG Lup [36], Geneva tracks contours. Best fits with Geneva (circle), Padova (square) 
and Granada (triangle) tracks are shown, as well as the best fit by Andersen {\al} (1993) 
(diamond, ACG93) with the Granada tracks. The data points in the HR diagram are 
from Andersen {\al} (1993).}
\label{fig:GGLup}
\end{figure}

\subsection{Systems with at least one star more massive than 3 {\MSun}}
\subsubsection{\object{GG Lup} [36] }
\label{section:GGLup}

For this system we use the Andersen {\al} (1993) data which update the 
A91 review. Even though the positions of the components in 
the HR diagram have not changed, the accuracy has increased two- or three-fold. 
This improvement results in smaller uncertainty regions\footnote{
The A91 parameters give 
$Z_{Geneva}$ $\in$ [0.011, 0.040], $Z_{Padova}$ $\in$ [0.010, 0.062]
and $Z_{Granada}$ $\in$ [0.010, 0.030], while the ones from Andersen {\al} (1993) 
produce $Z_{Geneva}$ $\in$ [0.015, 0.037],
$Z_{Padova}$ $\in$ [0.014, 0.035] and $Z_{Granada}$ $\in$ [0.015, 0.030].}, 
but still highly elongated along the $Z$ direction (Fig.~\ref{fig:GGLup}). 
 Andersen {\al} (1993) derived a best fit isochrone with 
($Z, \log t) =$ (0.015, 7.30) with the Granada models\footnote{P97 obtained similar 
solutions: Z$=$0.015$\pm$0.002, $\log t=$7.18$^{+0.14}_{-0.22}$. R00 obtained 
no solution in the ($Z$, $Y$) range covered by the CG models.}, claiming that
this  metallicity is in agreement with unpublished preliminary determinations
by Clausen. Our results are consistent with these values at
the 1$\sigma$ level, but a much wider range in $Z$ is allowed by the
evolutionary tracks. 
Basically all isochrones with ages younger than 5$\times 10^{7}$ yr and  
a metallicity between $Z$$=$0.014 and $Z$$=$0.037 can fit reasonably well this
system in the HR diagram.   
These results are unchanged if we consider the recent {\teff} revision 
from R00.  
On the other hand, Lastennet {\al} (1999a)  
provide a slightly hotter {\teff} (by $\sim$0.9\%) and a smaller uncertainty 
(by a factor of 3) on the coolest 
component\footnote{log {\teff}$=$4.045$^{+0.005}_{-0.011}$, to be compared with 
A91 or R00: log {\teff}$=$4.041$\pm$0.024.} which would imply 
smaller contours in Fig.~\ref{fig:GGLup}, excluding all the solutions with 
Z$\geq$0.025.  

\subsubsection{\object{V539 Ara} [41]} 
\label{section:V539Ara}

\begin{figure*}[htb]
\psfig{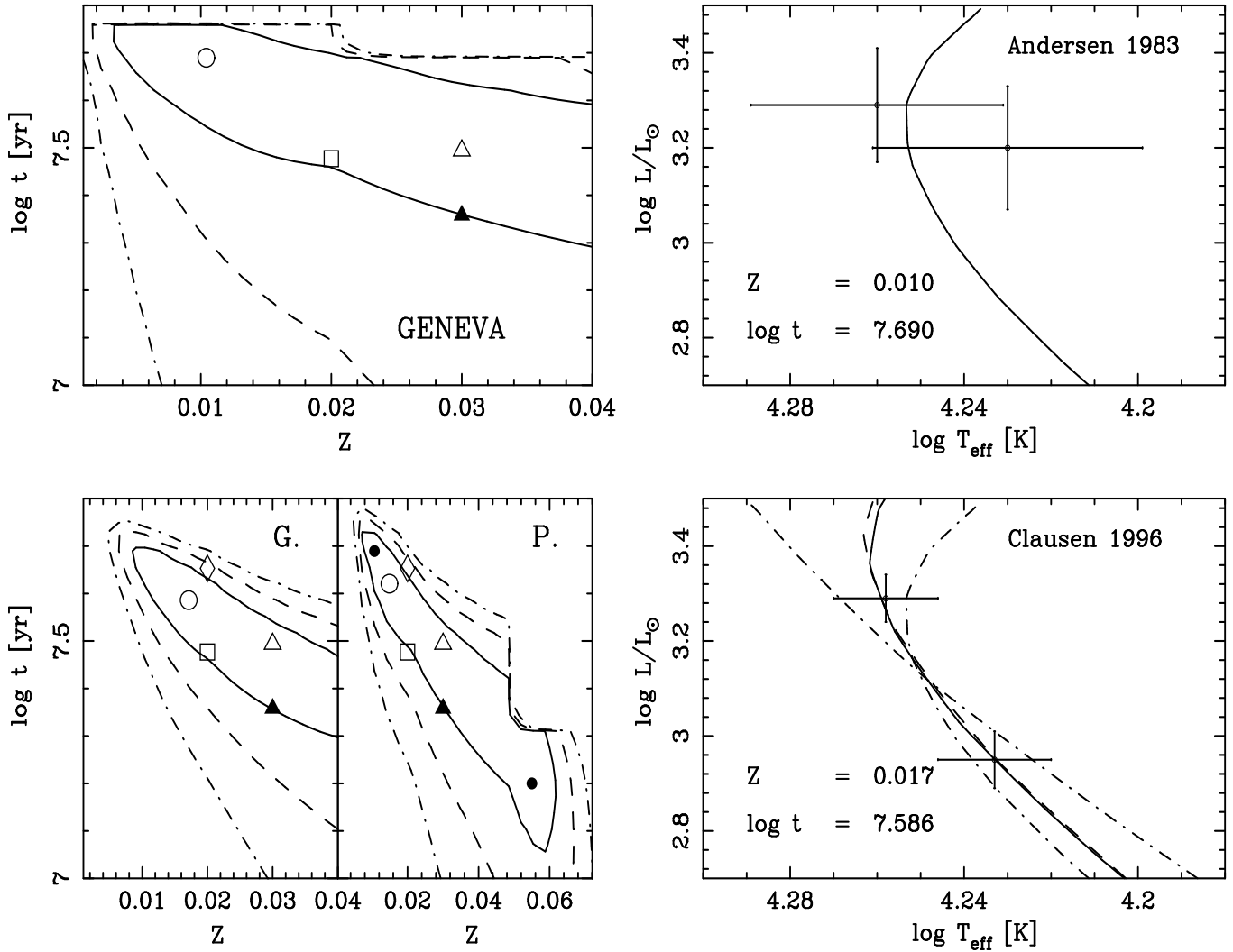}
\caption{V539 Ara system [41]. {\it Top panels: } metallicity--age contours
and HR diagram with the Andersen (1983) data. The best fitting isochrone with
the current Geneva tracks (open circle, at ($\log t, Z) =$ (7.7, 0.01)) gives a smaller 
$Z$ and larger age than previous determinations (square: Clausen 1979 [with the Hejlesen 
{\al} 1972 tracks]; open triangle : De Gr\`eve 1989 with overshooting; filled triangle : 
De Gr\`eve 1989 without overshooting [the latter two with the Prantzos
{\al} 1986 tracks]). {\it Lower panels: } The same as above, but with the
more recent values from Clausen (1996) and tracks from Geneva (G.) and Padova (P.).
The diamond represents the best fit value obtained by Clausen (1996) using 
the Granada tracks. The open circles (G. at ($\log t, Z) =$ (7.58, 0.017) and P. at 
($\log t, Z) =$ (7.62, 0.015)) give the best fits from Geneva (full line) 
and Padova (dashed), while typical Padova 1$\sigma$ isochrones (filled circles) are
also shown for comparison (dot-dashed lines). }
\label{fig:v539ara}
\end{figure*}

This system was studied by Andersen (1983), and more recently by Clausen (1996), 
allowing to study the influence of more accurate data on the best fit isochrones. 
Figure~\ref{fig:v539ara} uses the Andersen (1983) data in the top panels and the 
Clausen (1996) values in the lower panels.  
Both Geneva and Padova tracks give a smaller $Z$ and a larger age than previous 
determinations\footnote{e.g. Clausen (1979) with the Hejlesen {\al} (1972) tracks, 
De Gr\`eve (1989) with or without overshooting (tracks from Prantzos {\al} 1986).}, 
although they 
all lie on the 1$\sigma$ contours. The updated values decrease somewhat the uncertainty 
area in the metallicity--age plane, but a large range of $Z$ is still allowed (see 
Fig.~\ref{fig:v539ara}), despite the two-fold increase in the accuracy of the data. 
Unfortunately, as indicated by Clausen (1996), attempts to measure the metallicity are 
hampered by the relatively large rotational velocities of its components, respectively 
 75 km s$^{-1}$ (component A) and 48 km s$^{-1}$ 
(component B). \\
The best fit values obtained by Clausen (1996) using the Granada tracks are 
($\log t, Z, Y$) = (7.65, 0.02, 0.28), identical to our results (see Table 
\ref{tab:results}), but there are many other possible solutions given approximately by the  
confidence regions ($\log t, Z$) = ($7.65^{+0.08}_{-0.21}$ yr, $0.020^{+0.010}_{-0.008}$). 
This point emphasizes the power of the method used in the present work: 
the confidence regions are essential to properly assess and 
understand the results of the tests. 

\subsection{Very massive stars: M $>$ 10 {\MSun}}
As quoted in 1987 by Hilditch \& Bell, there were not many accurate systems 
in this mass range, and this is unfortunately still true: 
only 7 systems of our working sample fall in this category, namely (by 
increasing order of mass) \object{CW Cep} [43], \object{AH Cep} [51], 
\object{V478 Cyg} [44], \object{Y Cyg} [50], 
\object{EM Car} [45],  \object{V3903 Sgr} [52] and 
\object{DH Cep} [53]\footnote{As quoted in 
\S\ref{section:sample}, the last one does not match the 1-2\% 
level of accuracy of the core sample studied in this paper.}. 
The main reason is that the MS lifetime of such massive stars is
short, and so  these 
binaries are often observed during their interacting phase of evolution, excluding 
them from our sample of well-detached systems. 

All these systems have good fits, and the more massive ones  
($\geq$ 15 {\MSun}, which then excludes CW Cep and AH Cep) seem to
yield  low 
metallicities. 
As mentioned earlier on (\S\ref{section:sample}), DH Cep may have overflown its 
Roche lobe, and  we exclude it from further discussion (even though 
it follows this trend of lower $Z$ as well). 
Previous studies with different fitting methods also show the same trend: 
extrapolated solutions at $Z=$ 0.009 for V478 Cyg, Y Cyg and EM Car, 
with a possibly even lower $Z$ for Y Cyg according to its bad fit 
(Pols {\al}, 1997),  
$Z=$0.005 for EM Car and $Z=$0.010 for V3903 Sgr (Ribas {\al}, 2000).
   
Can new and carefully derived temperatures change this low metallicity trend? 
The revised temperatures of Ribas {\al} (2000) leave unchanged the {\teff} 
of V478 Cyg, EM Car, Y Cyg and V3903 Sgr and decrease by only 0.003 dex 
both components of CW Cep, therefore same results are expected with these 
new determinations.  
Obviously, results that can be derived for O-B binaries in the HR
diagram have to be 
taken cautiously due to the more uncertain {\teff}-scale for massive 
stars, and so definitive conclusions are premature.    
Moreover, because metal abundance is difficult to determine in hot stars, 
it becomes difficult to discriminate between models. 
For this reason, EM Car deserves more discussion because it is the 
only system with metallicity indications. 
Our theoretical predictions favour low metallicities: 
$Z$$=$0.003 (Geneva models) and $Z$$=$0.004 (Padova models), 
in agreement with Pols {\al} 1997 ($Z$$=$0.009, extrapolated solution), 
and Ribas {\al} 2000 
($Z$$=$0.005$\pm$0.002)\footnote{However all these stellar 
evolution models might be systematically wrong 
for very massive stars (rotation, mass loss rates, diffusion
effects).}. 

The metallicity constraint on  EM Car is
[Fe/H]$<$0.10, as inferred from the photometry of its secondary
component, and 
favours low metallicity solutions ([Fe/H]$=-$0.90, i.e. $Z$$\sim$0.004). 
This result is 
obtained from photometric calibration (BaSeL library) matching the very 
accurate surface gravity (log g$=$3.928) and the observed  dereddened colours
(b$-$y)$_0$, m$_0$, and c$_0$ (Lastennet {\al} 1999a),  and supports the
inferred low-$Z$ trend mentioned above.  

Alternative explanations for low metallicities in massive binaries may 
come from the fact that the models may be too dim, so low metallicties 
models can fit the data without a proper 
physical reason (Young {\al} 2001). 
However, while Young {\al} (2001) and our results do use the information 
on the luminosity, this do not explain the low-$Z$ results of Pols {\al} 
and Ribas {\al}. Therefore the heavy element abundance may be actually 
small for the systems of the sample with masses larger than $\sim$ 
15 {\MSun}. More accurate data for massive stars are needed to solve 
this issue.   

\section{General discussion}
\label{section:discussion}

We discuss here the implications of the careful determinations of
age and metallicity that we have obtained for most of the 
systems of our sample.

\begin{figure*}[htb]
\psfig{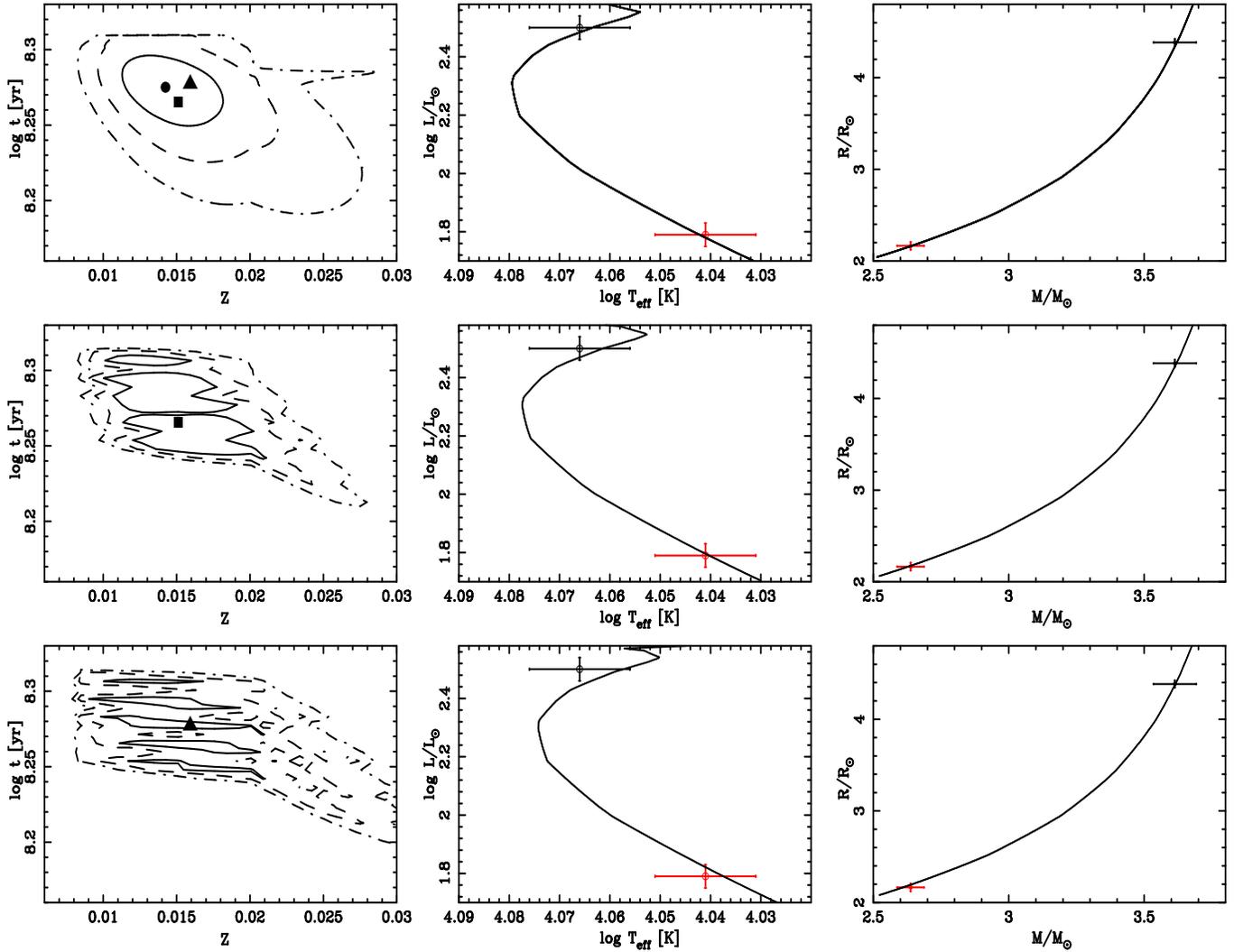}
\caption{
$\chi^2$ Hya system [34], Geneva tracks. 
We use 3 different $\chi^2$-estimators to fit both components : 
(1) a fit matching {\teff} and $L$ (our standard formulation) (upper panels), 
(2) a simultaneous fit of {\teff}, $L$, $M$ and $R$ (middle panels), 
and (3) a fit involving only $M$ and $R$ (lower panels). 
For each $\chi^2$ estimator the confidence regions in the
metallicity-age plane are given (left panels), along with 
the resulting HR diagrams (middle panels) and Mass-Radius diagrams (right
panels). For comparison, the minima obtained from each method (filled 
circle, square and triangle for methods (1), (2) and (3), respectively) 
are plotted on the upper left panel. 
}
\label{fig:estimators}
\end{figure*}

\subsection{Possible systematic trends in age and metallicity} 
\label{section:comparison}
Before attempting to draw general conclusions from these results, it 
is worth asking whether the inclusion of further information would
improve the determinations of age and metallicity. After all, as we
showed in Section \S~\ref{section:method}, when fitting a single
isochrone to the system, the degeneracy area in the
age-metallicity plane was in general reduced, due to the extra
information added (two luminosities and two effective temperatures).
Would the inclusion of mass and/or radius, into the $\chi^2$ functional
reduce the uncertainties in the final ages and metallicities? 
 We compare here three different formulations
for the $\chi^2$ functional : (1) our standard expression, based
only on {\teff} and $L$; (2) a functional minimizing the residuals
using all the information available, namely the four quantities 
{\teff}, $L$, $M$ and $R$; and (3) a functional using only the
most direct quantities $M$ and $R$. This last formulation allows us 
to check the predictive power of the theory on the other two
parameters, {\teff} and $L$.
As in \S\ref{section:method}, we will take the system $\chi^2$ Hya
[34] as an example.

Each formulation is illustrated on Fig.~\ref{fig:estimators} with a $Z$--$\tau$ 
diagram, a classical HR diagram and a $M-R$ diagram. It is immediately clear
that in the ($Z$, $\tau$) plane the confidence areas are roughly similar 
in shape and extension, although
method (2) provides slightly tighter constraints, as expected, at the 3$\sigma$
level. However in this case there are several ``islands'' at the
1$\sigma$ level, giving rise to multiple, unconnected solutions, which are
difficult to understand given Vogt's theorem. This multimodality also
indicates that 1$\sigma$ solutions may be vastly underestimated if a
local minimum (and not the absolute one) is found instead. 
These islands are even narrower when using $M$ and $R$ only (method 3). 
While method (3) makes a good fit to $R$ and $M$, it fails to provide
an excellent isochrone in the HR diagram. Likewise method (2), which has
no predictive power in this sense, yields somewhat worse fits in the
HR diagram than method (1) which is therefore our preferred one 
for providing the best possible fit in the HR diagram {\it and} for predicting 
properly masses and radii. 

 At first sight, these results might be surprising:  
the extra information 
added (method 2) does not yield a vast improvement on the solution, nor do the
more precise masses and radii (method 3). 
The addition of more information (method 2) imposes much tighter constraints 
on the models, and simply reveals their deficiencies in a better way. 
There are two possible additional reasons for this. 
First, all these quantities are correlated, and hence bring less information
than naively expected. When we added the two components in
\S\ref{section:method} fitting a single isochrone, 
the correlations between the parameters of one component and the 
parameters of the other
remain, but are in general much smaller than those within a
single component, and therefore yield better constraints.
Another possible reason is that these systems remain close to the
ZAMS, where there are well-defined mass-radius and mass-luminosity
relations, implying even stronger correlations, and thus even
less information. 
Systems with components in widely-separated stages of evolution would 
probably make much larger differences. 
With the current sample of binaries, however, these
different formulations yield somewhat similar results, with method (1)
being preferred for yielding the best fits to $L$ and {\teff}, and
predicting correctly $M$ and $R$. \\
 An important motivation for this choice is that for single stars and 
less constrained systems such as visual binaries, one does not have $M$ 
and $R$, whereas $L$ and {\teff} can be determined in many cases.

\begin{figure}[htb]
\psfig{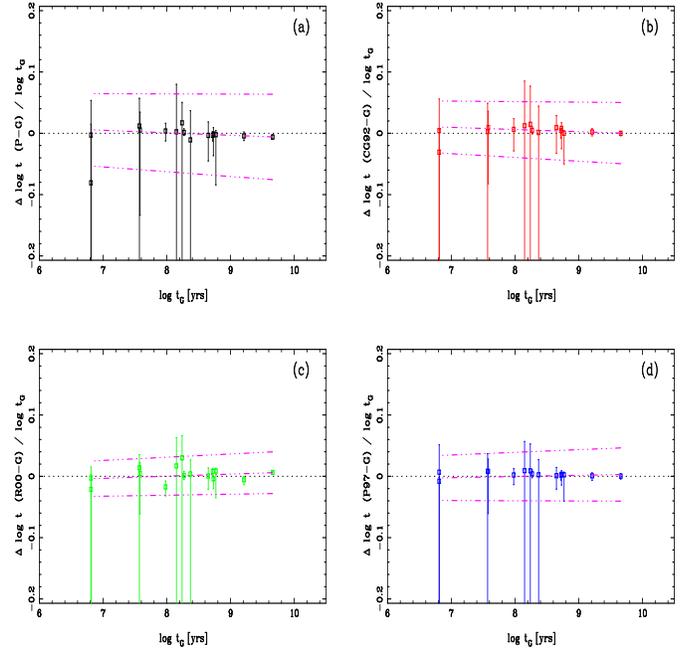} 
\caption{Relative errors in  age ($\log t$) derived with  
Padova models (a),  Granada tracks (b), R00 (c) and P97 (d), 
as a function of the age predicted by the Geneva tracks.
The dotted line is $\Delta$$\log t$ $=$ 0 and the dot-dashed lines 
are linear regressions to the data points.   
}
\label{fig:tcomp}
\end{figure}

Another question that arises is whether the solutions found here
are compatible with previous analyses, which used not only different
sets of tracks, but also different $\chi^2$ formulations. By and large,
the ($Z$,$\tau$) solutions are very similar and statistically
consistent with each other. But are there any systematic trends
between the different analyses? We compare here in detail our
results with those obtained by  Pols 
{\al} (1997) and by Ribas {\al} (2000).  We do not take into 
account the results from Young {\al} 2001 because they fixed all 
the metallicities to the solar value.  
The sub-sample in common  includes only 14 systems 
mainly because we are restricted to $Z$$<$0.03, the upper limit considered 
in P97 and R00.  
We plot the relative differences (in the sense results with Padova, Granada, 
R00 or P97 minus results derived from the Geneva tracks) in age on  
Fig.~\ref{fig:tcomp}. Not only the agreement between the 5 different
estimations are very good, but there are no systematic effects:  
the average relative error in age is $-$0.49\%, 0.32\%, 0.29\% and 0.35\%, 
with a dispersion of 2.2\%, 1.0\%, 1.3\% and 0.5\% for panels (a), (b), (c)
and (d), respectively. 
This is a very good result for the theoretical models considering the large 
individual error bars.  

\begin{figure}[htb]
\psfig{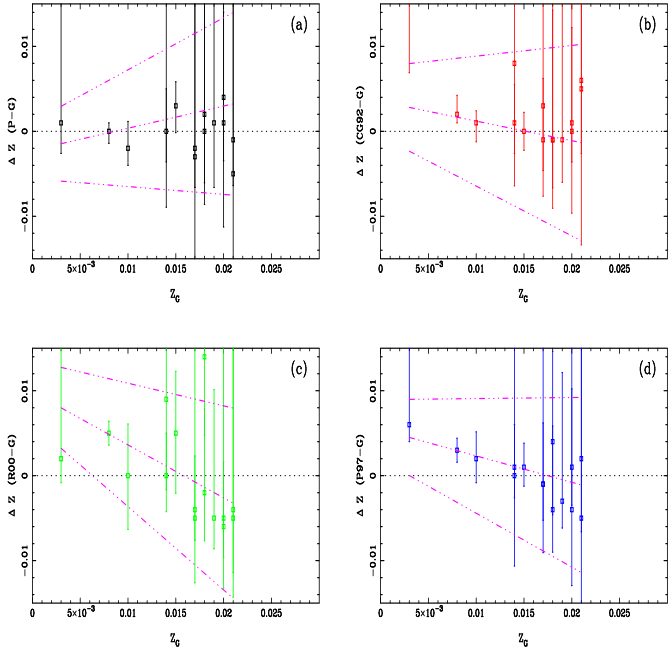}
\caption{Error in metallicity derived with Padova models (a), 
Granada tracks (b), R00 (c) and P97 (d), 
as a function of the metallicity predicted by the Geneva tracks. 
The dotted line is $\Delta$$Z$ $=$ 0 and the dot-dashed lines are 
linear regressions to the data points.
}
\label{fig:zcomp}
\end{figure}

The error in metallicity (Fig.~\ref{fig:zcomp}) is much larger, but 
results are consistent within 20\% for most cases, even though
uncertainties remain large. 
The metallicity provided by the Geneva tracks is compatible
with the one given by the Padova tracks (Fig.~\ref{fig:zcomp}a). 
There is a slight systematic trend of the Geneva (and hence
Padova) tracks with the other ones (Granada, P97 and R00) in the sense
that the low metallicity systems are given a smaller 
metallicity by the Geneva (and Padova tracks). However this trend is  
produced by two stars only ($\zeta$ Phe [35] and EM Car [45]) and
seems hardly significant because the P97 and R00 solutions are extrapolated 
for such low Z-values.  

\subsection{Metallicity constraints}
\label{section:feh}
Given that all these independent methods seem to predict consistent
ages, how do the metallicity values compare to the observed ones? 
Obviously the most discriminant systems are those, as pointed out 
by Andersen (1991) and earlier references therein, which have  
relative errors smaller than 2\% in  mass, less than 1\% in radius, 
less than 2\% for temperature and less than 25\% for [Fe/H]. 
Nearly all the binaries presented in this paper verify the two first criteria, 
but only 32 individual stars reach the target accuracy on {\teff}. 
In addition, we want both components in a system to be within these
limit of a 2\% error in {\teff}, so we have to exclude 4 systems ([18],
[35], [46], [59]). 
The revision proposed by Ribas {\al} (2000) affects some of these 
stars, so for extra precaution we also exclude the systems revised by 
more than 2\% ([7], [20], [24] and [33]). 
Among this reduced subsample of 10 systems\footnote{[2], [3], [4], [6], 
[8], [9], [11], [50], [56] and [57].}, we have to further exclude 
systems showing a bad fit in the HR diagram (see \S\ref{section:badfits})
or those for which models are unable to predict correctly the mass 
and/or radius (\S\ref{section:mass}). The 7  systems selected  
([3], [6], [8], [9], [11], [50], [56]) provide the best
possible tests for the comparison of [Fe/H].

Many attempts have been done in the past to measure [Fe/H] in
double-lined eclipsing binaries. 
 A summary is presented in Table \ref{tab:Zsummary}. It is somewhat
distressing to note that from all the systems listed in Table \ref{tab:Zsummary}
only UX Men [6] and DM Vir [8] belong to the sample selected
above with strict accuracy criteria.

\begin{table*}[h]
\caption{Summary of the observational metallicities 
from spectroscopic analysis (sp.) or photometric 
calibrations in the Str\"omgren (S) or Geneva (G) systems.
As in Table \ref{tab:results}, T (col. 7) refers to the tracks 
used to derive the values $Z_{tracks}$ and [Fe/H]$_{tracks}$ 
given in the 2 last columns. The high metallicity results ($Z_{tracks}$) 
for V1143 Cyg [7] are probably unreliable: applying the revised {\teff} 
from R00 would give a smaller value ($Z$ in the range 0.02-0.03).  
\label{tab:Zsummary} }
\begin{flushleft}
\begin{tabular}{llllllcll}
\hline\noalign{\smallskip}
 System  &   
\multicolumn{1}{c}{ [Fe/H] $^a$}   & 
\multicolumn{1}{c}{  $Z$ $^a$  }     &  
\multicolumn{1}{c}{ Method }       & 
\multicolumn{1}{c}{  Ref.  }       &  
\multicolumn{1}{c}{ $Z_{derived}$ $^b$ }  &
\multicolumn{1}{c}{ T }  &
\multicolumn{1}{c}{ $Z_{tracks}$ $^c$ }  & 
\multicolumn{1}{c}{ [Fe/H]$_{tracks}$ $^{b,c}$ }  \\ 

\noalign{\smallskip}
\hline\noalign{\smallskip}
$\beta$ Aur [27] &   +0.00 $^{d}$     &                         &  sp.      &  T69  & 
     0.017              & 1 & $0.014^{+0.006}_{-0.004}$  & $-0.08^{+0.16}_{-0.15}$ \\
         &     $-$0.03 $^{d}$   &                         &  sp.      &  T69   &  
     0.016              & 3 & $0.066^{+0.020}_{-0.051}$  & $+0.62^{+0.13}_{-1.00}$ \\
         &                 &                                    &                &          &
                        & 4 & $0.017^{+0.002}_{-0.005}$  & $+0.00^{+0.05}_{-0.15}$ \\

%

\noalign{\smallskip}
\hline\noalign{\smallskip}
 RZ Cha [10] & $-$0.02$\pm$0.15 & $0.020^{+0.008}_{-0.006}$   & 
S &  A75   &  $0.016^{+0.007}_{-0.004}$ & 1 &
$0.015^{+0.004}_{-0.003}$   & $-0.05^{+0.10}_{-0.10}$ \\
          &  $-$0.08$\pm$0.05    &           &   S
 & this work$^{\dag}$  & 0.014$\pm$0.002 & 3 & $0.018^{+0.008}_{-0.003}$   &
$+0.03^{+0.16}_{-0.08}$\\          &            &        &         &          
&             & 4 & $0.017^{+0.002}_{-0.004}$  & $+0.00^{+0.05}_{-0.12}$ \\

\noalign{\smallskip}
\hline\noalign{\smallskip}
 V1143 Cyg [7] &  $+$0.08         &                                    &  S &  A87   & 
  0.020                   & 1 & $0.040^{+0.002}_{-0.004}$   & $+0.38^{+0.03}_{-0.05}$ \\
         &                &                                    &   &     & 
                          & 3 & $0.050^{+0.004}_{-0.007}$  & $+0.49^{+0.04}_{-0.08}$ \\
         &                &                                    &          &   & 
                          & 4 & $0.030^{+0.000}_{-0.010}$  & $+0.25^{+0.00}_{-0.18}$ \\

\noalign{\smallskip} 
\hline\noalign{\smallskip}
 TZ For [18] &  +0.1$\pm$0.2 $^{e1}$   &             & sp.         & A91b  
 &  $0.021^{+0.012}_{-0.007}$ & 1 & $0.015^{+0.004}_{-0.001}$  & $-0.05^{+0.10}_{-0.3}$ \\
             & +0.14$\pm$0.15 $^{e2}$ &             & sp.         &  A91b   
 &  $0.023^{+0.009}_{-0.006}$ & 3 & $0.038^{+0.004}_{-0.017}$  & $+0.37^{+0.04}_{-0.27}$ \\
           & +0.1$\pm$0.15 $^{e3}$  & $0.021^{+0.009}_{-0.006}$  $^{f}$  &         &  A91b  
 &  $0.021^{+0.008}_{-0.006}$ & 4 & $0.016^{+0.014}_{-0.002}$  & $-0.03^{+0.28}_{-0.06}$ \\

\noalign{\smallskip}
\hline\noalign{\smallskip}
 UX Men [6] &  $-$0.15         &                                    &  S &  CG76  & 
  0.012                   & 1 & 0.031$\pm$0.009            & $+0.27^{+0.12}_{-0.15}$ \\
         &  +0.04$\pm$0.10 &  0.019$\pm$0.004 $^g$               &  sp.       &  A89   & 
$0.018^{+0.005}_{-0.004}$ & 2 & 0.032$\pm$0.008           & $+0.29^{+0.10}_{-0.13}$ \\
         & $-$0.05$\pm$0.15 &                                    &  S &  A89   & 
$0.015^{+0.006}_{-0.004}$ & 3 & $0.034^{+0.010}_{-0.009}$  & $+0.31^{+0.12}_{-0.13}$ \\
         &  $-$0.04         &                                    &  G        &  KN90  & 
 0.015                    & 4 & $0.027^{+0.003}_{-0.007}$  & $+0.21^{+0.05}_{-0.04}$ \\
         &  0.1           &  0.024                                  &         &  T97  & 
                          &  &     &  \\

\noalign{\smallskip}
\hline\noalign{\smallskip}
 AI Phe [5] &  +0.17$\pm$0.20    &                                    &  S &  VH85  & 
$0.025^{+0.014}_{-0.009}$ & 1 & $0.010^{+0.001}_{-0.002}$  &  $-0.23^{+0.04}_{-0.10}$ \\
            &  $-$0.14$\pm$0.1   &  0.012$\pm$0.003                   &  sp.      &  A88   & 
$0.012^{+0.003}_{-0.002}$ & 2 & $0.009^{+0.004}_{-0.001}$  &  $-0.28^{+0.16}_{-0.05}$ \\
            &  $-$0.1     &  0.012                         &                &  T97      & 
0.014                     & 3 & $0.008^{+0.003}_{-0.000}$  &  $-0.33^{+0.14}_{-0.00}$ \\
            &             &                                &                &          & 
                          & 4 & 0.011$\pm$0.001            &  $-$0.19$\pm$0.04        \\

\noalign{\smallskip} 
\hline\noalign{\smallskip}
 PV Pup [12] & $-$0.06  &   0.017   & S &  VA84  &  
 0.015       & 1 & $\geq$ 0.024         & $\geq$ 0.16  \\
             &          &           &               &          &                 
             & 3 & $0.040^{+0.032}_{-0.012}$   & $+0.39^{+0.28}_{-0.16}$ \\
             &          &           &               &          &       
             & 4 & $\geq$ 0.021                & $\geq$ 0.10 \\

\noalign{\smallskip}
\hline\noalign{\smallskip}
 V1647 Sgr [22] &    &    0.02-0.04    &  ?    &  AG85   & 
0.07-0.39 & 1 & $0.017^{+0.006}_{-0.003}$  & $+0.00^{+0.13}_{-0.09}$ \\
          &   &            &                   &           & 
          & 3 & $0.014^{+0.019}_{-0.002}$   & $-0.09^{+0.38}_{-0.06}$ \\
          &   &            &                   &           & 
          & 4 & $0.016^{+0.004}_{-0.002}$  & $-0.03^{+0.10}_{-0.06}$ \\

\noalign{\smallskip}
\hline\noalign{\smallskip}
 DM Vir [8] & 0.12$\pm$0.12    &    0.023$\pm$0.006     &  S 
 &  A84b   &  $0.022^{+0.007}_{-0.005}$  & 1 & $0.034^{+0.021}_{-0.011}$
 & $+0.31^{+0.22}_{-0.17}$ \\
            &  0.13$\pm$0.05   &                        &  S  
 & this work$^{\dag}$ &  $0.023^{+0.002}_{-0.003}$ & 3 &
$0.041^{+0.022}_{-0.013}$  & $+0.40^{+0.16}_{-0.19}$ \\
        &          &  &               &             &                 
      & 4 & $0.030^{+0.000}_{-0.011}$  & $+0.25^{+0.00}_{-0.20}$ \\

\noalign{\smallskip}
\hline
\noalign{\smallskip}
\noalign{\smallskip}
\end{tabular}
\\
$^{a}$
Values of $Z$ as given in the papers of column 5 (Ref.).
$^{b}$
Adopting $Z_{\odot} =$ 0.017  and using Eq. \ref{eqn:fit2pagel}.
$^{c}$ 
1 $\sigma$ confidence interval, fitting both components except 
for TZ For [18]: fitting only TZ For B with the CG92 tracks (T$=$4).
$^{d}$ 
[Fe/H] = 0.00 for the binary and [Fe/H] = $-$ 0.03 for the secondary
component.
$^{e1}$
Mc Donald observations (A91b).
$^{e2}$
ESO observations (A91b).
$^{e3}$
Adopted mean by A91b.
$^{f}$ 
Assuming  $Z_{\odot} =$ 0.0169 (VandenBerg 1985).
$^{g}$ 
Assuming  $Z_{\odot} =$ 0.0189 (Maeder \& Meynet 1988, 1989).
$^{\dag}$ Updade inferred from [Fe/H] $=$ [Fe/H]$_{\rm Hyades}$ $-$ 
[Fe/H]$_{\Delta m_1}$, by adopting [Fe/H]$_{\rm Hyades}$$=$0.14$\pm$0.05 
(Perryman {\al}, 1998). 
\\ \end{flushleft}
\end{table*}

An inspection of Table \ref{tab:Zsummary} shows that the derived and observed 
metallicities are not always in  perfect agreement but are consistent 
within the quoted error bars. 
The most significant discrepancy appears for the system UX Men [6], where 
the theoretical models predict too large  
$Z$-values\footnote{Even if Ribas {\al} (2000) mention that their solution is 
unreliable (probably due to the $Z$$=$0.03 upper limit of the
models they used),  it is worth noting that they also obtain an extrapolated large
metallicity ($Z$$=$0.039). This is the result we would have obtained assuming the 
lower {\teff}s they adopt.}, even if they are 
consistent at the 1-$\sigma$ level as shown in Fig.~\ref{fig:ux.men}. 
Whilst the disagreement may well come from the physical ingredients assumed in the tracks, 
we note that larger effective temperatures (by $\sim$100-200 K) for each component 
would give a good agreement. Assuming the [Fe/H] value from spectroscopy 
(Andersen {\al}, 1989) and E(b$-$y)$=$0.07 (the colour excess derived by 
Lastennet, Cuisinier \& Lejeune, 2002), the (b$-$y)-{\teff} calibration from the 
BaSeL models points to an increase by $\sim$250 K and 200 K 
for UX Men A and B respectively\footnote{These results are sensitive to the 
reddening. Assuming E(b$-$y)$=$0.04 would give an increase of about 70 K 
for both components.}. 
It is then likely that the most significant discrepancy shown in 
Table \ref{tab:Zsummary} might be accounted for by the {\teff} scaling.

\begin{table*}[h]
\caption{New Z determinations from photometric calibrations: BaSeL 
models (Lastennet {\al} 1999a) and R00 (Ribas {\al} 2000), both
based  on the Str\"omgren photometry of individual stars. 
The $Z$-values are derived according to the correction presented in \S\ref{section:Z}. 
We exclude from these studies the results based on components with 
atmospheric anomalies (mainly Am stars). 
The column ``Notes'' gives the component(s) used (A for primary, B for secondary 
component).
\label{tab:ZBaSeL} } 
\begin{flushleft}
\begin{center}
\begin{tabular}{rrrrlrrl}
\hline\noalign{\smallskip}
 System          & & \multicolumn{3}{c}{BaSeL}                  &       \multicolumn{3}{c}{R00}    \\ 
                 & & [Fe/H]            & $Z$ $^{\dag}$ &  Note  & [Fe/H]   & $Z$ $^{\dag}$    &  Note    \\
\noalign{\smallskip}
\hline\noalign{\smallskip}
 BW Aqr & [9]      & $-$0.20$\pm$0.25 & 0.011$^{+0.008}_{-0.004}$ & A \& B  &         &       &     \\
 AR Aur & [48]     & $>+$0.00          & $>$0.017      & A       &         &       &     \\
 GZ CMa & [23]     & $<+$0.00          & $<$0.017      & B       &         &       &      \\
 EM Car & [45]     & $<+$0.10          & $<$0.021      & B       &         &       &      \\
 YZ Cas & [25]     & no constraints    &               & B       & $+$0.03 & 0.018 & B     \\
 WX Cep$^{a}$ & [30] & $>+$0.05        & $>$0.019      & A       & $-$0.03 & 0.016 & A \& B  \\
 CW Cep & [43]     & $<-$0.15          & $<$0.012      & A       &         &       &     \\
 RZ Cha & [10]     & [$-$0.50,0.00]    & [0.006,0.017] & A \& B  &         &       &   \\
 V380 Cyg$^{b}$    &                   &               &         &         & $-$0.03 & 0.016 &   \\
 HS Hya$^{b}$      &                   &               &         &         & $-$0.17 & 0.012 &   \\
 KW Hya & [15]     & $+$0.00$\pm$0.10  & 0.017$\pm$0.004  & B       &         &       &   \\
 GG Lup & [36]     & $<-$0.20          & $<$0.011      & A       &         &       &   \\
 TZ Men & [29]     & $-$0.20$\pm$0.40    &  0.011$^{+0.016}_{-0.006}$  & A       & $-$0.03 & 0.016 & B   \\
 V451 Oph & [31]   & $<+$0.10          & $<$0.021      & B       &         &       &    \\
 V1031 Ori & [28]  & $<+$0.25          & $<$0.029      & B       & $-$0.25 & 0.010 & A \& B \\
 IQ Per & [33]     & $<+$0.25          & $<$0.029      & B       &         &       &  \\
 AI Phe & [5]      & [$-$0.35,$-$0.15] & [0.008,0.012] & B       &         &       &  \\
 $\zeta$ Phe & [35]& $<+$0.20          & $<$0.027      & B       &         &       &  \\
 PV Pup & [12]     & $<-$0.40          & $<$0.008      & A       &         &       &  \\
\noalign{\smallskip}
\hline
\noalign{\smallskip}
\noalign{\smallskip}
\end{tabular}
\end{center} 
$^{\dag}$ This work, assuming $Z_{\odot}$$=$0.017 and $X_{\odot}$$=$0.713
(Grevesse, 1997, priv. comm.).
$^{a}$ According to L99a (cf. their Tab. 2 and Fig. 3). 
$^{b}$ Spectrophotometric determination, see Ribas {\al} (2000) 
and references therein. 
\\
\end{flushleft}
\end{table*}

Due to the intrinsic difficulty to derive  metallicities from spectroscopic analyses, 
photometric calibrations can be very helpful. 
Lastennet {\al} (1999a) presented some metallicity-dependent 
temperatures determinations of 20 EBs from Str\"omgren synthetic photometry 
(BaSeL models), and Table 5 of Ribas {\al} (2000) give
determinations from other photometric and spectrophotometric methods.  
We exclude from these studies the results based on components with 
atmospheric anomalies (mainly Am stars) and present a summary
in Table~\ref{tab:ZBaSeL}. 
There are only 3 systems in common: \object{WX Cep} [30], \object{TZ
Men} [29] and 
\object{V1031 Ori} [28]. The agreement is good for [28] and [29] (mainly 
because the BaSeL results are weakly constraining for these systems), 
but not for WX Cep [30]. However, inspection of Fig. 4 of Lastennet 
{\al} (1999a) shows that the BaSeL models are consistent with the R00 result 
at the 3-$\sigma$ level for this system. \\ 
Another special case is the binary GZ
CMa [23] whose primary component is an Am star. 
As we mentioned in \S\ref{section:temperature}, the difference between
the revised R00 {\teff}s and the A91 {\teff}s which we adopt for this 
system is larger than 4\%, so we consider our results
([23] in Tab.~\ref{tab:results}) to be less reliable.  
This gives rise to different interpretations even if our results are 
marginally consistent with their lower limit: we predict
(Tab.~\ref{tab:results}) basically solar abundance ($Z$$\sim$0.018), 
while Ribas {\al}
obtain a metal-rich solution ($Z$$\sim$0.032$\pm$0.006). 
As shown in Table~\ref{tab:gzcma}, both the R00 and L99a {\teff}s 
are cooler than the values of A91. Adopting these {\teff}s would give 
a more metal-rich solution in Tab.~\ref{tab:results}.   

\begin{table}[hbt]
\caption[]{Comparison of {\teff} determinations for GZ CMa [23].} 
\label{tab:gzcma}
\begin{flushleft}
\begin{center}
\small
\begin{tabular}{lccc}
\hline\noalign{\smallskip}
        &  \multicolumn{3}{c}{log {\teff} (K)} \\
 GZ CMa & A91 & L99a & R00  \\
  \noalign{\smallskip}
\hline\noalign{\smallskip}
 A &  3.945$\pm$0.017 & 3.928$\pm$0.008 & 3.927$\pm$0.017 \\
 B &  3.931$\pm$0.017 & 3.922$\pm$0.005 & 3.914$\pm$0.017 \\
\noalign{\smallskip}  
\hline
\noalign{\smallskip}
\end{tabular}
\end{center}
\end{flushleft}
\end{table}

 The photometric calibration used by  Ribas {\al} 
strongly supports a large value for the metallicity. Again, the
temperature scale proves to be a critical issue, and yet 
this comparison may not be entirely relevant because the 
photometrically-derived $Z$ is inferred from the primary 
component which is an Am star. 
An estimation from the secondary component would be far more secure, 
and the BaSeL models (Tab.~\ref{tab:ZBaSeL}) indicate a strict
upper limit of $Z<$0.017. 
There is clearly an anomaly: on the one hand, a metal-rich solution 
is obtained when adopting the most recent revised {\teff}s, and on 
the other hand photometric estimates from the secondary component yield a 
solar value. 

This discussion is relevant for the problem of the chemical enrichment ratio 
$\Delta$Y/$\Delta$Z carefully derived by Ribas {\al} (2000), because 
GZ CMa is the only binary of their sample at $Z$$>$0.026 (cf. their 
Figure 7). 
Since we suggest to shift the metallicity of GZ CMa to solar values 
(or even slightly less), 
the data points shown in their Fig. 7 would cluster even more. 
Although this would not change much their linear relationship between initial
metal and helium abundances,  the $Z$ range would be much reduced  and
the results would be less constraining for chemical evolution.  
A detailed spectroscopic analysis of GZ CMa, 
and particularly of the secondary component, a (normal) A-type star
is much needed to resolve this issue.

\subsection{On the relative and absolute dating of stars} 

Since the ages derived by the analyses done here seem to agree with 
each other and with other determinations (\S\ref{section:comparison})
and that, in addition, the predicted metallicities are roughly
consistent with their observed values (\S \ref{section:feh}), we may
consider that at least the relative dating of these stars is well 
rooted.

This has reassuring consequences for the chemical evolution of the
Galactic disc. 
The detailed study made by Edvardsson {\al} (1993) presented precise 
photometric and spectroscopic data for a sample of 189 field stars, 
and showed that there is no tight age-metallicity relation. 
This was recently confirmed with a larger sample (Feltzing {\al} 2001): 
at a given age, the scatter in metallicity is very large.  

In this paper we derived simultaneous metallicity and age estimates 
(listed in Table \ref{tab:results}) for 60 eclipsing binaries, 
most of them being representative of single stars.  Of course this
sample is highly biased and cannot be used for chemical evolution
studies, but it is interesting to see, given the very precise dating
that is achieved by this analysis, whether these stars follow the
same trend. 
The ages derived by Edvardsson {\al} from isochrone fitting 
range from about $\tau = \log t =$ 9.2 to 10.25. 
We have no stars older than $\log t =$10.25, the oldest being 
$\sim$9.5 Gyrs old (HS Aur [2]), but many more younger than their
limit. 
After excluding stars with  weak constraints on their age and/or $Z$, 
well-fitted stars with ages between 10$^8$ and 10$^9$ yrs present
a spread in $Z$  between 
0.05 and 0.01, i.e [Fe/H] between $\sim$$+$0.49 and $-$0.23 
(or even below if the low metallicity of VV Pyx [19] is
real, see \S\ref{section:vv.pyx}). This trend suggests that the 
dispersion observed by Edvardsson {\al} may well continue at younger ages.

\section{Conclusions}
We have developed a method to derive simultaneously ages and metallicities 
from binary stars and single stars. Instead of giving one solution (i.e. one 
isochrone defined by a unique $Z$-$\tau$ pair), 
this method provides a {\it set} of possible  
solutions, as defined by $\chi^2$ contours. Within them, Vogt's theorem
is (operationally) violated: stars of different masses and
chemical compositions lie on the same position in the HR diagram.
We have applied this method to the largest sample of well-detached binary 
systems for which most stellar parameters are extremely well known, 
to test the power of the method
which could be applied to  samples with looser constraints.  In
addition the method allows to test the predictive power of stellar
evolution
theory, by taking only two of the well-constrained quantities, and
predicting the others. The choice of luminosity and effective
temperature as primary quantities turned out to be the best one, 
with accurate determination of the parameters of the isochrone and 
also with excellent predictions for masses and radii (dispersion in their
relative errors smaller than 2\% and 4\%, respectively). 
The validation of the method 
through these systems makes it possible to apply it to other, less 
constrained systems such as visual binaries, for which catalogues are 
growing much faster. 

 The critical issue of systematic errors due to the use of a particular
set of evolutionary tracks turned out not to be very important, at
least for the sample used here. While the absolute results can be different 
in individual cases, all three sets of tracks used here 
(Geneva, Padova and Granada) yield - in a statistical sense - consistent 
values and similar age-metallicity degeneracies.

We have made a detailed  comparison of our results with previous
 studies whenever possible.   
The main inconsistencies found and discussed in this paper are 
the following: 
\\
a) We have shown that the {\teff}s of YY Gem [1a] may still need to be 
revised to slightly cooler temperatures.  
The Brocato {\al} (1998) models are not able to fit the CM Dra [1b]
components (both masses below 0.25 {\MSun}) in the HR diagram for the
expected solar metallicity, while the Baraffe {\al} (1995) models
succeed. 
\\
b) We obtain (bad) fits which suggest large metallicities for 4
 systems
which have  at 
least one component in the 0.7-1.1 {\MSun} mass range. This  
supports the results found by Popper (1997) and  Clausen {\al} (1999) 
which indicated this peculiarity. 
Whilst some tentative explanations are discussed ({\teff}-scaling, activity, mass transfer, 
etc...), it appears that no Geneva isochrone can fit simultaneously both 
components of HS Aur [2] in the HR diagram.   
The best fit obtained with Padova tracks yields  
a (questionable) metal rich solution, about twice solar. This provides
 an 
ideal observational test  to disentangle these sets of tracks. 
\\
c) If the metal content of DM Vir [8] is confirmed to be similar to
the one of the Hyades, the Padova models would fail to fit
the system, while the Geneva and Granada ones would succeed.
This system  is another test  case to check 
stellar evolution theory in stars of masses of about 1.4 \MSun. 
\\
d) For RZ Cha [10], we have updated the not very constraining [Fe/H] value
obtained by J{\o}rgensen \& Gyldenkerne (1975) (the Hyades metallicity
they adopted was an overestimation), giving a very good agreement with
the Geneva tracks, and a good agreement with the Granada and Padova tracks.
\\
e) The three sets of theoretical tracks suggest a metal rich solution
for the binary PV Pup [12] ($\simeq$ 1.55 \MSun\ for both components)
which is in disagreement with indications of a solar metallicity
($Z$$\simeq$0.017, Lastennet {\al} 1999a), even if the 
solar metallicity remains possible at the 2-$\sigma$ level. 
The new {\teff} determinations of Lastennet {\al} (1999a) provides 
a slightly better agreement with solar metallicity values.      
\\
f) For the $\sim$2 {\MSun} evolved components of TZ For [18],
the agreement of a metallicity around the spectroscopic value (nearly solar) 
is satisfactory for the Geneva tracks, but seems  difficult 
to reconcile with the Padova models. 
In spite of the better agreement obtained with the Geneva models, 
we have shown that if the mass of the tracks is fixed to the  
measured value of TZ For B, none of the tracks succeeds in fitting 
this star, independently of  the metallicity used. 
Alternatively, if the metallicity is fixed to its spectroscopic value, 
the mass of the tracks must be reduced by 5$\sigma$ (0.11 {\MSun}) 
to reproduce the mass of TZ For B. 
\\
g) If the lower limit of a solar metallicity is confirmed observationally 
for V1647 Sgr [22], the Granada tracks may not be able to fit the
system, although the effects of rotation must be assessed first for
 an unambiguous test.
\\
h) The disagreement between determinations of metallicity in GZ CMa
[23] may bear some consequences on the chemical enrichment ratio 
$\Delta$Y/$\Delta$Z carefully derived by Ribas {\al} (2000). 
While they find that GZ CMa is the more metal rich binary of their 
sample ($Z$$=$0.032), the BaSeL models suggest a solar or sub-solar 
composition (L99a).
\\
i) Our analysis predict that the heavy element abundance may 
be small ($Z$$<$0.010) for systems  with components of masses larger than 
$\sim$ 15 {\MSun}. The photometrically derived value for EM Car [45] supports 
this pattern.
\\   

We summarize the general results obtained as follows.
 \\
1. There is a wide age--metallicity degeneracy in the fitting of
double-lined eclipsing binaries, as well as for their individual
components and therefore also for stars in the field, no matter
how accurate the measures are. 
\\
2. This degeneracy limits, in practice, the empirical validity of
Vogt's theorem, assuming its premises give sufficient and necessary
conditions for the existence of unique solutions to the equations
of stellar structure.
\\              
3. The systematic errors introduced by the use of a given set of tracks
have been assessed by comparing the results obtained for three different
sets (Geneva, Padova, Granada). The systematics are rather small, and in
most cases the solutions are statistically compatible at the 1$\sigma$
level. This may be due, in part, to their similar
 metallicity-correlated He abundance. 
\\            
4. In spite of the high precision measures that lead to very small errors
in the astrophysical parameters of these systems, the lack of metallicity
determinations limits a precise measure (0.1 dex) of the ages,
except in a few odd cases. 
Since photometric determinations provide useful constraints for 
only few systems, 
the spectroscopic measure of the metallicity in the
components of these systems is essential for this technique to be an
efficient test of stellar evolution.\\  
 
Another way to bypass the difficulty to determine the metallicity 
of a particular binary would be to select binaries which belong to 
star clusters with a very well measured metallicity (as proposed by 
Clausen \& Gim\'enez 1996; see Lastennet {\al} 1999b for an application to
Hyades binaries,  and Lastennet, Valls-Gabaud, Oblak 2000). 
This is a promising area given the difficulties to derive accurate 
metallicities from detailed spectroscopic analyses of some eclipsing binaries. 
\\

\begin{acknowledgements}
EL gratefully acknowledges financial support from the FCT 
({\it Bolsa de Investiga\c c\~ao} BPD/5556/2001), CNPq and CDS. 
We are grateful to A. Claret for providing some of his models in electronic
form, to F. James to discussions on the advanced use of {\tt MINUIT},
and to C. Corbally, J. Fernandes and B. Nordstr\"om for sending preprints in advance
of publication.
We thank J.V. Clausen for useful comments on an earlier version of this paper  
and the referee, G. Torres, for constructive comments and detailed
 suggestions.  
This research has made use of the SIMBAD database operated at CDS, Strasbourg, France, 
and of NASA's Astrophysics Data System Abstract Service. 
\end{acknowledgements}          

\newpage  
%
 \tabcolsep=0.8mm 
 \begin{table*}[t]
 \small 
 \begin{flushleft} 
    \caption[]{Ages and metallicities for the best fit isochrones to the sample of 58 double-lined eclipsing
binaries for the different sets of tracks used$^\dag$. 
 Systems YY Gem [1a] and 
CM Dra [1b] are discussed in \S\ref{section:yygem}. 
T refers to the tracks used to derive the fit : (1) Geneva with overshooting,
(2) Geneva without overshooting, (3) Padova, and (4) CG92. Maximum and
minimum values are given at 68\% c.l. ($\pm$1 $\sigma$) for the fits of individual
components, while the values for the combined system are at $\pm n \sigma$.
Ages consistent with the ZAMS are indicated by 'ZA', that is, $t=t_{\rm ZAMS}$ in
Eq. \ref{eq:age}.} 
    \label{tab:results} 
    \begin{flushleft}
  \begin{tabular}{|r|l|c|rrr|lll|rrr|lll|rrr|lll|c|}           \hline  
 \multicolumn{3}{|c}{  } & 
 \multicolumn{6}{|c|}{Component A } & 
 \multicolumn{6}{c|}{Component B } & 
 \multicolumn{7}{c|}{System A$+$B           \strutup \strutdown} \\  \cline{4-22}
 \multicolumn{1}{|c}{ \raisebox{1.5ex}[0pt]{\#} } & 
 \multicolumn{1}{c}{ \raisebox{1.5ex}[0pt]{Name} } & 
 \multicolumn{1}{c|}{ \raisebox{1.5ex}[0pt]{T} } & 
 \multicolumn{1}{|c}{ $\tau$ } & 
 \multicolumn{1}{c}{ $\tau_{\rm min}$ } & 
 \multicolumn{1}{c}{ $\tau_{\rm max} $ } & 
 \multicolumn{1}{c}{$ Z $}  & 
 \multicolumn{1}{c}{$ Z_{\rm min} $}  & 
 \multicolumn{1}{c|}{$ Z_{\rm max} $}  & 
 \multicolumn{1}{|c}{ $\tau$} & 
 \multicolumn{1}{c}{ $\tau_{\rm min}$ } & 
 \multicolumn{1}{c}{ $\tau_{\rm max} $ } & 
 \multicolumn{1}{c}{$ Z $}  & 
 \multicolumn{1}{c}{$ Z_{\rm min} $}  & 
 \multicolumn{1}{c|}{$ Z_{\rm max} $}  & 
 \multicolumn{1}{|c}{ $\tau$} & 
 \multicolumn{1}{c}{ $\tau_{\rm min}$ } & 
 \multicolumn{1}{c}{ $\tau_{\rm max} $ } & 
 \multicolumn{1}{c}{$ Z $}  & 
 \multicolumn{1}{c}{$ Z_{\rm min} $}  & 
 \multicolumn{1}{c}{$ Z_{\rm max} $}  & 
  $n$  \strutdown \strutup \\ \hline 
  2 & HS Aur       & 1 & 10.06 &  9.95 & 10.15 &  0.040 &  0.020 &  0.040 &  9.86 &  9.24 & 10.10 &  0.040 &  0.017 &  0.040 &  9.98 &  9.94 & 10.02 &  0.040 &  0.035 &  0.040\strutup & 3  \\ 
        &   & 2 & 10.13 &  9.93 & 10.19 &  0.040 &  0.020 &  0.040 &  9.96 &  9.28 & 10.06 &  0.039 &  0.017 &  0.040 &  9.98 &  9.92 & 10.02 &  0.040 &  0.035 &  0.040 & 3 \\ 
        &   & 3 & 10.05 &  7.43 & 10.14 &  0.049 &  0.022 &  0.100 &  8.64 &  ZA    & 10.05 &  0.073 &  0.016 &  0.100 &  9.98 &  9.87 & 10.06 &  0.049 &  0.021 &  0.050 \strutdown  & 3 \\  \hline  
  3 & EW Ori       & 1 &  8.39 &  ZA    &  9.15 &  0.040 &  0.024 &  0.040 &  8.49 &  ZA    &  9.22 &  0.040 &  0.028 &  0.040 &  8.42 &  ZA    &  8.79 &  0.040 &  0.036 &  0.040\strutup & 1  \\ 
        &   & 2 &  7.35 &  ZA    &  9.11 &  0.040 &  0.024 &  0.040 &  7.27 &  ZA    &  9.15 &  0.040 &  0.025 &  0.040 &  7.30 &  ZA    &  8.75 &  0.040 &  0.032 &  0.040 & 1 \\ 
        &   & 3 &  ZA    &  ZA    &  9.00 &  0.051 &  0.023 &  0.069 &  ZA    &  ZA    &  9.14 &  0.053 &  0.024 &  0.079 &  ZA    &  ZA    &  8.80 &  0.051 &  0.032 &  0.065 & 1 \\ 
        &   & 4 &  5.91 &  ZA    &  9.10 &  0.030 &  0.017 &  0.030 &  5.91 &  ZA    &  9.16 &  0.030 &  0.018 &  0.030 &  9.60 &  ZA    &  9.60 &  0.030 &  0.021 &  0.030 \strutdown  & 1 \\  \hline  
  4 & FL Lyr       & 1 &  9.36 &  8.77 &  9.55 &  0.040 &  0.020 &  0.040 &  9.43 &  ZA    &  9.75 &  0.040 &  0.021 &  0.040 &  9.38 &  9.19 &  9.50 &  0.040 &  0.027 &  0.040\strutup & 1  \\ 
        &   & 2 &  9.37 &  8.67 &  9.55 &  0.040 &  0.021 &  0.040 &  9.55 &  ZA    &  9.78 &  0.040 &  0.018 &  0.040 &  9.40 &  9.09 &  9.50 &  0.040 &  0.028 &  0.040 & 1 \\ 
        &   & 3 &  7.73 &  ZA    &  9.50 &  0.062 &  0.020 &  0.086 &  3.79 &  ZA    &  9.93 &  0.081 &  0.029 &  0.100 &  7.69 &  ZA    &  9.41 &  0.065 &  0.021 &  0.084 \strutdown  & 1 \\  \hline  
  5 & AI Phe       & 1 &  9.66 &  9.64 &  9.76 &  0.009 &  0.009 &  0.028 &  9.60 &  9.55 &  9.68 &  0.008 &  0.004 &  0.012 &  9.66 &  9.64 &  9.67 &  0.010 &  0.008 &  0.011\strutup & 1  \\ 
        &   & 2 &  9.71 &  9.60 &  9.73 &  0.025 &  0.008 &  0.040 &  9.60 &  9.59 &  9.66 &  0.006 &  0.006 &  0.013 &  9.64 &  9.61 &  9.65 &  0.009 &  0.008 &  0.013 & 1 \\ 
        &   & 3 &  9.68 &  9.67 &  9.70 &  0.027 &  0.022 &  0.035 &  9.60 &  9.54 &  9.70 &  0.007 &  0.006 &  0.013 &  9.60 &  9.59 &  9.63 &  0.008 &  0.008 &  0.011 & 1 \\ 
        &   & 4 &  9.67 &  9.64 &  9.68 &  0.011 &  0.010 &  0.015 &  9.65 &  9.63 &  9.68 &  0.010 &  0.010 &  0.013 &  9.66 &  9.64 &  9.67 &  0.011 &  0.010 &  0.012 \strutdown  & 1 \\  \hline  
  6 & UX Men       & 1 &  9.42 &  8.90 &  9.49 &  0.033 &  0.021 &  0.060 &  9.44 &  8.80 &  9.49 &  0.030 &  0.018 &  0.062 &  9.45 &  8.95 &  9.56 &  0.031 &  0.022 &  0.056\strutup & 1  \\ 
        &   & 2 &  9.43 &  9.15 &  9.55 &  0.031 &  0.021 &  0.040 &  9.47 &  9.10 &  9.58 &  0.027 &  0.019 &  0.040 &  9.45 &  9.29 &  9.52 &  0.032 &  0.024 &  0.040 & 1 \\ 
        &   & 3 &  8.83 &  6.86 &  9.20 &  0.060 &  0.024 &  0.092 &  7.97 &  7.33 &  9.33 &  0.069 &  0.020 &  0.087 &  9.26 &  9.15 &  9.38 &  0.034 &  0.025 &  0.044 & 1 \\ 
        &   & 4 &  9.38 &  9.08 &  9.57 &  0.030 &  0.017 &  0.030 &  9.44 &  9.12 &  9.60 &  0.024 &  0.016 &  0.030 &  9.40 &  9.27 &  9.51 &  0.027 &  0.020 &  0.030 \strutdown  & 1 \\  \hline    
       \end{tabular} 
     \end{flushleft} 
   \end{flushleft} 
$^{\dag}$ The full table
is only available in electronic form. The sample printed here is for
orientation on contents and format only.
 \end{table*}




\end{document}